%% file: Ramification.tex
\documentclass[11pt,letterpaper]{JHEP3}
\pdfoutput=1

\usepackage{latexsym,amsmath,amsfonts,amssymb}
\usepackage{graphicx}
\usepackage{subfig}
\usepackage{url}

\usepackage{enumerate}
\usepackage{youngtab}

\title{Ramification Points of Seiberg-Witten Curves}
\author{Chan Y. Park\\
	\href{mailto:splendid@caltech.edu}{\tt splendid@caltech.edu}\\
	\textit{California Institute of Technology}\\
	\textit{Pasadena, CA 91125, USA}}

\abstract{When the Seiberg-Witten curve of a four-dimensional $\mathcal{N}=2$ supersymmetric gauge theory wraps a Riemann surface as a multi-sheeted cover, a topological constraint requires that in general the curve should develop ramification points. We show that, while some of the branch points of the covering map can be identified with the punctures that appear in the work of Gaiotto, the ramification points give us additional branch points whose locations on the Riemann surface can have dependence not only on gauge coupling parameters but on Coulomb branch parameters and mass parameters of the theory. We describe how these branch points can help us to understand interesting physics in various limits of the parameters, including Argyres-Seiberg duality and Argyres-Douglas fixed points.
} 

\keywords{Supersymmetric Gauge Theory, Duality in Gauge Field Theories, M-Theory, Differential and Algebraic Geometry}

\preprint{CALT-68-2839}

\begin{document}
  


\section{Introduction}
\label{section:introduction}
In \cite{Witten:1997sc}, it was shown that we can describe the Seiberg-Witten curve \cite{Seiberg:1994rs, Seiberg:1994aj} of a four-dimensional $\mathcal{N}=2$ supersymmetric field theory by a complex algebraic curve with various parameters of the theory as the coefficients of a polynomial that defines the curve. For example, an $\mathcal{N}=2$ supersymmetric gauge theory with gauge group $SU(2)$ and four massless hypermultiplets is a superconformal field theory (SCFT) whose Seiberg-Witten curve $C_{\text{SW}}$ is defined as the zero locus of
\begin{align}
	(t-1)(t-t_1)v^2 - ut, \label{eq:SU(2) N_f=4 SW-curve}
\end{align}
where $(t,v)$ is a coordinate of $\mathbb{C}^* \times \mathbb{C}$ that contains $C_{\text{SW}}$, $t_1$ is related to the marginal gauge coupling parameter of the theory, and $u$ is the Coulomb branch parameter. 

In \cite{Gaiotto:2009we}, Gaiotto showed that by wrapping $N$ M5-branes over a Riemann surface with punctures, we can get a four-dimensional gauge theory with $\mathcal{N}=2$ supersymmetry. The locations of the punctures on the Riemann surface describe the gauge coupling parameters of the theory, and each puncture is characterized with a Young tableau of $N$ boxes. 

In much the same spirit, we can think of a Seiberg-Witten curve $C_{\text{SW}}$ wrapping a Riemann surface $C_{\text{B}}$ in the following way. For $C_{\text{SW}}$ that Eq. (\ref{eq:SU(2) N_f=4 SW-curve}) defines, consider $t$ as a coordinate for a base $C_{\text{B}}$, which is a Riemann sphere in this case, and $v$ as a coordinate normal to $C_{\text{B}}$. Then a projection $(v,t) \mapsto t$ gives us the required covering map from $C_{\text{SW}}$ to $C_{\text{B}}$. When we generalize this geometric picture to the case of $C_{\text{SW}}$ wrapping $C_{\text{B}}$ $N$ times, one natural way of thinking why each puncture has its Young tableau is to consider a puncture as a branch point of the projection $\pi$, which is now an $N$-sheeted covering map from $C_{\text{SW}}$ onto $C_{\text{B}}$. Then the partition associated to the Young tableau of a puncture shows how the branching of the $N$ sheets occurs there. 

Now we can ask a question: for the Seiberg-Witten curve $C_{\text{SW}}$ of a four-dimensional $\mathcal{N}=2$ supersymmetric gauge theory, can we identify every branch point on $C_{\text{B}}$ of the covering map from $C_{\text{SW}}$ to $C_{\text{B}}$ with a puncture of \cite{Gaiotto:2009we}? To answer this question we will investigate several examples, which will lead us to the conclusion that, in addition to the branch points that are identified with the punctures, there are in general other branch points that are not directly related to the punctures. The locations of these additional branch points on $C_{\text{B}}$ are related in general to every parameter of the theory, including not only gauge coupling parameters but Coulomb branch parameters and mass parameters, unlike the punctures whose positions on $C_{\text{B}}$ are characterized by the gauge coupling parameters only. We will illustrate how these branch points can be utilized to explore interesting limits of the various parameters of the theory.

We start in Section \ref{section:SU(2) SCFT} with $SU(2)$ SCFT to explain how the covering map $\pi$ provides the ramification of the Seiberg-Witten curve $C_{\text{SW}}$ of the theory over a Riemann sphere $C_{\text{B}}$. In Section \ref{section:SU(2) times SU(2) SCFT}, we repeat the analysis of Section \ref{section:SU(2) SCFT} to study $SU(2) \times SU(2)$ SCFT, where we find a branch point that is not identified with a puncture of \cite{Gaiotto:2009we}. Its location on $C_{\text{B}}$ depends on the Coulomb branch parameters of the theory, which enables us to investigate how the branch point behaves under various limits of the Coulomb branch parameters. In Section \ref{section:SU(3) SCFT} we study $SU(3)$ SCFT and how the branch points behave under the limit of the Argyres-Seiberg duality \cite{Argyres:2007cn}. In Section \ref{section:SU(3) pure gauge theory}, we extend the analysis to $SU(3$) pure gauge theory that is not a SCFT. There we will see how the branch points help us to identify interesting limits of the Coulomb branch parameters of the theory, the Argyres-Douglas fixed points \cite{Argyres:1995jj}. In Section \ref{section:SU(2) with massive matter} we consider $SU(2)$ gauge theories with massive hypermultiplets and illustrate how mass parameters are incorporated in the geometric description of the ramification of $C_{\text{SW}}$.
Appendices contain the details of the mathematical procedures and the calculations of the main text.


\section{$SU(2)$ SCFT and the ramification of the Seiberg-Witten curve}
\label{section:SU(2) SCFT}

\begin{figure}[ht]
	\begin{center}
		\includegraphics{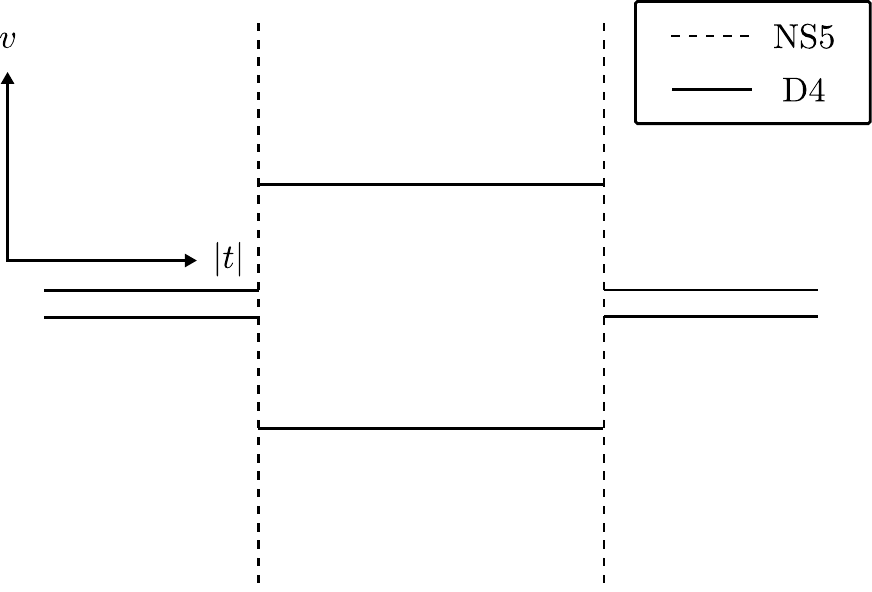}
		\caption{Brane configuration of $SU(2)$ SCFT}
		\label{figure:SU_2_SCFT_brane}
	\end{center}
\end{figure}

The first example is a four-dimensional $\mathcal{N}=2$ superconformal $SU(2)$ gauge theory. The corresponding brane configuration in the type IIA theory \cite{Witten:1997sc} is shown in Figure \ref{figure:SU_2_SCFT_brane}.

After the M-theory lift \cite{Witten:1997sc} this brane system becomes an M5-brane that fills the four dimensional spacetime, where the gauge theory lives, and wraps the Seiberg-Witten curve, which is the zero locus of 
\begin{align}
	f(t,v) = (t-1)(t-t_1)v^2 - ut. \label{eq:C_SW for SU(2) SCFT}
\end{align}
This is a smooth, non-compact Riemann surface in $\mathbb{C}^2$. Note that by construction the following four points
\begin{align*}
	\mathcal{I} = \{(t,v) \in \mathbb{C}^2\ |\ (0, 0),\ (1, \infty),\ (t_1, \infty),\ (\infty, 0)\},
\end{align*}
are not included in $C_{\text{SW}}$. 

It would be preferable if we can find a compact Riemann surface that describes the same physics as $C_{\text{SW}}$. One natural way to compactify $C_{\text{SW}}$ is embedding it into $\mathbb{CP}^2$ to get a compact algebraic curve $\bar{C}_{\text{SW}}$ defined as the zero locus of
\begin{align*}
	F(X,Y,Z) = (X-Z)(X-t_1 Z)Y^2 - u X Z^3,
\end{align*}
which we will call $\bar{C}_{\text{SW}}$. The four points of $\mathcal{I}$ are now mapped to
\begin{align*}
	\{[X, Y, Z] \in \mathbb{CP}^2\ |\ [0,0,1],\ [0,1,0],\ [1,0,0]\}.
\end{align*}
$\bar{C}_{\text{SW}}$ obtained this way is guaranteed to be smooth except at the points we added for the compactification, where it can have singularities \cite{Kirwan}. Indeed $\bar{C}_{\text{SW}}$ is singular at $[0,1,0]$ and $[1,0,0]$,
which implies that $\bar{C}_{\text{SW}}$ is not a Riemann surface. The singularity at $[0,1,0]$ corresponds to having two different tangents there.
The other singularity at $[1,0,0]$ corresponds to a cusp. 

\begin{figure}[t]
	\begin{center}
		\includegraphics{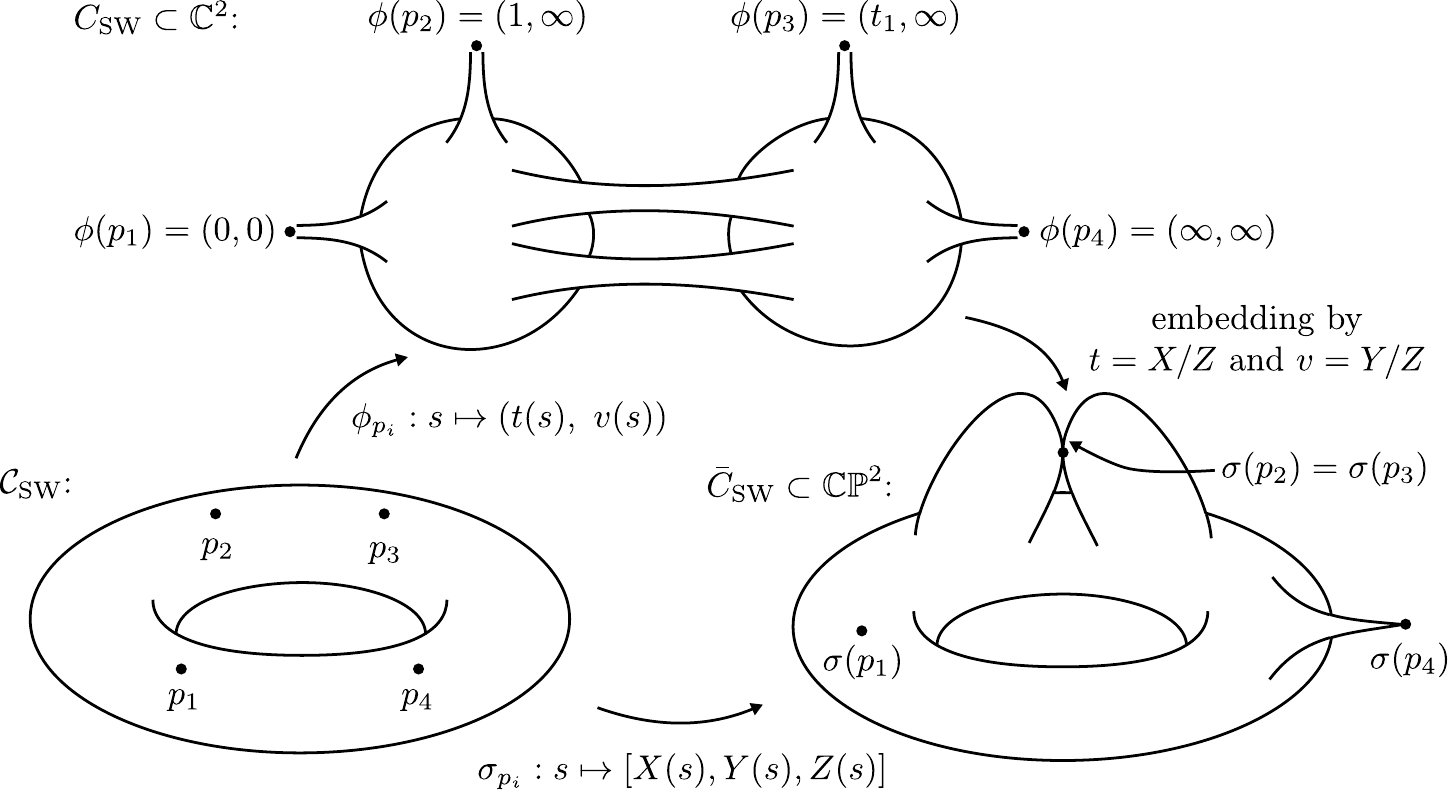}
		\caption{Schematic description of the compactification and the normalization of a Seiberg-Witten curve}
		\label{figure:compactification_and_normalization}
	\end{center}
\end{figure}

Smoothing out a singular algebraic curve to find the corresponding Riemann surface can be done by normalization \cite{Kirwan, Griffiths}. This means finding a smooth Riemann surface $\mathcal{C}_{\text{SW}}$ and a holomorphic map $\sigma : \mathcal{C}_{\text{SW}} \to \bar{C}_{\text{SW}}$. Appendix \ref{appendix:normalization} illustrates how we can get a normalization of a singular curve. After the normalization we can find, for every point $s_i \in \mathcal{C}_{\text{SW}}$, the local normalization map
\begin{align*}
	\sigma_{s_i}: \mathcal{N}_{s_i} \to \mathbb{CP}^2,\ s \mapsto [X(s), Y(s), Z(s)],
\end{align*}
where $s \in \mathbb{C}$ is a local coordinate such that $s_i = 0$. Figure \ref{figure:compactification_and_normalization} illustrates how we get from the noncompact Seiberg-Witten curve $C_{\text{SW}}$ its compactification $\bar{C}_{\text{SW}}$ and the compact Riemann surface $\mathcal{C}_{\text{SW}}$, together the relations among them. Here we use the normalization map $\sigma$ to build a map $\phi: \mathcal{C}_{\text{SW}} \to \{ C_{\text{SW}} \cup \mathcal{I} \}$, whose local description near a point $s_i \in \mathcal{C}_{\text{SW}}$ is
\begin{align*}
	\phi_{s_i}: \mathcal{N}_{s_i} \to \mathbb{C}^2,\ s \mapsto \left(t(s),v(s)\right) = \left(\frac{X(s)}{Z(s)}, \frac{Y(s)}{Z(s)}\right),
\end{align*}
where $s \in \mathbb{C}$ is a local coordinate such that $s_i = 0$.

The compactification of a Seiberg-Witten curve to a Riemann surface is discussed previously in \cite{Witten:1997sc}. It is also mentioned in \cite{Gaiotto:2009hg} from the viewpoint of seeing a Seiberg-Witten curve as a cycle embedded in the cotangent bundle $T^*C_{\text{B}}$ of the base $C_{\text{B}}$. 

Whether $\mathcal{C}_{\text{SW}}$ gives the same physics as $C_{\text{SW}}$ is a challenging question, whose answer will depend on what we mean by ``the same physics.'' For example, it is argued in \cite{Witten:1997sc} and is illustrated with great detail in \cite{Gaiotto:2009hg} that the the low-energy effective theory of an M5-brane wrapping $C_{\text{SW}}$ is described by the Jacobian of $\mathcal{C}_{\text{SW}}$. Extending those arguments is a very intriguing task but we will not try to address it here.

\begin{figure}[ht]
	\begin{center}
		\includegraphics{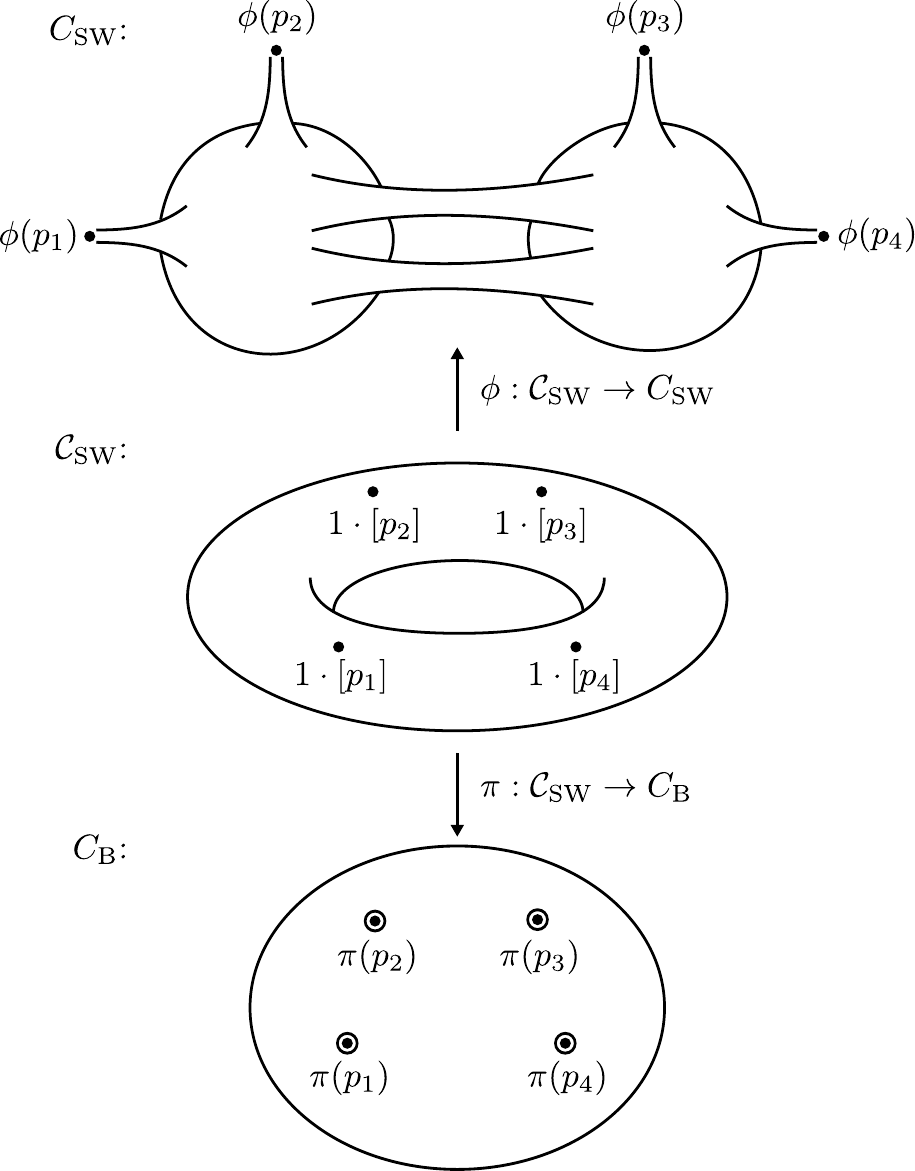}
		\caption{Summary of how to obtain $\mathcal{C}_{\text{SW}}$ and $C_{\text{B}}$ from $C_{\text{SW}}$}
		\label{figure:summary_SU_2_SCFT}
	\end{center}
\end{figure}

Now that we have a smooth Riemann surface $\mathcal{C}_{\text{SW}}$, we want to wrap it over a Riemann surface, $C_{\text{B}}$. Note that for the current example we want $C_{\text{B}}$ to be a Riemann sphere, or $\mathbb{CP}^1$, because the corresponding four-dimensional gauge theory comes from a linear quiver brane configuration \cite{Gaiotto:2009we}. To implement the wrapping, or the projection, from $\mathcal{C}_{\text{SW}}$ to $\mathbb{CP}^1$, we use $\phi$ to define a meromorphic function $\pi$ on $\mathcal{C}_{\text{SW}}$ such that its restriction to the neighborhood of $s_i \in \mathcal{C}_{\text{SW}}$ is
\begin{align*}
	\pi_{s_i}(s) = t(s) = \frac{X(s)}{Z(s)}, 
\end{align*}
where $t(s)$ is the value of the $t$-coordinate of $\{ C_{\text{SW}} \cup \mathcal{I} \}$ at $\phi(s)$ and therefore has the range of $\mathbb{CP}^1$.\footnote{Note that $t:\mathcal{C}_{\text{SW}} \to \mathbb{CP}^1$ is well-defined over $\mathcal{C}_{\text{SW}}$, although $X/Z : \bar{C}_{\text{SW}} \to \mathbb{CP}^1$ is not well-defined at $[0,1,0] \in \bar{C}_{\text{SW}}$ because it maps the point on $\bar{C}_{\text{SW}}$ to two different points on $\mathbb{CP}^1$, $1$ and $t_1$. This ill-definedness arises because we compactify $C_{\text{SW}}$ by embedding it into $\mathbb{CP}^2$, which maps two different points on $C_{\text{SW}}$, $(1,\infty)$ and $(t_1, \infty)$, to one point in $\mathbb{CP}^2$, $[0,1,0]$, and therefore is the artifact of our embedding scheme. Normalization separates the two and resolves this difficulty, after which $t$ is a well-defined function over all $\mathcal{C}_{\text{SW}}$.} This $\pi$ is in general a many-to-one (two-to-one for the current example) mapping, therefore it realizes the required wrapping of $\mathcal{C}_{\text{SW}}$, or its ramification, over $\mathbb{CP}^1$. Figure \ref{figure:summary_SU_2_SCFT} summarizes the whole procedure of getting from $C_{\text{SW}}$ the normalization $\mathcal{C}_{\text{SW}}$ of $\bar{C}_{\text{SW}}$ and finding the ramification of $\mathcal{C}_{\text{SW}}$ over $C_{\text{B}}$.

To analyze the ramification it is convenient to introduce a ramification divisor $R_\pi$ \cite{Griffiths},
\begin{align*}
	R_\pi = \sum_{s \in \mathcal{C}_{\text{SW}}}(\nu_s(\pi) - 1)[s] = \sum_i(\nu_{s_i}(\pi) - 1)[s_i].
\end{align*}
Here $\nu_s(\pi) \in \mathbb{Z}$ is the ramification index of $s \in \mathcal{C}_{\text{SW}}$, $s_i \in \mathcal{C}_{\text{SW}}$ is a point where $\nu_{s_i}(\pi) > 1$, and $[s_i]$ is the corresponding divisor\footnote{A divisor is a formal representation of a complex-one-codimension object, a point in this case.} of $\mathcal{C}_{\text{SW}}$. In colloquial language, having a ramification index $\nu_s(\pi)$ at $s \in \mathcal{C}_{\text{SW}}$ means that $\nu_s(\pi)$ sheets over $C_{\text{B}}$ come together at $\pi(s)$. When $\nu_s(\pi) > 1$ we say $s$ is a ramification point on $\mathcal{C}_{\text{SW}}$, $\pi(s)$ is a branch point on $C_{\text{B}}$, and $\pi : \mathcal{C}_{\text{SW}} \to C_{\text{B}}$ has a ramification at $\pi(s)$.

The Riemann-Hurwitz formula \cite{Griffiths} provides a relation between $\pi$, $R_\pi$, and the genus of $\mathcal{C}_{\text{SW}}$, $g(\mathcal{C}_{\text{SW}})$.
\begin{align}
	\chi_{\mathcal{C}_{\text{SW}}} = \deg(\pi) \cdot \chi_{\mathbb{CP}^1} - \deg(R_\pi) \Leftrightarrow \deg(R_\pi) = 2(g(\mathcal{C}_{\text{SW}}) + \deg(\pi) -1).\label{eq:riemann-hurwitz formula}
\end{align}
Here $\chi_C$ is the Euler characteristic of $C$, and $\deg(\pi)$ is the number of intersections of $\mathcal{C}_{\text{SW}}$ and $\pi^{-1}(t_0)$ for a general $t_0 \in \mathbb{CP}^1$. In the current example where $C_{\text{SW}}$ is the zero locus of Eq. (\ref{eq:C_SW for SU(2) SCFT}), it is easy to see that $\deg(\pi) = 2$ because the equation is quadratic in $v$. Using this Riemann-Hurwitz formula, we can check if we have found all ramification points that are needed to describe the wrapping of $\mathcal{C}_{\text{SW}}$ over $C_{\text{B}}$.

\begin{figure}[ht]
	\begin{center}
		\includegraphics{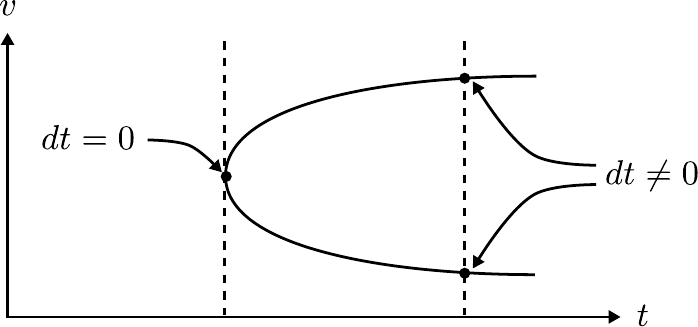}
		\caption{Why a nontrivial ramification occurs at $dt=0$}
		\label{figure:ramification_at_dt_0}
	\end{center}
\end{figure}

What we want to know is where the ramification points of $\mathcal{C}_{\text{SW}}$ are and what ramification indices they have. We will try to guess where they are by investigating every point $s \in \mathcal{C}_{\text{SW}}$ that might have a nontrivial behavior under $\pi$. The candidates of such points are 
\begin{enumerate}[(1)]
	\item $\{p_i \in \mathcal{C}_{\text{SW}}\ |\ \phi(p_i) \in \mathcal{I} \}$,
	\item $\{q_i \in \mathcal{C}_{\text{SW}}\ |\ dt(q_i) = 0\} \Leftrightarrow \{q_i \in \mathcal{C}_{\text{SW}}\ |\ (\partial f / \partial v)(t(q_i),v(q_i)) = 0\}$.
\end{enumerate}
We check the ramification of the points of (1) because at $t(p_i)$ some branches of $v(t)$ meet ``at infinity.''\footnote{This qualification is because it is not true in $t$-coordinate. For example, $\phi(p_1)$ is not at infinity, because the $t$-coordinate is in fact the exponentiation of the spacetime coordinate, $t = \exp(-(x^6 + i x^{10}))$ \cite{Witten:1997sc}. By ``at infinity'' we imply that the point is at infinity of the ten- or eleven-dimensional spacetime that contains the brane configuration.} Note that $\phi(p_i) = (t(p_i), v(p_i))$ is a point where $\lambda = \frac{v}{t}dt$, the Seiberg-Witten differential \cite{Fayyazuddin:1997by, Henningson:1997hy, Mikhailov:1997jv}, is singular, and therefore each $\pi(p_i)$ corresponds to a puncture of \cite{Gaiotto:2009we}. The reason why the points of (2) correspond to nontrivial ramifications can be illustrated as in Figure \ref{figure:ramification_at_dt_0}, which shows the real slice of $C_{\text{SW}}$ near $\phi(q_i) = (t(q_i), v(q_i))$ when two branches of $v(t)$ meet each other at $\phi(q_i)$. 

Using a local normalization map defined around each of these points, we can find the explicit form of $\pi$ at the neighborhood of the point. If $\pi$ is just a nice one-to-one mapping near the point, then we can forget about the point. But if $\pi$ shows a nontrivial ramification at the point, we can describe the ramification of $\mathcal{C}_{\text{SW}}$ near the point explicitly and calculate its ramification index.

\begin{figure}
	\centering
	\begin{minipage}{150pt}
	\subfloat[2 sheets]
		{\includegraphics{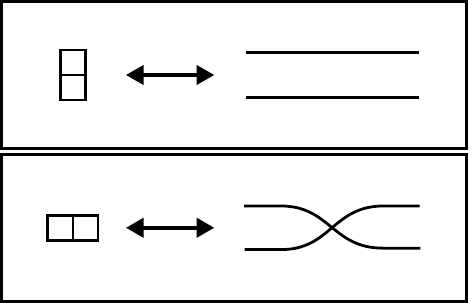}}\\
	\subfloat[3 sheets]
		{\includegraphics{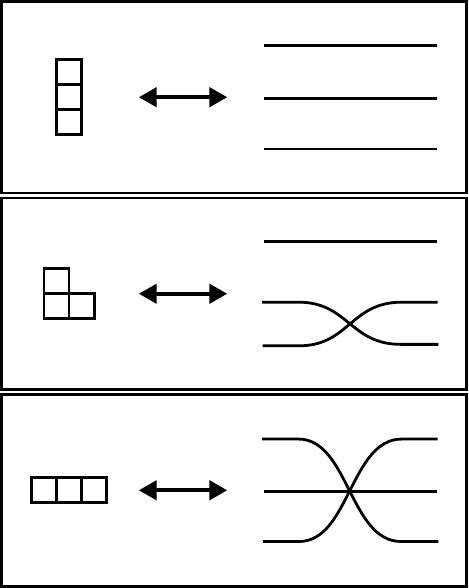}}
	\end{minipage}
	\begin{minipage}{150pt}
	\subfloat[4 sheets]
		{\includegraphics{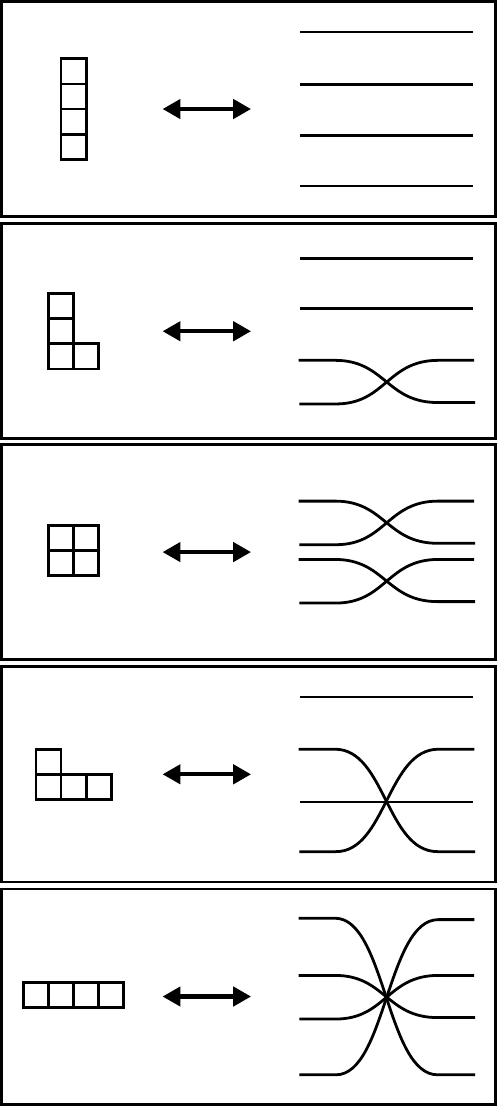}}
	\end{minipage}
	\caption{Young tableaux and the corresponding ramification structures}
	\label{figure:Young_tableaux}
\end{figure}

To represent what ramification structure each branch point on $C_{\text{B}}$ has, we will decorate it with a Young tableau, which will be constructed in the following way: start with $N = \deg(\pi)$ boxes. Collect the ramification points that are mapped to the same branch point, and put as many boxes as the ramification index of a ramification point in a row. Repeat this to form a row of boxes for each ramification point. Then stack these rows of boxes in an appropriate manner. If we run out of boxes then we are done. If not, then each remaining box is a row by itself, and we stack them too. Figure \ref{figure:Young_tableaux} shows several examples of Young tableaux constructed in this way for various ramification structures.

\begin{figure}[ht]
	\begin{center}
		\includegraphics{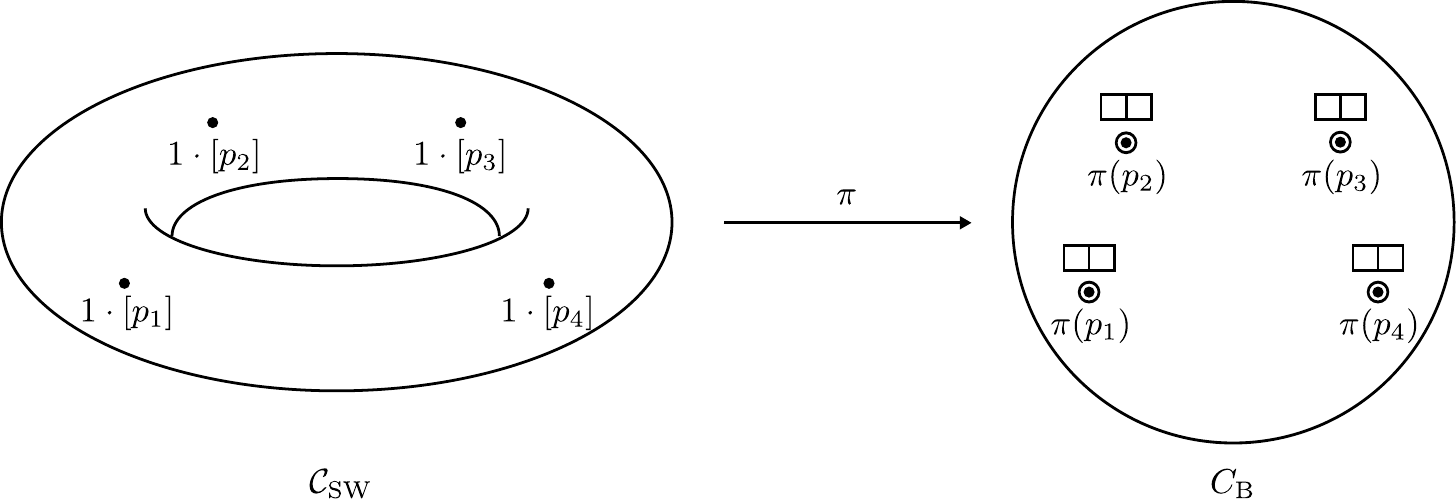}
		\caption{$\mathcal{C}_{\text{SW}}$ and $C_{\text{B}}$ for $SU(2)$ SCFT}
		\label{figure:SU_2_SCFT}
	\end{center}
\end{figure}

For the example we are considering now, (1) gives us $\{p_1,\ \ldots,\ p_4\}$ such that
\begin{align*}
	\phi(p_1) = (0, 0),\ \phi(p_2) = (1, \infty),\ \phi(p_3) = (t_1, \infty),\ \phi(p_4) = (\infty, 0),
\end{align*}
and (2) does not give any new point other than (1) provides, so we have $\{p_i\}$ as the candidates to check if $\mathcal{C}_{\text{SW}}$ has nontrivial ramifications at the points. The local normalization near each $p_i$ is calculated in Appendix \ref{appendix:SU(2)}. From the local normalizations we get $\pi$, which maps $\{p_i\}$ to
\begin{align*}
	\{\pi(p_1) = 0,\ \pi(p_2) = 1,\ \pi(p_3) = t_1,\ \pi(p_4) = \infty\}.
\end{align*}
The ramification divisor of $\pi$ is also calculated in Appendix \ref{appendix:SU(2)},
\begin{align}
R_\pi = 1 \cdot [p_1] + 1 \cdot [p_2] + 1 \cdot [p_3] + 1 \cdot [p_4], \label{eq:SU(2) SCFT ramification divisor}
\end{align}
which shows that every $p_i$ has a nontrivial ramification index of 2, and this is consistent with the Riemann-Hurwitz formula, Eq. (\ref{eq:riemann-hurwitz formula}),
\begin{align*}
	\deg(R_\pi) = 1 + 1 + 1 + 1 = 4 = 2(g(\mathcal{C}_{\text{SW}}) + \deg(\pi) -1),
\end{align*}
considering $\deg(\pi) = 2$ and $g(\mathcal{C}_{\text{SW}}) = 1$. In the current example, where $\mathcal{C}_{\text{SW}}$ is an elliptic curve, the result of Eq. (\ref{eq:SU(2) SCFT ramification divisor}) can be expected because an elliptic curve, when considered as a 2-sheeted cover over $\mathbb{CP}^1$, has four ramification points of index 2. Figure \ref{figure:SU_2_SCFT} shows how $\pi$ maps $R_\pi$ of $\mathcal{C}_{\text{SW}}$ to the branch points of $C_{\text{B}}$. 
For this example, all of the branch points are the images of the points $\{p_i\}$,
therefore each branch point corresponds to a puncture of \cite{Gaiotto:2009we}. This example provides a geometric explanation of why each puncture can be labeled with its Young tableau. 

The wrapping of the noncompact Seiberg-Witten curve $C_{\text{SW}}$ over $C_{\text{B}}$ is described by the composition of $\phi^{-1}: C_{\text{SW}} \to \mathcal{C}_{\text{SW}}\backslash\{p_i\}$ and $\pi$,
\begin{align*}
	\pi \circ \phi^{-1}: C_{\text{SW}} \to C_{\text{B}}\backslash\{\pi(p_i)\},\ (t,v) \mapsto t,
\end{align*}
which is the projection we discussed in Section \ref{section:introduction}. Note that the noncompact Seiberg-Witten curve $C_{\text{SW}}$ does not contain $\{\phi(p_i)\} = \mathcal{I}$. Therefore $C_{\text{SW}}$ has no ramification point, unlike the compact Riemann surface $\mathcal{C}_{\text{SW}}$. That is, the two branches of $C_{\text{SW}}$ only meet ``at infinity,'' and all branch points on $C_{\text{B}}$, $\{\pi(p_i)\}$, are from the points ``at infinity.''

After embedding $C_{\text{SW}}$ into $\mathbb{CP}^2$, the Seiberg-Witten differential form $\lambda$, 
\begin{align*}
	\lambda = \frac{v}{t} dt,
\end{align*}
which is a meromorphic 1-form on $C_{\text{SW}}$, becomes\footnote{Whether this embedding of $\lambda$ is justifiable is a part of the question that the embedding of $C_{\text{SW}}$ into $\mathbb{CP}^2$ gives the same physics as $C_{\text{SW}}$ does or not.}
\begin{align*}
	\lambda = \frac{Y}{X} d\left(\frac{X}{Z}\right), 
\end{align*}
which defines a meromorphic 1-form on $\bar{C}_{\text{SW}}$.  We pull $\lambda$ back to $\omega = \sigma^* (\lambda)$, which defines a meromorphic 1-form on $\mathcal{C}_{\text{SW}}$ and therefore should satisfy the Poincar\'{e}-Hopf theorem \cite{Griffiths}
\begin{align}
	\deg[(\omega)] = 2(g(\mathcal{C}_{\text{SW}}) - 1), \label{eq:poincare-hopf theorem}
\end{align}
where $(\omega)$ is a divisor of $\omega$ on $\mathcal{C}_{\text{SW}}$, which is defined as
\begin{align*}
	(\omega) = \sum_{s \in \mathcal{C}_{\text{SW}}} \nu_{s}(\omega)[s],
\end{align*}
where $\nu_{s}(\omega) \in \mathbb{Z}$ is the order\footnote{When $\omega$ has a pole at $s$, the pole is of order $-\nu_{s}(\omega)$; when $\omega$ has a zero at $s$, the zero is of order $\nu_{s}(\omega)$; otherwise $\nu_{s}(\omega) = 0$.} of $\omega$ at $s$.

We want to see if Eq. (\ref{eq:poincare-hopf theorem}) holds for this example as a consistency check. In order to do that, we need to find out every $s \in \mathcal{C}_{\text{SW}}$ that has a nonzero value of $\nu_{s}(\omega)$. Considering that $\omega$ is a pullback of $\lambda$, the candidates of such points are
\begin{enumerate}[(1)]
	\item $\{p_i \in \mathcal{C}_{\text{SW}}\ |\ \phi(p_i) \in \mathcal{I}\}$,
	\item $\{q_i \in \mathcal{C}_{\text{SW}}\ |\ dt(q_i) = 0\} \Leftrightarrow \{q_i \in \mathcal{C}_{\text{SW}}\ |\ (\partial f / \partial v)(t(q_i),v(q_i)) = 0\}$,
	\item $\{r_i \in \mathcal{C}_{\text{SW}}\ |\ v(r_i) = 0\}$.
\end{enumerate}
We check (1) because $\lambda$ is singular at $\phi(p_i)$ and therefore $\omega$ may have a pole at $p_i$. We also check (2) and (3) because $\lambda$ vanishes at $\phi(q_i)$ and $\phi(r_i)$ and therefore $\omega$ may have a zero at $q_i$ or $r_i$. For this example (2) and (3) do not give us any additional point other than the points from (1). Therefore the candidates are $\{p_1,\ \ldots,\ p_4\}$, the same set of points we have met when calculating $R_\pi$. Using the local normalizations near these points described in Appendix \ref{appendix:SU(2)}, we get 
\begin{align*}
	(\omega) = 0, 
\end{align*}
which means $\omega$ has neither zero nor pole over $\mathcal{C}_{\text{SW}}$. This is an expected result, since we can find a globally well-defined coordinate $z$ of the elliptic curve $\mathcal{C}_{\text{SW}}$ such that $\sigma^*(\lambda) = dz$. 

The result is consistent with the Poincar\'{e}-Hopf theorem, Eq. (\ref{eq:poincare-hopf theorem}), 
\begin{align*}
	\deg[(\omega)] = 0 = 2(g(\mathcal{C}_{\text{SW}}) - 1),
\end{align*}
considering $g(\mathcal{C}_{\text{SW}}) = 1$.


\section{$SU(2) \times SU(2)$ SCFT and the ramification point}
\label{section:SU(2) times SU(2) SCFT}
In Section \ref{section:SU(2) SCFT} we have studied the Seiberg-Witten curve of a four-dimensional $\mathcal{N} = 2$ $SU(2)$ SCFT to identify how the wrapping of the curve over a Riemann sphere can be described by a covering map. In this section we apply the same analysis to the Seiberg-Witten curve of a four-dimensional $\mathcal{N}=2$ $SU(2) \times SU(2)$ SCFT. From this example, we will learn that on the curve there is a ramification point whose image under the covering map cannot be identified with one of the punctures of \cite{Gaiotto:2009we}.

\begin{figure}[ht]
	\begin{center}
		\includegraphics{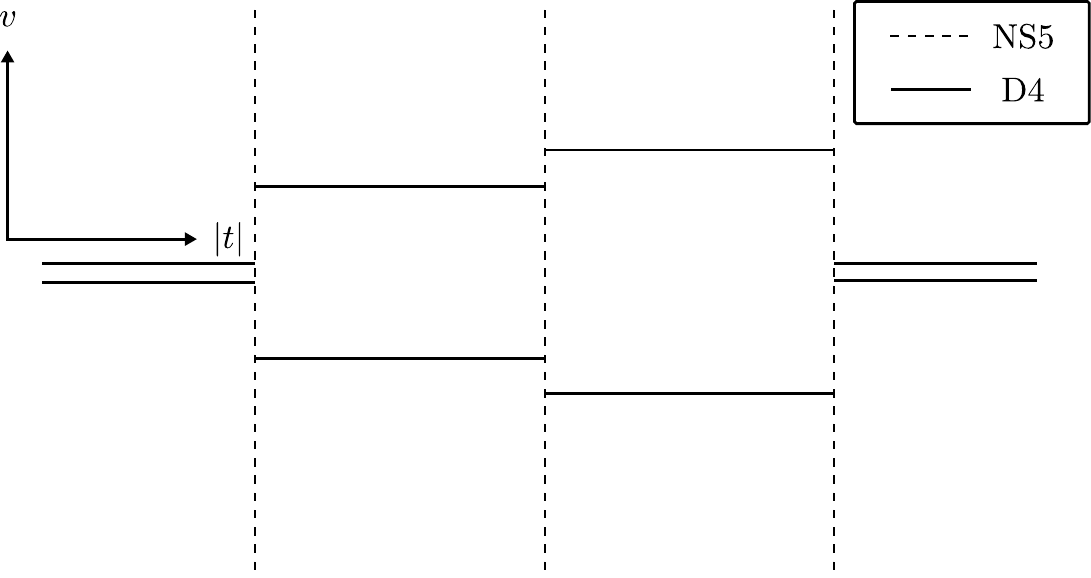}
		\caption{Brane configuration of $SU(2) \times SU(2)$ SCFT}
		\label{figure:SU_2_X_SU_2_SCFT_brane}
	\end{center}
\end{figure}

\begin{figure}[ht]
	\begin{center}
		\includegraphics{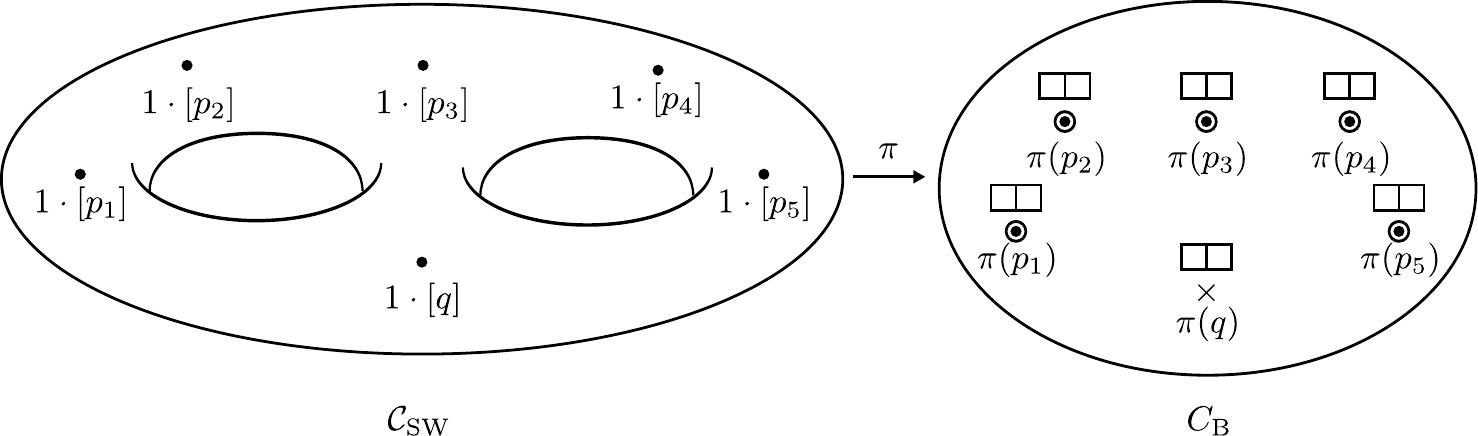}
		\caption{$\mathcal{C}_{\text{SW}}$ and $C_{\text{B}}$ for $SU(2) \times SU(2)$ SCFT}
		\label{figure:SU_2_X_SU_2_SCFT}
	\end{center}
\end{figure}

The brane configuration of Figure \ref{figure:SU_2_X_SU_2_SCFT_brane} gives a four-dimensional $\mathcal{N}=2$ $SU(2) \times SU(2)$ SCFT. 
The corresponding Seiberg-Witten curve $C_{\text{SW}}$ is the zero locus of
\begin{align}
	f(t,v) = (t-1) (t-t_1) (t-t_2) v^2 - u_1 t^2 - u_2 t. \label{eq:C_SW of SU(2) times SU(2) SCFT}
\end{align}
Considering a normalization $\sigma: \mathcal{C}_{\text{SW}} \to \bar{C}_{\text{SW}}$ and a meromorphic function $\pi: \mathcal{C}_{\text{SW}} \to \mathbb{CP}^1$, we can introduce a ramification divisor $R_\pi = \sum_s (\nu_{s}(\pi) - 1)[s]$. Nontrivial ramifications may occur at
\begin{enumerate}[(1)]
	\item $\{p_i \in \mathcal{C}_{\text{SW}}\}$, where $\{\phi(p_i)\}$ are the points we add to $C_{\text{SW}}$ to compactify it,
	\item $\{q_i \in \mathcal{C}_{\text{SW}}\ |\ dt(q_i) = 0\} \Leftrightarrow \{q_i \in \mathcal{C}_{\text{SW}}\ |\ (\partial f / \partial v)(t(q_i),v(q_i)) = 0\}$.
\end{enumerate}
(1) gives us $\{p_1,\ \ldots,\ p_5\}$ such that
\begin{align*}
	\phi(p_1) = (0, 0),\ \phi(p_2) = (1, \infty),\ \phi(p_3) = (t_1, \infty),\ \phi(p_4) = (t_2, \infty),\ \phi(p_5) = (\infty, 0),
\end{align*}
and from (2) we get $\{q\}$ such that
\begin{align*}
	\phi(q) = (\rho, 0),\ \rho = -u_2/u_1.
\end{align*}
Using the local normalizations calculated in Appendix \ref{appendix:SU(2) times SU(2)}, we get
\begin{align*}
	R_\pi = 1 \cdot [p_1] + 1 \cdot [p_2] + 1 \cdot [p_3] + 1 \cdot [p_4] + 1 \cdot [p_5] + 1 \cdot [q],
\end{align*}
and
\begin{align*}
	\deg(R_\pi) = 1 + 1 + 1 + 1 + 1 + 1 = 6,
\end{align*}
which is consistent with the Riemann-Hurwitz formula, Eq. (\ref{eq:riemann-hurwitz formula}), considering $\deg(\pi) = 2$ and $g(\mathcal{C}_{\text{SW}}) = 2$. Figure \ref{figure:SU_2_X_SU_2_SCFT} shows how $\pi$ maps $\mathcal{C}_{\text{SW}}$ with its ramification points to $C_{\text{B}}$ with its branch points.

Again $R_\pi$ has a divisor $[p_i]$ whose image under $\pi$ can be identified with a puncture of \cite{Gaiotto:2009we}. However, $R_\pi$ also contains $[q]$, which means that ramification occurs also at $q$. The location of $\pi(q)$ on $C_{\text{B}}$ depends on the Coulomb branch parameters $u_1$ and $u_2$, unlike $\{\pi(p_i)\}$ whose locations depend only on the gauge coupling parameters $t_1$ and $t_2$. In Figure \ref{figure:SU_2_X_SU_2_SCFT} we denoted $\pi(q)$ with a symbol different from that of $\{\pi(p_i)\}$ to distinguish between the two. In this example, two sheets are coming together at both $\{\pi(p_i)\}$ and $\pi(q)$, and therefore each of them has the same Young tableau correspoding to the ramification structure. 

However note that the noncompact Seiberg-Witten curve $C_{\text{SW}}$ does not contain $\{\phi(p_i)\}$ but contains $\phi(q)$ only, therefore it is the only ramification point that exists in $C_{\text{SW}}$. That is, the branch point $\pi(q)$ comes from the ramification point of $C_{\text{SW}}$, whereas the other branch points $\{\pi(p_i)\}$ that are identified with the punctures are from the points ``at infinity.'' 
\begin{figure}[ht]
	\begin{center}
		\includegraphics{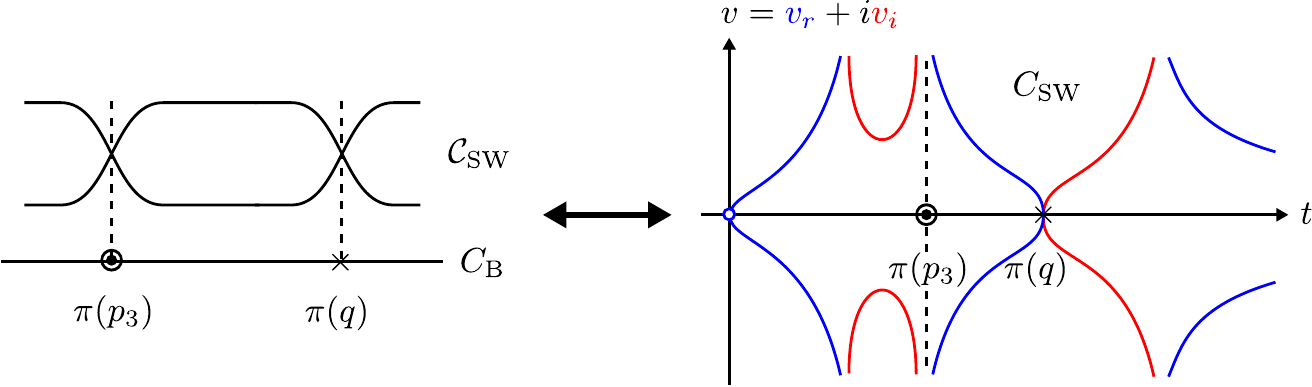}
		\caption{Two branch points of different kinds: $\pi(p_3)$ from a point at $v = \infty$, $\pi(q)$ from the ramification point of $C_{\text{SW}}$.}
		\label{figure:two_different_ramifications}
	\end{center}
\end{figure}
Figure \ref{figure:two_different_ramifications} shows the schematic cross-section of the compact Riemann surface $\mathcal{C}_{\text{SW}}$ near $p_3$ and $q$ on the left side, and the real (and imaginary) slice of the noncompact Seiberg-Witten curve $C_{\text{SW}}$ on the right side. This illustrates the difference between the two kinds of branch points.

\begin{figure}[ht]
	\begin{center}
		\includegraphics{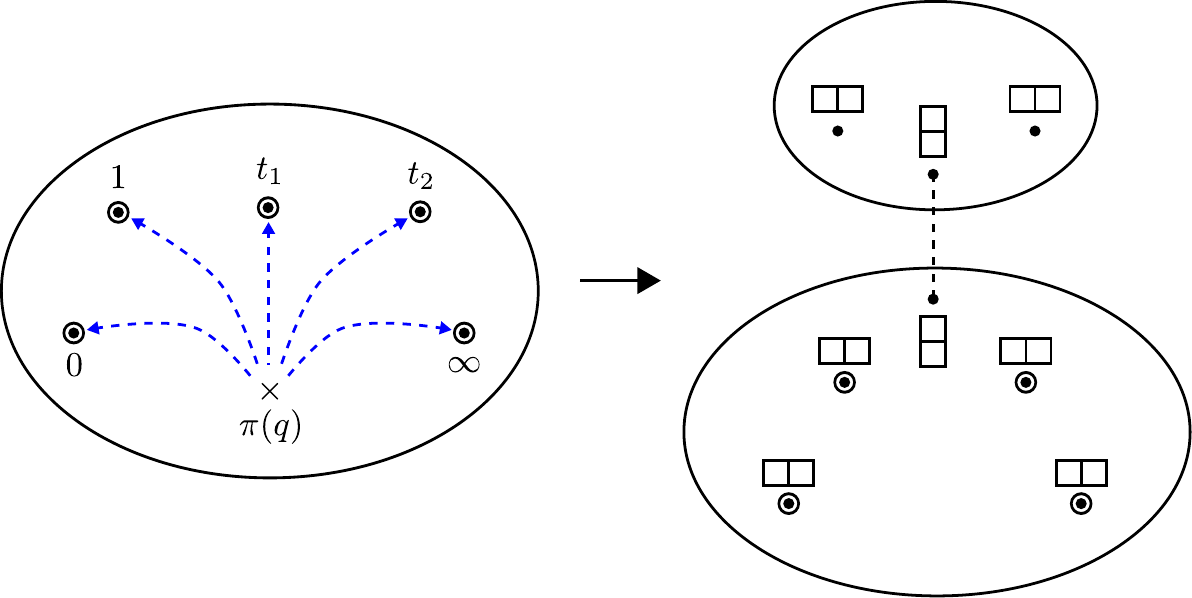}
		\caption{Branch point $\pi(q)$ under various limits of Coulomb branch parameters}
		\label{figure:SU_2_X_SU_2_SCFT_limits}
	\end{center}
\end{figure}

Taking various limits of the Coulomb branch parameters corresponds to moving $\pi(q)$ on $C_{\text{B}}$ in various ways, as shown in Figure \ref{figure:SU_2_X_SU_2_SCFT_limits}. 
When $\pi(q)$ is infinitesimally away from one of $\{\pi(p_i)\}$, imagine cutting out a part of the Seiberg-Witten curve around the preimages of the two branch points. As there is no monodromy of $v(t)$ when going around a route that encircles the two branch points, we can fill the excised area topologically with two points, the result of which is shown in the lower right side of Figure \ref{figure:SU_2_X_SU_2_SCFT_limits}. This corresponds to the Seiberg-Witten curve of $SU(2)$ SCFT that we have investigated in Section \ref{section:SU(2) SCFT}. And the excised part of the Seiberg-Witten curve separates itself from the rest of the curve to form another curve which has the topology of a sphere. This is shown in the upper right side of Figure \ref{figure:SU_2_X_SU_2_SCFT_limits}, where we represented only the ramification structure of each branch point. This can also be checked by taking the limits of the Coulomb branch parameters of Eq. (\ref{eq:C_SW of SU(2) times SU(2) SCFT}), which will result in a reducible curve with two components, one being the curve of $SU(2)$ SCFT and the other a Riemann sphere.

Now we repeat the same analysis of the Seiberg-Witten differential $\lambda = \frac{v}{t} dt$ that we did in Section \ref{section:SU(2) SCFT}. The candidates for the points on $\mathcal{C}_{\text{SW}}$ where $\omega$ has nonzero order are
\begin{enumerate}[(1)]
	\item $\{p_i \in \mathcal{C}_{\text{SW}}\}$, where $\{\phi(p_i)\}$ are the points we add to $C_{\text{SW}}$ to compactify it,
	\item $\{q_i \in \mathcal{C}_{\text{SW}}|dt(q_i) = 0\} \Leftrightarrow \{q_i \in \mathcal{C}_{\text{SW}}\ |\ (\partial f / \partial v)(t(q_i),v(q_i)) = 0\}$,
	\item $\{r_i \in \mathcal{C}_{\text{SW}}|v(r_i) = 0\}$.
\end{enumerate}
(3) does not give us any new point other than the points from (1) and (2) for this example, so the candidates are $\{p_1,\ \ldots,\ p_5\}$ and $\{q\}$. Again we can analyze how $\omega$ behaves near those points by using the local normalizations calculated in Appendix \ref{appendix:SU(2) times SU(2)}, which gives 
\begin{align*}
	(\omega) = 2 \cdot [q],
\end{align*}
and
\begin{align*}
	\deg[(\omega)] = 2 = 2(g(\mathcal{C}_{\text{SW}}) -1).
\end{align*}
This result is consistent with the Poincar\'{e}-Hopf theorem, Eq. (\ref{eq:poincare-hopf theorem}), considering $g(\mathcal{C}_{\text{SW}}) = 2$. 


\section{$SU(3)$ SCFT and Argyres-Seiberg duality}
\label{section:SU(3) SCFT}
In Section \ref{section:SU(2) times SU(2) SCFT} we have found a branch point on $C_{\text{B}}$ that comes from the ramification point of the Seiberg-Witten curve and cannot be identified with a puncture. The location of this branch point on $C_{\text{B}}$ depends on the Coulomb branch parameters, which enables us to use it as a tool to describe various limits of the parameters. In this section, we do the same analysis for the example of a four-dimensional $\mathcal{N}=2$ $SU(3)$ SCFT to find the branch points from the ramification points of its Seiberg-Witten curve, this time their locations on $C_{\text{B}}$ depending on both the gauge coupling parameter and the Coulomb branch parameters. And we will see how these branch points help us to illustrate the interesting limit of the theory studied by Argyres and Seiberg \cite{Argyres:2007cn}.

\begin{figure}[ht]
	\begin{center}
		\includegraphics{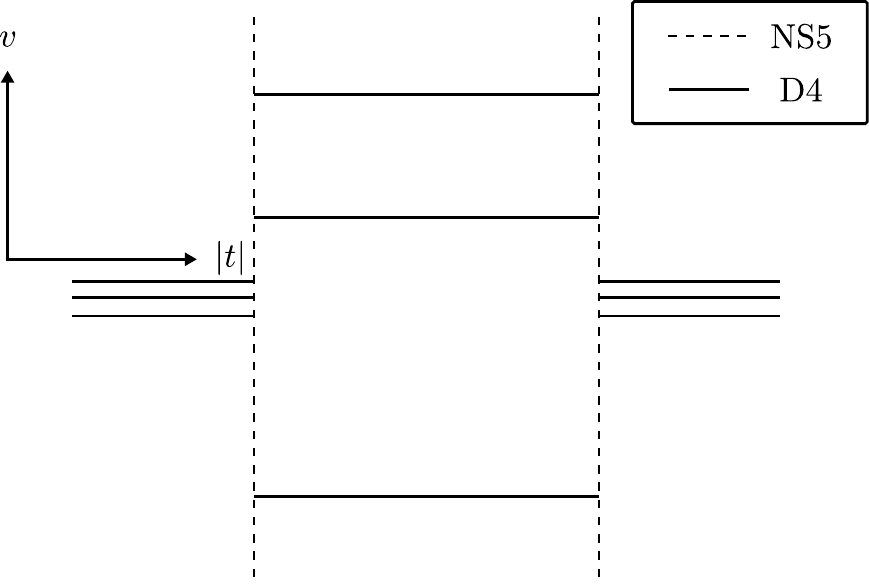}
		\caption{Brane configuration of $SU(3)$ SCFT}
		\label{figure:SU_3_SCFT_brane}
	\end{center}
\end{figure}

The starting point is a four-dimensional $\mathcal{N}=2$ $SU(3)$ SCFT associated to the brane configuration of Figure \ref{figure:SU_3_SCFT_brane}.

The corresponding Seiberg-Witten curve $C_{\text{SW}}$ is the zero locus of
\begin{align}
	f(t,v) = (t-1)(t-t_1)v^3 - u_2 t v - u_3 t. \label{eq:SU(3) SCFT SW curve}
\end{align}
Considering a normalization $\sigma: \mathcal{C}_{\text{SW}} \to \bar{C}_{\text{SW}}$ and a meromorphic function $\pi: \mathcal{C}_{\text{SW}} \to \mathbb{CP}^1$, we can introduce a ramification divisor $R_\pi = \sum_s (\nu_{s}(\pi) - 1)[s]$. Nontrivial ramifications may occur at
\begin{enumerate}[(1)]
	\item $\{p_i \in \mathcal{C}_{\text{SW}}\}$, where $\{\phi(p_i)\}$ are the points we add to $C_{\text{SW}}$ to compactify it,
	\item $\{q_i \in \mathcal{C}_{\text{SW}}\ |\ dt(q_i) = 0\} \Leftrightarrow \{q_i \in \mathcal{C}_{\text{SW}}\ |\ (\partial f / \partial v)(t(q_i),v(q_i)) = 0\}$.
\end{enumerate}
From (1) we get $\{p_1,\ \ldots,\ p_4 \}$ such that
\begin{align*}
	\phi(p_1) = (0,0),\ \phi(p_2) = (1,\infty),\ \phi(p_3) = (t_1,\infty),\ \phi(p_4) = (\infty,0).
\end{align*}
(2) gives us $\{q_+,\ q_-\}$ such that
\begin{align*}
	\phi(q_\pm) = (t_\pm, v_0),
\end{align*}	
where 
\begin{align*}
	v_0 = -\frac{(u_3 / 2)}{(u_2 / 3)}
\end{align*}	
and $t_\pm$ are the two roots of $f(t, v_0) = 0$,
\begin{align*}
	t_\pm = \frac{1 + t_1 + \rho}{2} \pm \sqrt{\left(\frac{1 + t_1 + \rho}{2}\right)^2 - t_1},\ \rho = \frac{(u_2 / 3)^3}{(u_3 / 2)^2}.
\end{align*}

\begin{figure}[ht]
	\begin{center}
		\includegraphics{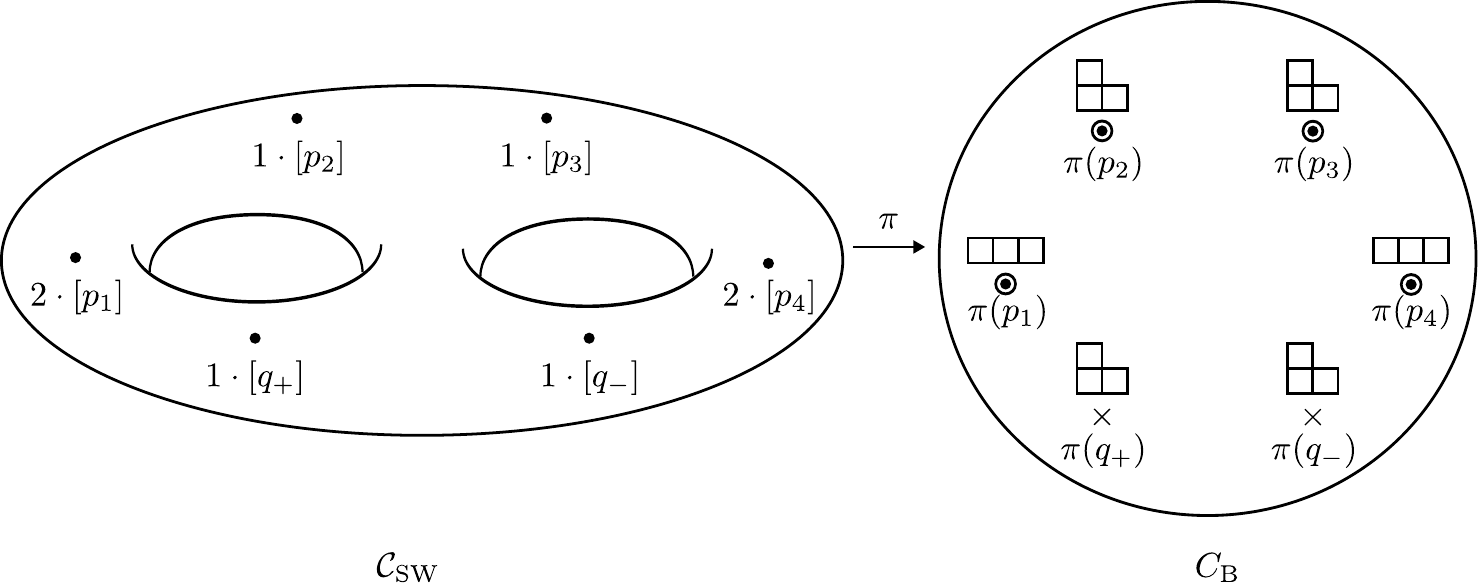}
		\caption{$\mathcal{C}_{\text{SW}}$ and $C_{\text{B}}$ for $SU(3)$ SCFT}
		\label{figure:SU_3_SCFT}
	\end{center}
\end{figure}

Calculations for the local normalizations near the points are given in Appendix \ref{appendix:SU(3)}, from which we get the ramification divisor of $\pi$ as 
\begin{align*}
	R_\pi = 2\cdot[p_1] + 1\cdot[p_2] + 1\cdot[p_3] + 2\cdot[p_4] + 1\cdot[q_+] + 1\cdot[q_-],
\end{align*}
and	this satisfies
\begin{align*}
	\deg(R_\pi) = 2 + 1 + 1 + 2 + 1 + 1 = 8 = 2(g(\mathcal{C}_{\text{SW}}) + \deg(\pi) -1), 
\end{align*}
which is consistent with the Riemann-Hurwitz formula, Eq. (\ref{eq:riemann-hurwitz formula}), considering $\deg(\pi) = 3$ and $g(\mathcal{C}_{\text{SW}}) = 2$. Figure \ref{figure:SU_3_SCFT} shows how $\pi$ works.

Considering that $\pi$ is in general three-to-one mapping, the fact that $R_\pi$ has degree 2 at $p_1$ implies that the three sheets are coming together at $\pi(p_1)$, which corresponds to a Young tableau {\tiny\yng(3)}. And $R_\pi$ having degree 1 at $p_2$ is translated into only two out of three sheets coming together at $\pi(p_2)$, which corresponds to a Young tableau {\tiny\yng(1,2)}. These $\{\pi(p_i)\}$ are identified with the punctures of \cite{Gaiotto:2009we}.\footnote{Note that at $t = \pi(p_2)$ and at $t = \pi(p_3)$ only two among the three branches have the divergent $v(t)$, and therefore $\lambda$ is divergent along only the two branches. This means that our analysis corresponds to that of \cite{Gaiotto:2009we} before making a shift of $v$. In \cite{Gaiotto:2009we} every branch has the divergence after the shift in $v$ so that the flavor symmetry at the puncture is evident. Here we prefer not to shift $v$ so that we can analyze the Seiberg-Witten curve as an algebraic curve studied in \cite{Witten:1997sc}.} 

However $R_\pi$ also contains $[q_\pm]$, which means that ramifications of $\mathcal{C}_{\text{SW}}$ occur also at $q_\pm$. These are the points of $\mathcal{C}_{\text{SW}}$ where $dt=0$ along $\mathcal{C}_{\text{SW}}$. The locations of $\pi(q_\pm)$ on $C_{\text{B}}$ depend on both the gauge coupling parameter $t_1$ and the Coulomb branch parameters $u_2$ and $u_3$, unlike $\{\pi(p_i)\}$ whose locations depend only on $t_1$. Therefore $\{ \pi(q_\pm) \}$ are the branch points that are not identified with the punctures. 

Again note that $\{\pi(q_\pm)\}$ are distinguished from $\{\pi(p_i)\}$ in that they are from the ramification points of the noncompact Seiberg-Witten curve $C_{\text{SW}}$. That is, $\{\phi(q_\pm)\}$ are the only ramification points of $C_{\text{SW}}$, whereas $\{\phi(p_i)\}$ are the points ``at infinity.''

\begin{figure}[ht]
	\begin{center}
		\includegraphics{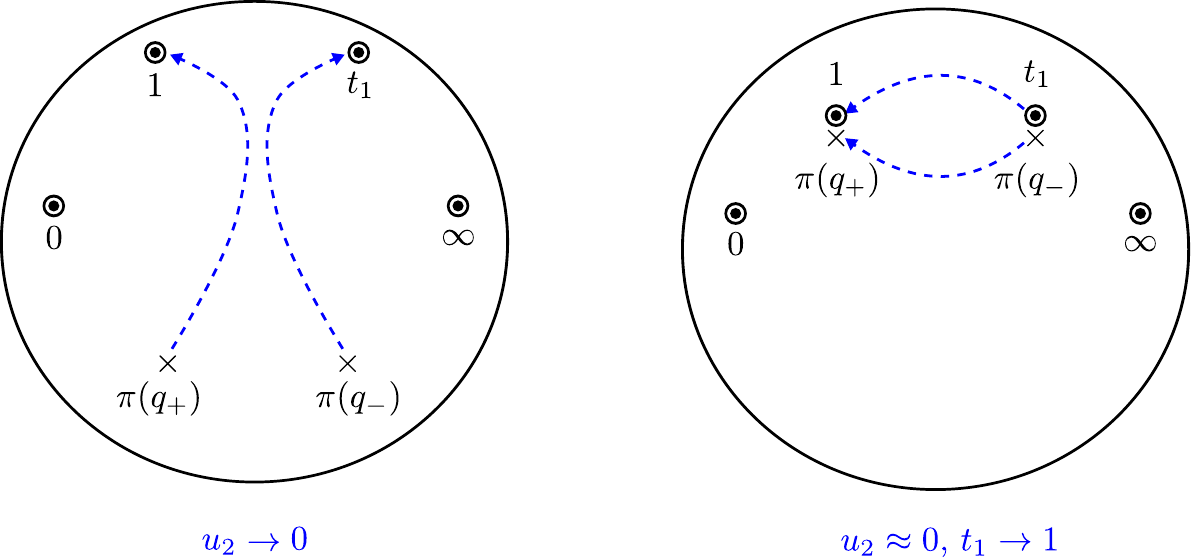}
		\caption{Behaviors of the branch points under the limit $u_2 \to 0$ and $t_1 \to 1$}
		\label{figure:SU_3_SCFT_AS_Limit_u2}
	\end{center}
\end{figure}

\begin{figure}[ht]
	\begin{center}
		\includegraphics{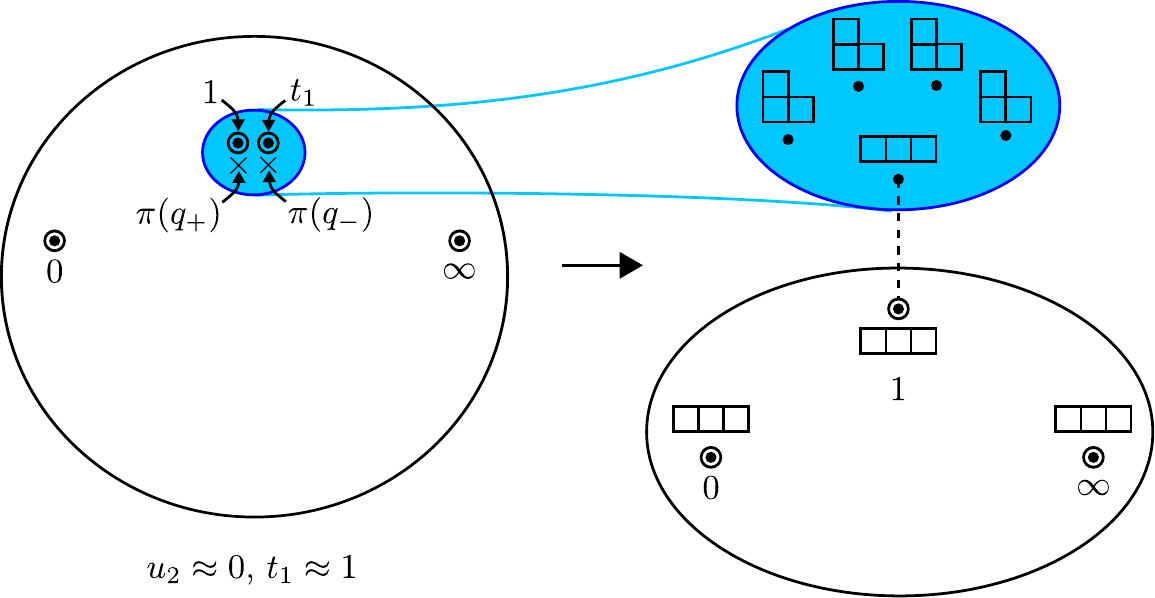}
		\caption{Appearance of $E_6$ curve under the limit $u_2 \to 0$ and $t_1 \to 1$}
		\label{figure:SU_3_SCFT_AS_Limit_E6}
	\end{center}
\end{figure}

To see how the Argyres-Seiberg duality \cite{Argyres:2007cn} is illustrated by the branch points, we take the corresponding limits for the Coulomb branch parameters and the gauge coupling parameter. When we take $u_2 \to 0$, $\pi(q_+)$ and $\pi(q_-)$ move toward $\pi(p_2) = 1$ and $\pi(p_3) = t_1$, respectively. In addition we take the limit of $t_1 \to 1$, and the four branch points come together. Figure \ref{figure:SU_3_SCFT_AS_Limit_u2} shows the behavior of the branch points under the limit of the parameters. 
When we are near the limit of $u_2 = 0$ and $t_1 = 1$, the four branch points become infinitesimally separated from one another and we can imagine cutting out a part of the Seiberg-Witten curve around the preimages of the four branch points, separating the original curve into two parts. As the monodromy of $v(t)$ around the four branch points corresponds to a point of ramification index 3, we can see that one part becomes a genus 1 curve and the other becomes another genus 1 curve, considering the ramification structure of each of them. Figure \ref{figure:SU_3_SCFT_AS_Limit_E6} illustrates this. 
The genus 1 curve with three branch points of ramification index 3 corresponds to the zero locus of
\begin{align*}
	(t - 1)^2 v^3 - u_3 t,
\end{align*}
which is from Eq. (\ref{eq:SU(3) SCFT SW curve}) by setting $t_1 = 1$ and $u_2 = 0$. This curve can be identified with the Seiberg-Witten curve of $E_6$ theory \cite{Gaiotto:2009we, Argyres:2007cn}. The other genus 1 curve is a small torus, which reminds us of the weakly gauged $SU(2)$ theory coupled to the $E_6$ theory that appears in \cite{Gaiotto:2009we, Argyres:2007cn}.

\begin{figure}[ht]
	\begin{center}
		\includegraphics{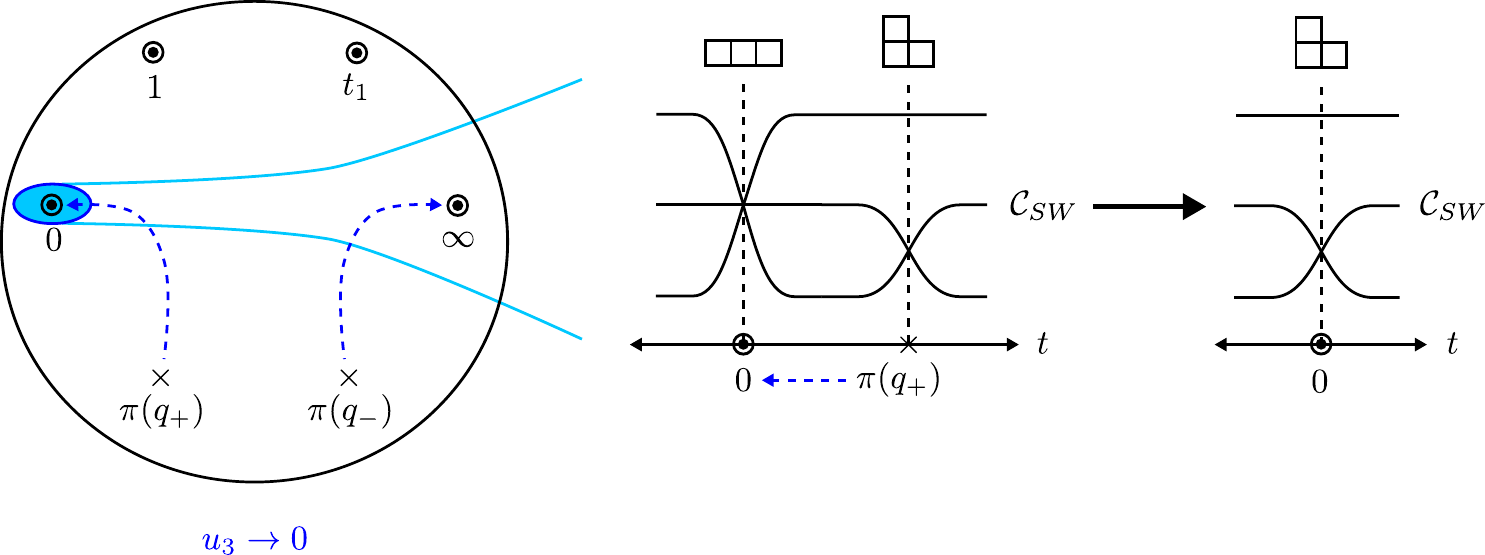}
		\caption{Behaviors of the branch points under the limit $u_3 \to 0$}
		\label{figure:SU_3_SCFT_AS_Limit_u3}
	\end{center}
\end{figure}

When we take $u_3 \to 0$ limit, $\pi(q_+)$ and $\pi(q_-)$ move toward $\pi(p_1) = 0$ and $\pi(p_4) = \infty$, respectively. The collision of $\pi(q_+)$ with $\pi(p_1)$ partially unravels the ramification over the two branch points, which results in one branch point with the corresponding ramification point having index 2, and the third sheet falling apart from the branch point. The same thing happens at $t = \infty$, so the result of the limit is a reducible curve with two components, one component being the same $SU(2)$ SCFT curve that we have investigated in Section \ref{section:SU(2) SCFT} and the other a Riemann sphere. This can also be checked by setting $u_3 = 0$ in Eq. (\ref{eq:SU(3) SCFT SW curve}), which gives us an $SU(2)$ SCFT curve and a Riemann sphere. Figure \ref{figure:SU_3_SCFT_AS_Limit_u3} illustrates the limit and the partial unraveling of the ramification.

Let's proceed to the calculation of $(\omega) = \sum_s \nu_s(\omega)[s]$. The candidates for the points on $\mathcal{C}_{\text{SW}}$ where $\omega$ has nonzero order are
\begin{enumerate}[(1)]
	\item $\{p_i \in \mathcal{C}_{\text{SW}}\}$, where $\{\phi(p_i)\}$ are the points we add to $C_{\text{SW}}$ to compactify it,
	\item $\{q_i \in \mathcal{C}_{\text{SW}}|dt(q_i) = 0\} \Leftrightarrow \{q_i \in \mathcal{C}_{\text{SW}}\ |\ (\partial f / \partial v)(t(q_i),v(q_i)) = 0\}$,
	\item $\{r_i \in \mathcal{C}_{\text{SW}}|v(r_i) = 0\}$.
\end{enumerate}
(1) and (2) give us $\{p_1,\ \ldots,\ p_4\}$ and $\{q_\pm\}$, respectively. (3) does not result in any additional point. Using the local normalizations calculated in Appendix \ref{appendix:SU(3)}, we can get 
\begin{align*}
	(\omega) = 1 \cdot [q_+] + 1 \cdot [q_-],
\end{align*}
which is consistent with the Poincar\'{e}-Hopf theorem, Eq. (\ref{eq:poincare-hopf theorem}),
\begin{align*}
	\deg[(\omega)] = 1 + 1 = 2 = 2(g(\mathcal{C}_{\text{SW}}) - 1),
\end{align*}
considering $g(\mathcal{C}_{\text{SW}}) = 2$.


\section{$SU(3)$ pure gauge theory and Argyres-Douglas fixed points}
\label{section:SU(3) pure gauge theory}
What is interesting about the branch points we have found in Sections \ref{section:SU(2) times SU(2) SCFT} and \ref{section:SU(3) SCFT}, the images of the ramification points of the Seiberg-Witten curve under the covering map, is that their locations on $C_{\text{B}}$ depend in general on every parameter of the Seiberg-Witten curve, including both gauge coupling parameters and Coulomb branch parameters. Therefore they can be useful in analyzing how a Seiberg-Witten curve behaves as we take various limits for the parameters. 

Furthermore, considering that branch points are important in understanding various noncontractible 1-cycles of a curve and that each such cycle on a Seiberg-Witten curve corresponds to a BPS state with its mass given by the integration of the Seiberg-Witten differential along the cycle \cite{Seiberg:1994rs, Seiberg:1994aj}, the behaviors of branch points under the various limits of the parameters tell us some information regarding the BPS states.

To expand on these ideas, we will investigate in this section the case of a four-dimensional $\mathcal{N}=2$ $SU(3)$ pure gauge theory, which has the special limits of the Coulomb branch parameters, the Argyres-Douglas fixed points \cite{Argyres:1995jj}. We will describe how the branch points from the ramification points of the Seiberg-Witten curve of the theory help us to identify the small torus that arises at the fixed points. 

\begin{figure}[ht]
	\begin{center}
		\includegraphics{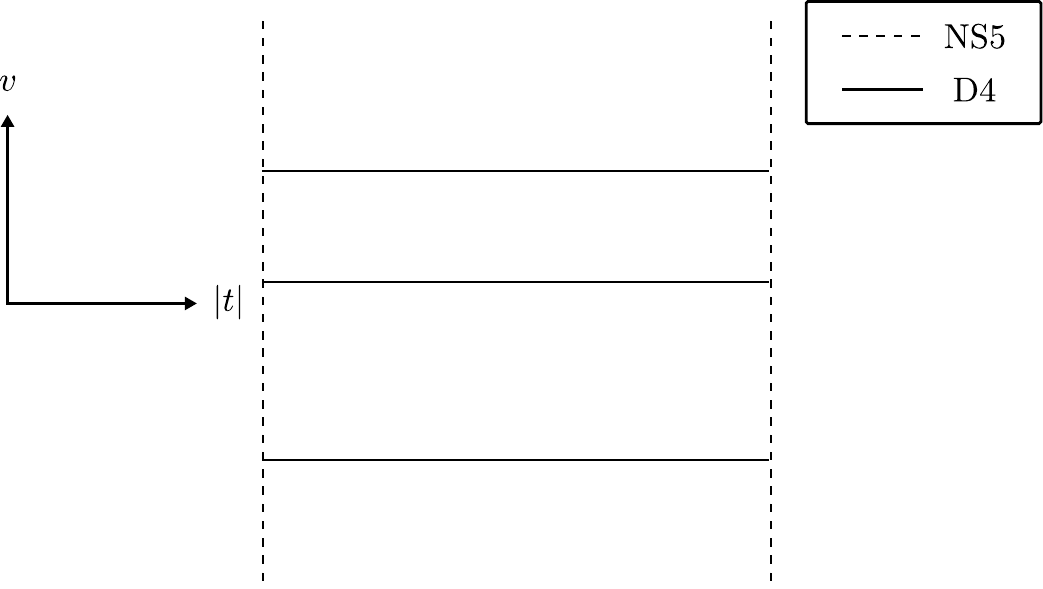}
		\caption{Brane configuration of $SU(3)$ pure gauge theory}
		\label{figure:SU_3_no_matter_brane}
	\end{center}
\end{figure}

Here the starting point is a four-dimensional $\mathcal{N}=2$ $SU(3)$ pure gauge theory associated to the brane configuration of Figure \ref{figure:SU_3_no_matter_brane}.
The corresponding Seiberg-Witten curve $C_{\text{SW}}$ is the zero locus of 
\begin{align*}
	f(t,v) = t^2 + (v^3 - u_2 v - u_3) t + \Lambda^6,
\end{align*} 
where $\Lambda$ is the dynamically generated scale of the four-dimensional theory. This is different from the previous examples, where the corresponding four-dimensional theories are conformal and therefore are scale-free. 

\begin{figure}[ht]
	\begin{center}
		\includegraphics{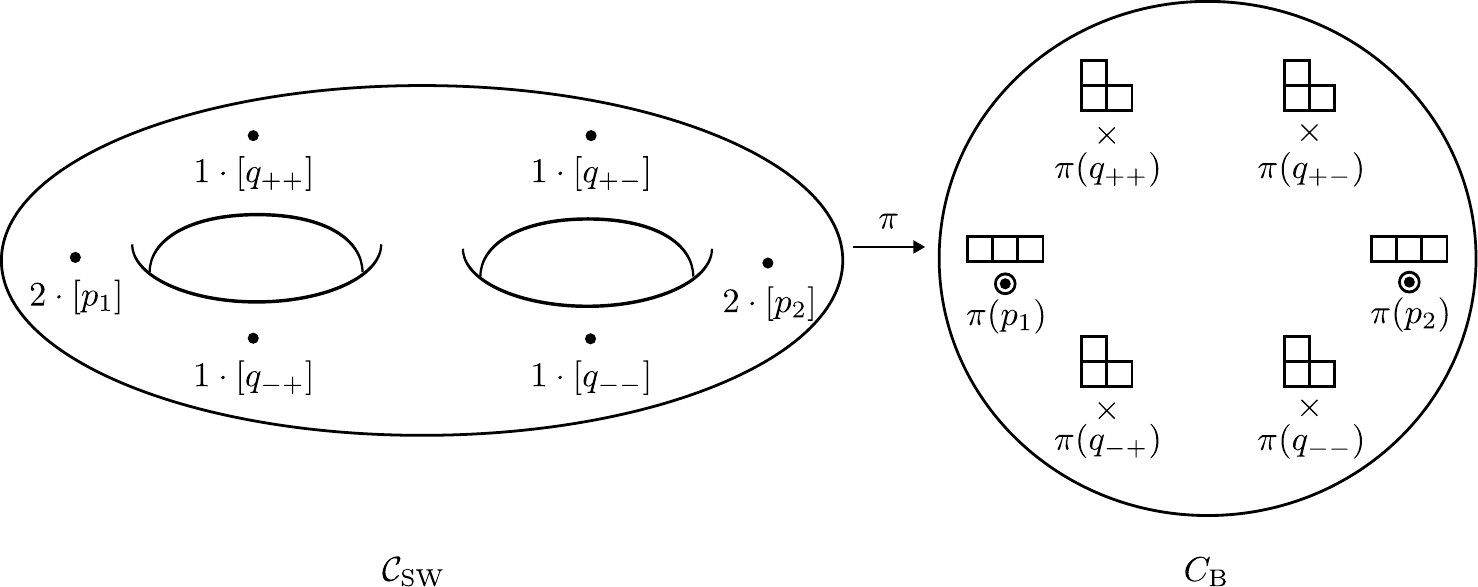}
		\caption{$\mathcal{C}_{\text{SW}}$ and $C_{\text{B}}$ for $SU(3)$ pure gauge theory}
		\label{figure:SU_3_no_matter}
	\end{center}
\end{figure}

Considering a normalization $\sigma: \mathcal{C}_{\text{SW}} \to \bar{C}_{\text{SW}}$ and a meromorphic function $\pi: \mathcal{C}_{\text{SW}} \to \mathbb{CP}^1$, we can introduce a ramification divisor $R_\pi = \sum_s (\nu_{s}(\pi) - 1)[s]$. Nontrivial ramifications may occur at
\begin{enumerate}[(1)]
	\item $\{p_i \in \mathcal{C}_{\text{SW}}\}$, where $\{\phi(p_i)\}$ are the points we add to $C_{\text{SW}}$ to compactify it,
	\item $\{q_i \in \mathcal{C}_{\text{SW}}\ |\ dt(q_i) = 0\} \Leftrightarrow \{q_i \in \mathcal{C}_{\text{SW}}\ |\ (\partial f / \partial v)(t(q_i),v(q_i)) = 0\}$.
\end{enumerate}
(1) gives us $\{p_1,\ p_2\}$ such that
\begin{align*}
	\phi(p_1) = (0,\infty),\	\phi(p_2) = (\infty, \infty),
\end{align*}
and (2) gives us $\{q_{++},\ q_{+-},\ q_{-+},\ q_{--}\}$ such that
\begin{align*}
	\phi(q_{ab}) = (t_{2ab}, v_{2a}),
\end{align*}
where $a,b = \pm1$, $v_{2a} = a\sqrt{u_2/3}$, and $t_{2ab}$ are the two roots of $f(t, v_{2a})=0$,
\begin{align*}
	t_{2ab} = \left( {v_{2a}}^3 + \frac{u_3}{2} \right) + b \sqrt{\left( {v_{2a}}^3 + \frac{u_3}{2} \right)^2 - \Lambda^6}.
\end{align*}
Using the local normalizations calculated in Section \ref{appendix:SU(3) pure gauge theory}, we get
\begin{align*}
	R_\pi = 2 \cdot [p_1] + 2 \cdot [p_2] + 1 \cdot [q_{++}] + 1 \cdot [q_{+-}] + 1 \cdot [q_{-+}] + 1 \cdot [q_{--}],
\end{align*}
and considering $\deg(\pi) = 3$ and $g(\mathcal{C}_{\text{SW}}) = 2$,
\begin{align*}
	\deg(R_\pi) = 8 = 2(g(\mathcal{C}_{\text{SW}}) + \deg(\pi) -1)
\end{align*}
is consistent with the Riemann-Hurwitz formula, Eq. (\ref{eq:riemann-hurwitz formula}). Figure \ref{figure:SU_3_no_matter} illustrates how $\pi$ works for this example. The appearance of the four branch points, $\{\pi(q_{\pm\pm})\}$, in addition to the branch points $\{\pi(p_i)\}$ that are identified with the punctures of \cite{Gaiotto:2009we}, was previously observed in \cite{Hollowood:1997pp}. 

Again, $\{\phi(q_{\pm\pm})\}$ are the ramification points of the noncompact Seiberg-Witten curve $C_{\text{SW}}$, whereas $\{\phi(p_i)\}$ are the points ``at infinity,'' therefore $\{\pi(q_{\pm\pm})\}$ are from the ramification points of $C_{\text{SW}}$.

The divisor of $\omega = \sigma^*(\lambda)$ is $(\omega) = \sum_s \nu_{s}(\omega)[s]$, and the candidates for the points on $\mathcal{C}_{\text{SW}}$ where $\omega$ has nonzero order are
\begin{enumerate}[(1)]
	\item $\{p_i \in \mathcal{C}_{\text{SW}}\}$, where $\{\phi(p_i)\}$ are the points we add to $C_{\text{SW}}$ to compactify it,
	\item $\{q_i \in \mathcal{C}_{\text{SW}}|dt(q_i) = 0\} \Leftrightarrow \{q_i \in \mathcal{C}_{\text{SW}}\ |\ (\partial f / \partial v)(t(q_i),v(q_i)) = 0\}$,
	\item $\{r_i \in \mathcal{C}_{\text{SW}}|v(r_i) = 0\}$.
\end{enumerate}
(1) and (2) result in $\{p_1, p_2\}$ and $\{q_{ab}\}$, respectively. (3) gives us $\{r_\pm\}$ such that
\begin{align*}
	\phi(r_{\pm}) = (t_{3\pm}, 0),
\end{align*}
where $t_{3\pm}$ are the two roots of $f(t,0)=0$. Using the local normalizations calculated in Appendix \ref{appendix:SU(3) pure gauge theory}, we can get 
\begin{align*}
	(\omega) = -2 \cdot [p_1] -2 \cdot [p_2] + 1 \cdot [q_{++}] + 1 \cdot [q_{+-}] + 1 \cdot [q_{-+}] + 1 \cdot [q_{--}] + 1 \cdot [r_+] + 1 \cdot [r_-],
\end{align*}
which is consistent with the Poincar\'{e}-Hopf theorem, Eq. (\ref{eq:poincare-hopf theorem}),
\begin{align*}
	\deg[(\omega)] = 2 = 2(g(\mathcal{C}_{\text{SW}}) - 1),
\end{align*}
considering $g(\mathcal{C}_{\text{SW}}) = 2$.

\begin{figure}[ht]
	\begin{center}
		\includegraphics{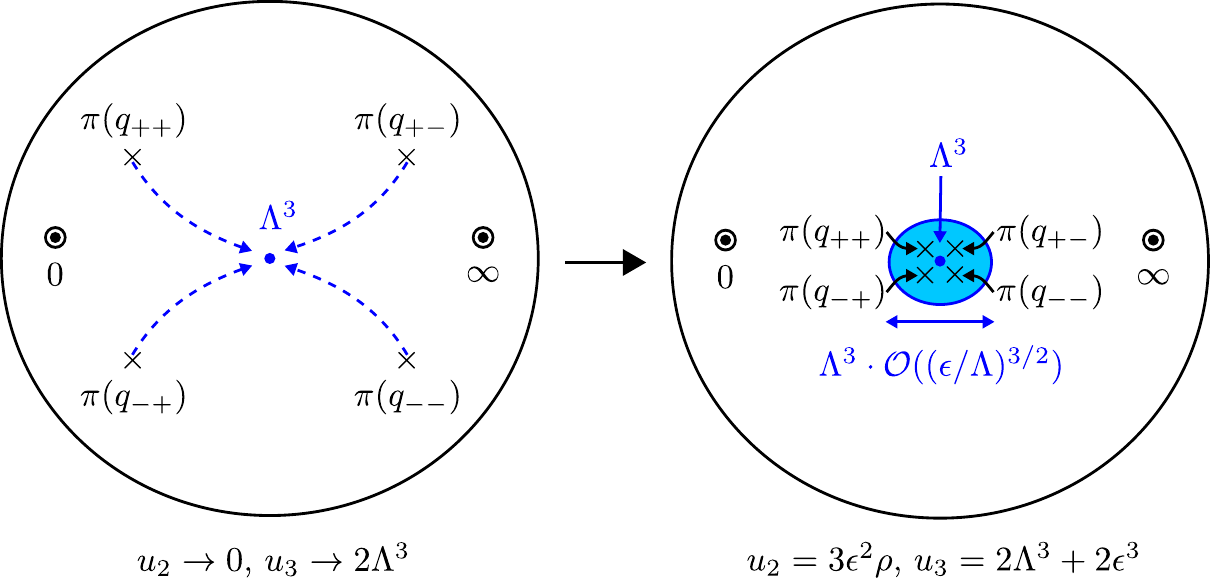}
		\caption{Behaviors of the branch points near the Argyres-Douglas fixed point}
		\label{figure:SU_3_no_matter_AD_limit}
	\end{center}
\end{figure}

Now let's consider how the branch points behave as we approach the Argyres-Douglas fixed points. As the fixed points are at $u_2 = 0$ and $u_3 = \pm 2 \Lambda^3$, let's denote the small deviations from one of the two fixed points by
\begin{align}
	u_2 &= 0 + \delta u_2 = 3 \epsilon^2 \rho, \label{eq:u_2}\\
	u_3 &= 2 \Lambda^3 + \delta u_3 = 2 \Lambda^3 + 2 \epsilon^3, \label{eq:u_3}
\end{align}
where we picked $u_3 = 2 \Lambda^3$. When $\epsilon \ll \Lambda$, 
\begin{align}
	\pi(q_{ab}) = t_{2ab} \approx \Lambda^3 \left[ 1 + b\sqrt{ 2 (1 + a \rho^{3/2}) \left( \frac{\epsilon}{\Lambda}\right)^3} \right]. \label{eq:t_2ab approx}
\end{align} 
That is, $\{\pi(q_{ab})\}$ gather together near $t = \Lambda^3$, away from $\{\pi(p_i)\}$. The four values of $t_{2ab}$ are away from $t = \Lambda^3$ by the distance of order $\Lambda^3 \cdot \mathcal{O}((\epsilon/\Lambda)^{3/2})$. Figure \ref{figure:SU_3_no_matter_AD_limit} illustrates this Coulomb branch limit.

\begin{figure}[ht]
	\begin{center}
		\includegraphics{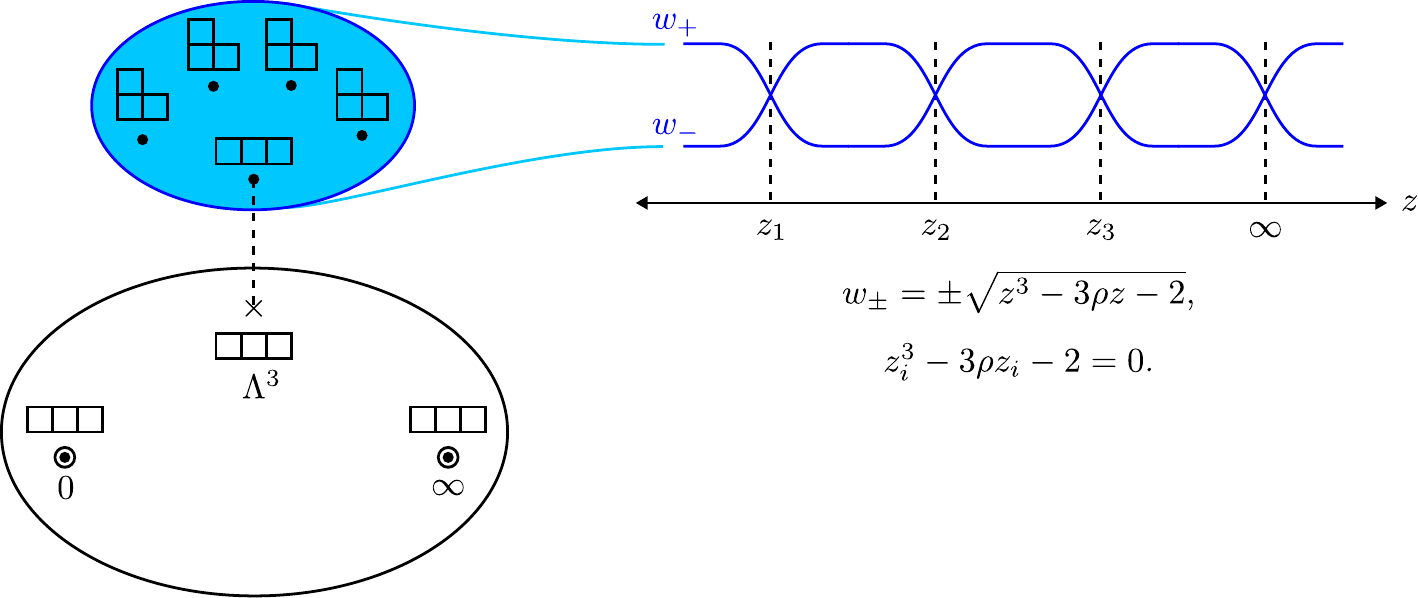}
		\caption{Appearance of a small torus at the Argyres-Douglas fixed points}
		\label{figure:SU_3_no_matter_AD_small_torus}
	\end{center}
\end{figure}

From the viewpoint of the ramification structure of the Seiberg-Witten curve, this is a similar situation to one that we have seen in Section \ref{section:SU(3) SCFT}, where we cut a Seiberg-Witten curve into two parts, giving each of them an additional point of ramification index 3. We do the same thing here, thereby getting a genus 1 curve, which is a small torus, and another genus 1 curve whose Seiberg-Witten curve is the zero locus of
\begin{align*}
	v^3 t + (t - \Lambda^3)^2,
\end{align*}
which is the curve with three branch points of ramification index 3. But this time we will try to find out the algebraic equation that describes the small torus. For that purpose it is tempting to zoom in on the part of $C_{\text{B}}$ near $t = \Lambda^3$, in such a way that every parameter has an appropriate dependence on $\epsilon$ so that we can cancel out $\epsilon$ from all of them. Considering (\ref{eq:u_2}), (\ref{eq:u_3}), (\ref{eq:t_2ab approx}), and the dimension of each parameter, a natural way to scale out $\epsilon$ is to redefine the variables as
\begin{align*}
	v &= \epsilon z, \\
	u_2 &= 0 + 3 \epsilon^2 \rho, \\
	u_3 &= 2 \Lambda^3 + 2 \epsilon^3 \\
	t &= \Lambda^3 + i (\epsilon \Lambda) ^{3/2} w.
\end{align*}
Then $f(t,v)$ becomes
\begin{align*}
	f(t, v) &= (t-\Lambda^3)^2 + \epsilon^3 (z^3 - 3 \rho z -2)t \\
		&\approx \Lambda^6 (-w^2 + z^3 - 3 \rho z -2)  ({\epsilon}/{\Lambda})^3 + \mathcal{O}(({\epsilon}/{\Lambda})^{9/2}),
\end{align*}
where we can identify a torus given by $w^2 = z^3 - 3 \rho z -2$, the same torus that appears at the Argyres-Douglas fixed points \cite{Argyres:1995jj}. Figure \ref{figure:SU_3_no_matter_AD_small_torus} illustrates this procedure.

We can also calculate the Seiberg-Witten differential $\lambda = \frac{v}{t} dt$ on the small torus,
\begin{align*}
	\lambda = \frac{v}{t}{dt} \approx \frac{\epsilon z}{\Lambda^3} \cdot i (\epsilon \Lambda)^{3/2} dw = i \frac{\epsilon^{5/2}}{\Lambda^{3/2}} z dw \propto \frac{\epsilon^{5/2}}{\Lambda^{3/2}} \frac{z (z^2 - \rho)}{w} dz,
\end{align*}
which agrees with the Seiberg-Witten differential calculated in \cite{Argyres:1995jj}.


\section{$SU(2)$ gauge theory with massive matter}
\label{section:SU(2) with massive matter}
In this section we will take a look at the cases of four-dimensional $\mathcal{N}=2$ $SU(2)$ gauge theories with massive hypermultiplets, where we can observe interesting limits of the Coulomb branch parameters and the mass parameters \cite{Argyres:1995xn}.

\subsection{$SU(2)$ gauge theory with four massive hypermultiplets}
In section \ref{section:SU(2) SCFT} we analyzed a four-dimensional $\mathcal{N}=2$ $SU(2)$ SCFT, which has four massless hypermultiplets. Here we examine a gauge theory with the same amount of supersymmetry and the same gauge group but with massive hypermultiplets, and see how mass parameters change the ramification structure of the Seiberg-Witten curve.

\begin{figure}[ht]
	\begin{center}
		\includegraphics{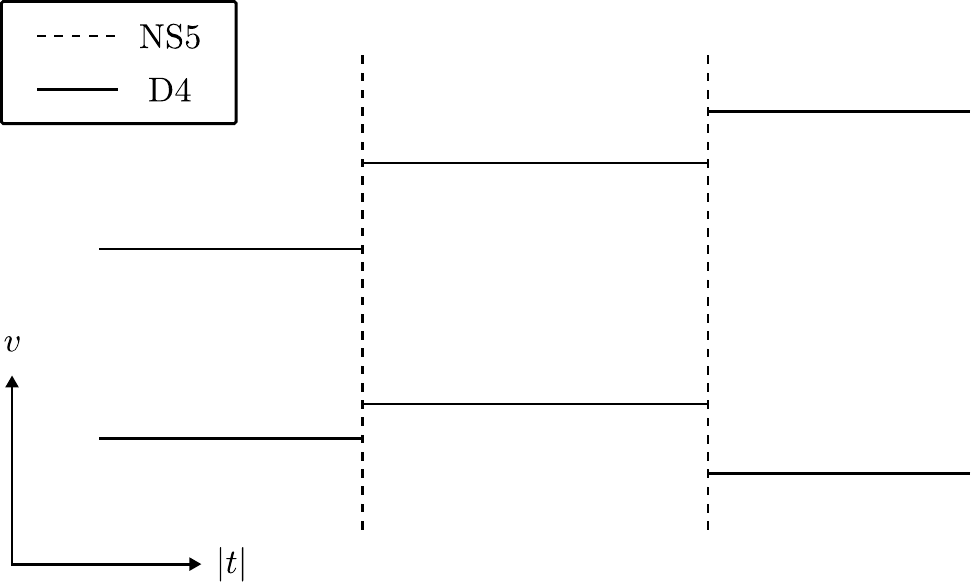}
		\caption{Brane configuration of $SU(2)$ gauge theory with four massive hypermultiplets}
		\label{figure:SU_2_four_massive_hypers_brane}
	\end{center}
\end{figure}

This gauge theory is associated to the brane configuration of Figure \ref{figure:SU_2_four_massive_hypers_brane}.
The corresponding Seiberg-Witten curve $C_{\text{SW}}$ is the zero locus of 
\begin{align}
	f(t,v) = \left(v-m_1\right)\left(v-m_3\right)t^2-\left(v^2-u_2\right)t+\left(v-m_2\right)\left(v-m_4\right)c_4, \label{eq:SU(2) with four massive hypers}
\end{align}
where $m_1$ and $m_3$ are the mass parameters of the hypermultiplets at $t = \infty$, $m_2$ and $m_4$ are the mass parameters of the hypermultiplets at $t = 0$, $u_2$ is the Coulomb branch parameter, and $c_4$ corresponds to the dimensionless gauge coupling parameter that cannot be absorbed by rescaling $t$ and $v$ \cite{Witten:1997sc}.

\begin{figure}[ht]
	\begin{center}
		\includegraphics{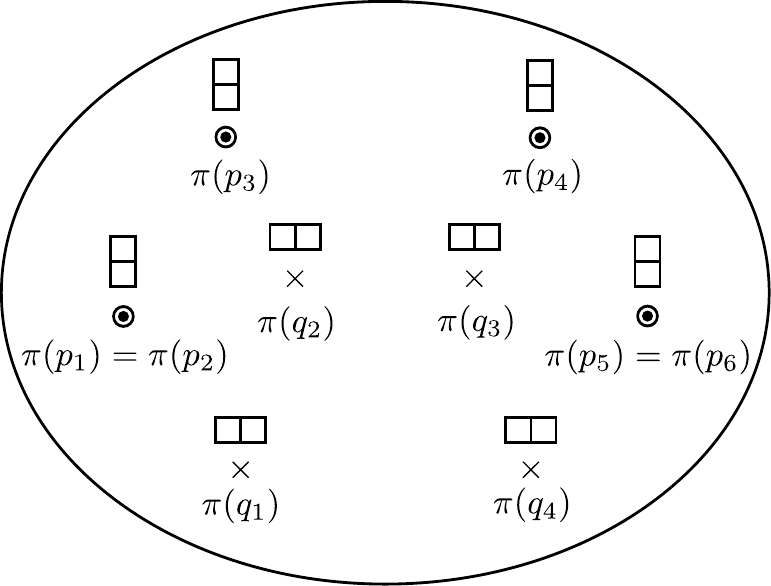}
		\caption{$C_{\text{B}}$ for $SU(2)$ gauge theory with four massive hypermultiplets}
		\label{figure:SU_2_four_massive_hypers}
	\end{center}
\end{figure}

From the usual analysis we get $C_{\text{B}}$ as shown in Figure \ref{figure:SU_2_four_massive_hypers}.
Here $\{p_i\}$ are the points on $\mathcal{C}_{\text{SW}}$  such that
\begin{align*}
	&\phi(p_1) = (0, m_2),\ \phi(p_2) = (0, m_4),\ \phi(p_3) = (t_-, \infty),\ \phi(p_4) = (t_+, \infty),\\
	&\quad \phi(p_5) = (\infty, m_1),\ \phi(p_6) = (\infty, m_3),\ t_\pm = \frac{1}{2}\left(1 \pm \sqrt{1 - 4 c_4} \right)
\end{align*}
are the points we add to $C_{\text{SW}}$ to compactify it, and $\{q_i\}$ are where $dt=0$ and whose images under $\pi$ are the four roots of 
\begin{align*}
	&\frac{1}{4} \left(m_1-m_3\right)^2 t^4 + \left(m_1 m_3 - u_2\right) t^3 +\\
	&\quad + \frac{1}{2} \left[c_4 \left(m_1 m_2 + m_2 m_3 + m_3 m_4 + m_4 m_1 - 2m_1 m_3 -2 m_2 m_4\right) + 2 u_2\right] t^2 + \\
	&\quad + c_4 \left(m_2 m_4 - u_2\right) t + \frac{1}{4} {c_4}^2 \left(m_2-m_4\right)^2.
\end{align*}
In \cite{Gaiotto:2009hg} there also appears a similar picture of branch points in the analysis of the gauge theory from the same brane configuration. Note that we made a choice among the various brane configurations that give the same four-dimensional $SU(2)$ gauge theory with four massive hypermultiplets, because each brane configuration in general results in a different ramification structure. So the choice does matter in our analysis and also when comparing our result with that of \cite{Gaiotto:2009hg}.

One notable difference from the previous examples is that $\{\pi(p_i)\}$ are not branch points. Instead we have four branch points $\{\pi(q_i)\}$ which furnish the required ramification structure. We can see that the locations of the branch points now depend also on the mass parameters in addition to the gauge coupling parameter and the Coulomb branch parameter. Note that all of the four branch points are from the ramification points of the noncompact Seiberg-Witten curve $C_{\text{SW}}$, because here the two branches of $v(t)$ do not meet ``at infinity'' with each other.

\begin{figure}[ht]
	\begin{center}
		\includegraphics{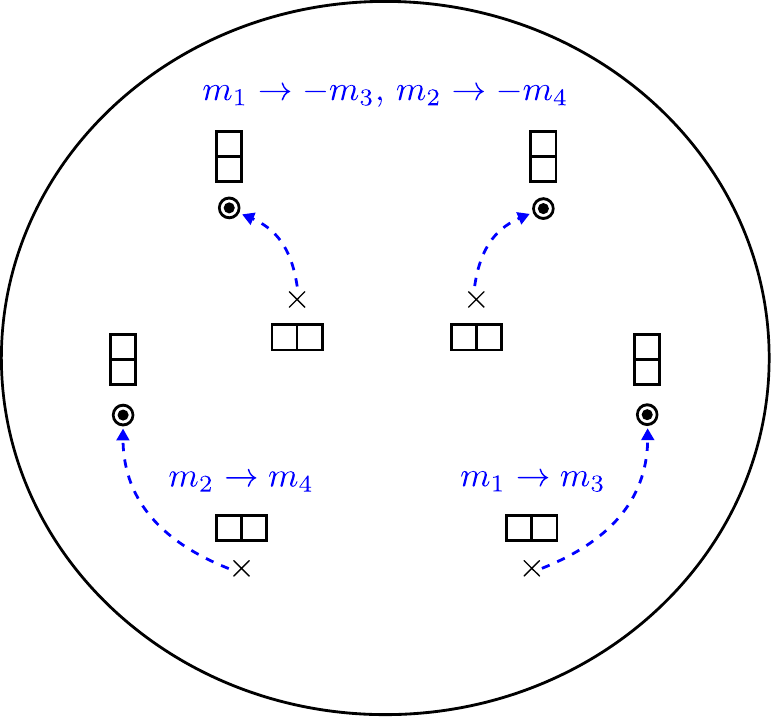}
		\caption{Behaviors of the branch points under various limits of mass parameters}
		\label{figure:SU_2_four_massive_hypers_limits}
	\end{center}
\end{figure}

This theory has four more parameters, $\{m_i\}$, when compared to $SU(2)$ SCFT. In some sense, these mass parameters represent the possible deformations of the Seiberg-Witten curve of $SU(2)$ SCFT. To understand what the deformations are, let's first see how $\{\pi(q_i)\}$ move when we take various limits of the mass parameters.
\begin{enumerate}
	\item When $m_1 \to m_3$, one of $\{\pi(q_i)\}$, say $\pi(q_4)$, moves to $t = \infty = \pi(p_5) = \pi(p_6)$.
	\item When $m_2 \to m_4$, one of $\{\pi(q_i)\}$, say $\pi(q_1)$, moves to $t = 0 = \pi(p_1) = \pi(p_2)$.
	\item When $m_1 \to -m_3$ and at the same time $m_2 \to -m_4$, $\pi(q_2)$ moves to $t = t_- = \pi(p_3)$ and $\pi(q_3)$ moves to $t = t_+ = \pi(p_4)$.
\end{enumerate}
The first limit corresponds to bringing the two points of $\mathcal{C}_{\text{SW}}$, $p_5$ and $p_6$, together to one point, thereby developing a ramification point of index 2 there. The others can also be understood in a similar way. Figure \ref{figure:SU_2_four_massive_hypers_limits} illustrates these limits.

\begin{figure}[ht]
	\begin{center}
		\includegraphics{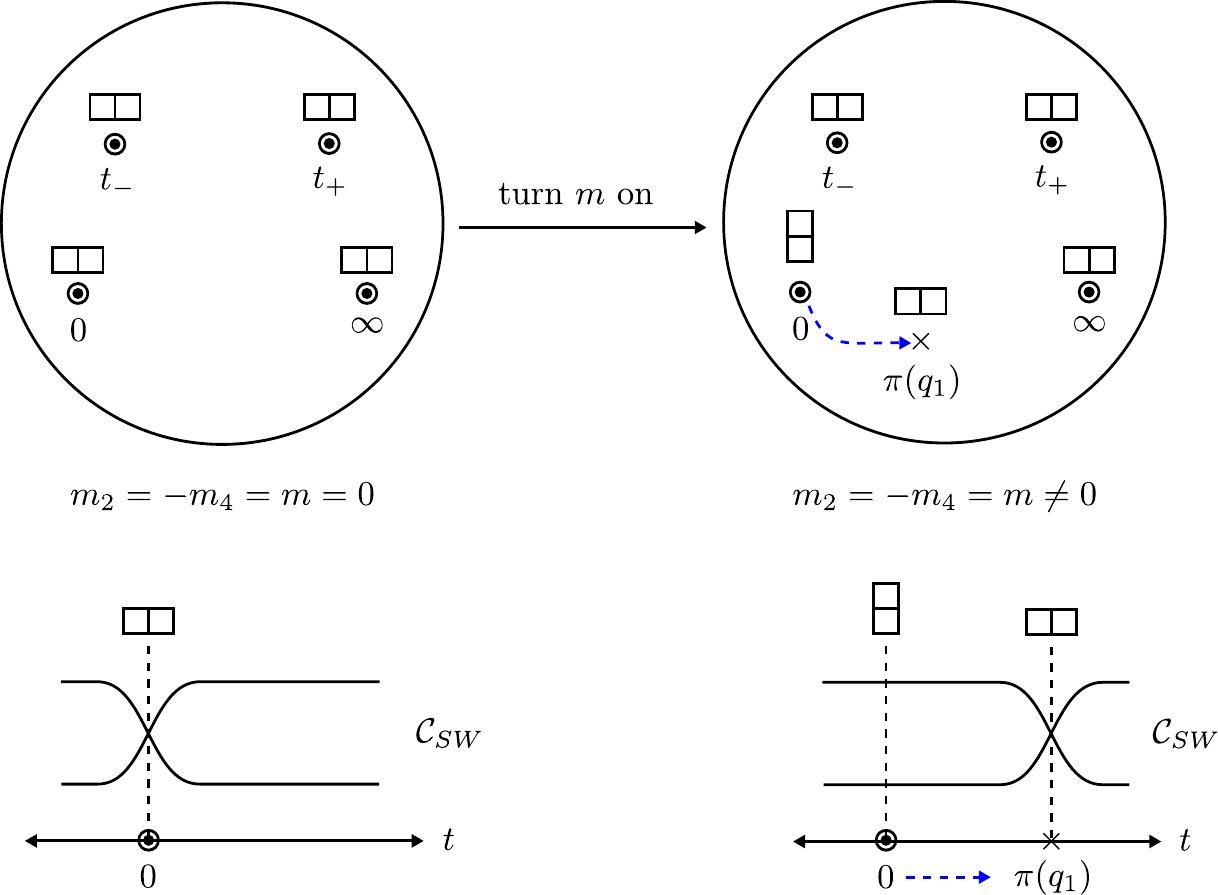}
		\caption{Removal of the branch point at $t=0$ when we turn $m_2 = -m_4 = m$ on}
		\label{figure:SU_2_four_massive_hypers_mass_on}
	\end{center}
\end{figure}

Note that we can get the Seiberg-Witten curve of $SU(2)$ SCFT by setting all the mass parameters of Eq. (\ref{eq:SU(2) with four massive hypers}) to zero, which corresponds to taking all of the limits at the same time, thereby sending each $\pi(q_i)$ to one of $\{\pi(p_i)\}$ and turning $\{\pi(p_i)\}$ into four branch points as expected.

Now we turn the previous arguments on its head and see how we can deform the Seiberg-Witten curve of $SU(2)$ SCFT by turning on mass parameters. As an example, let's consider turning on $m_2 = - m_4 = m$. When $m=0$, there is a branch point at $t=0$. Now we turn $m$ on, then this separates the two sheets at $t=0$, and $t=0$ is no longer a branch point. But the topological constraint by Riemann-Hurwitz formula requires four branch points to exist, and indeed a new branch point that corresponds to $\pi(q_1)$ develops. Figure \ref{figure:SU_2_four_massive_hypers_mass_on} illustrates this deformation.

The other mass parameters can also be understood in a similar way as deformations that detach the sheets meeting at the branch points from each other, and the most general deformation will result in the Seiberg-Witten curve of $SU(2)$ gauge theory with four massive hypermultiplets, the theory we started our analysis here. 


\subsection{$SU(2)$ gauge theory with two massive hypermultiplets}
\begin{figure}[ht]
	\begin{center}
		\includegraphics{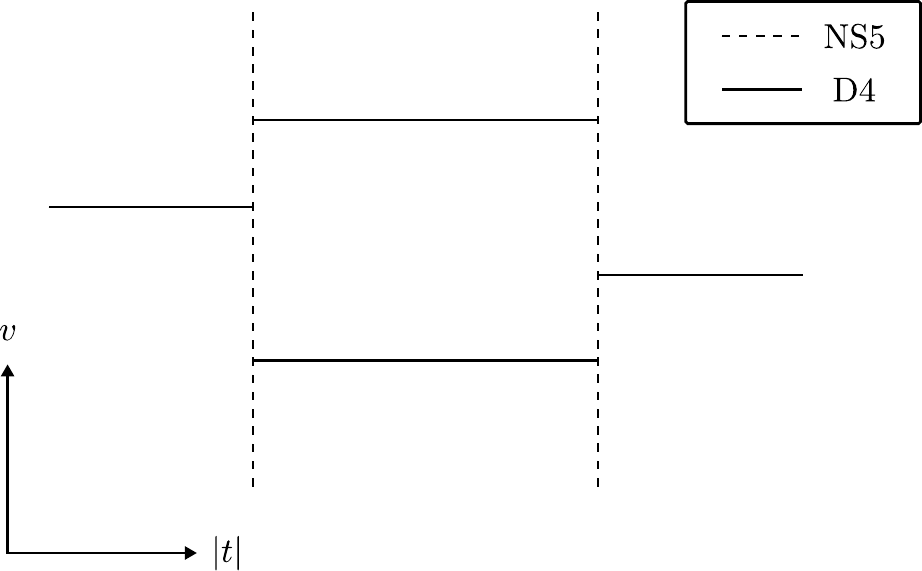}
		\caption{Brane configuration of $SU(2)$ gauge theory with two massive hypermultiplets, with symmetric distribution of D4-branes}
		\label{figure:SU_2_two_massive_hypers_symmetric_brane}
	\end{center}
\end{figure}

Now we examine the example of a four-dimensional $\mathcal{N}=2$ supersymmetric $SU(2)$ gauge theory with two massive hypermultiplets. As mentioned earlier, there are various ways in constructing the brane configuration associated to the four-dimensional theory. One possible brane configuration is shown in Figure \ref{figure:SU_2_two_massive_hypers_symmetric_brane}, where two D4-branes that provide the massive hypermultiplets are distributed symmetrically on both sides.

The corresponding Seiberg-Witten curve $C_{\text{SW}}$ is the zero locus of
\begin{align}
	f(t,v) = (v - m_1) t^2 - (v^2 - u_2) t + (v - m_2) {\Lambda}^2, \label{eq:SU(2) N_f=2 symmetric SW curve}
\end{align}
where $u_2$ is the Coulomb branch parameter, $m_1$ and $m_2$ are the mass parameters, and $\Lambda$ is the dynamically generated scale of the theory.

\begin{figure}[ht]
	\begin{center}
		\includegraphics{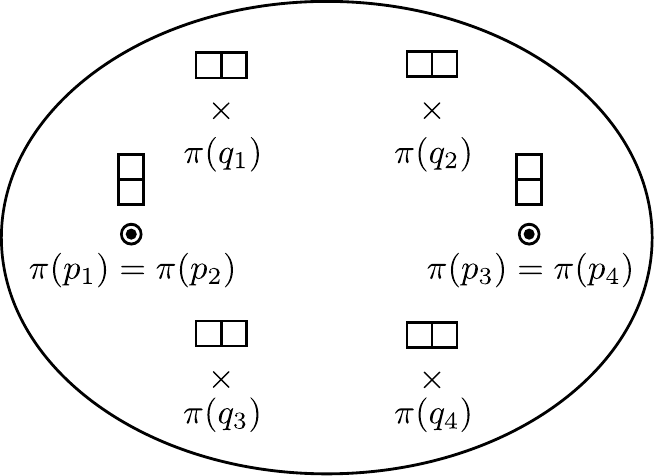}
		\caption{$C_{\text{B}}$ for $SU(2)$ gauge theory with two massive hypermultiplets when the brane configuration is symmetric}
		\label{figure:SU_2_two_massive_hypers_symmetric}
	\end{center}
\end{figure}

The usual analysis gives $C_{\text{B}}$ as shown in Figure \ref{figure:SU_2_two_massive_hypers_symmetric}. $\{p_i\}$ are the points on $\mathcal{C}_{\text{SW}}$ such that 
\begin{align*}
	\phi(p_1) = (0, m_2),\ \phi(p_2) = (0, \infty),\ \phi(p_3) = (\infty, m_1),\ \phi(p_4) = (\infty, \infty) 
\end{align*}
are the points we add to $C_{\text{SW}}$ to compactify it. Note that here $\pi(p_1) = \pi(p_2) = 0$ and $\pi(p_3) = \pi(p_4) = \infty$ are not branch points. There are four branch points $\{\pi(q_i)\}$ whose locations on $C_{\text{B}}$ are given by the four roots $\{t_i\}$ of the following equation.
\begin{align*}
	\frac{1}{4} t^4 - m_1 t^3 + \left(u_2 + \frac{{\Lambda}^2}{2}\right) t^2 - m_2 {\Lambda}^2 t + \frac{{\Lambda}^4}{4}. 
\end{align*}
We can see that the locations of $\{\pi(q_i)\}$ now depend also on the mass parameters in addition to the Coulomb branch parameter and the scale. Again the branch points come from the ramification points of the noncompact Seiberg-Witten curve $C_{\text{SW}}$. In \cite{Gaiotto:2009hg} there also appears a similar picture of branch points in the analysis of the gauge theory from the same symmetric brane configuration.

\begin{figure}[ht]
	\begin{center}
		\includegraphics{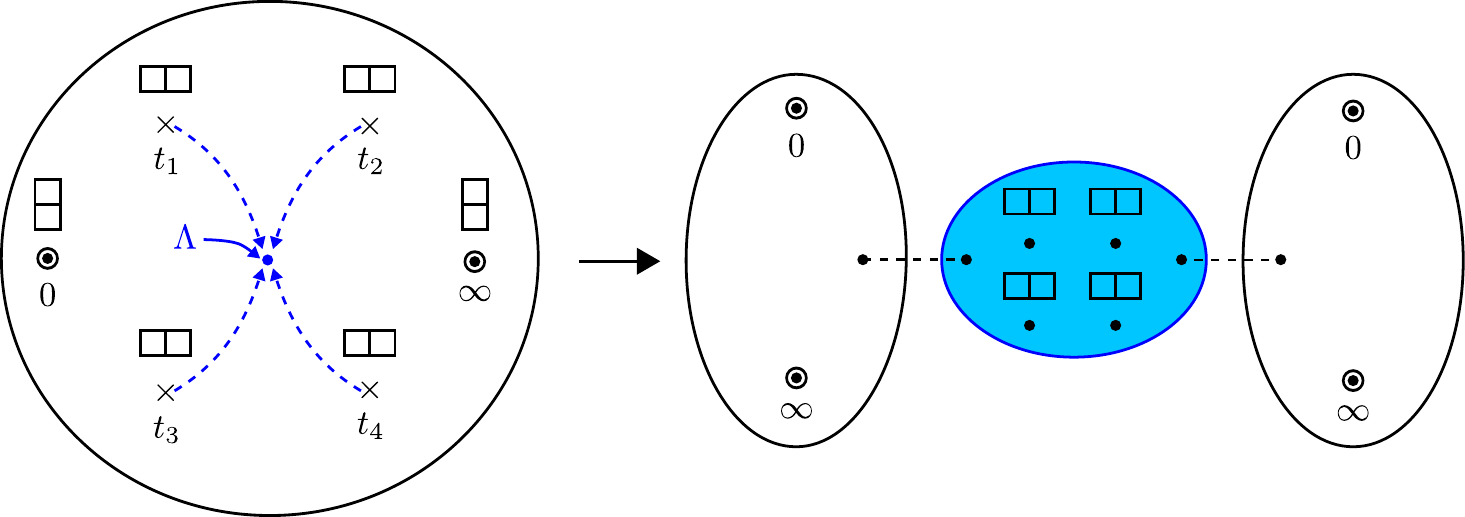}
		\caption{Behaviors of the branch points when $m_1 = m_2 \to {\Lambda}$, $u_2 \to {\Lambda}^2$}
		\label{figure:SU_2_two_massive_hypers_symmetric_limit}
	\end{center}
\end{figure}
When we take the limit of $m_1 = m_2 \to {\Lambda}$ and $u_2 \to {\Lambda}^2$, the four branch points approach $t = \Lambda$. Figure \ref{figure:SU_2_two_massive_hypers_symmetric_limit} illustrates the behavior of the branch points under the limit. This is a similar situation of four branch points of index 2 gathering together around a point as we have seen in Sections \ref{section:SU(3) SCFT} and \ref{section:SU(3) pure gauge theory}. Imagine cutting off a small region of the Seiberg-Witten curve around the preimages of the branch points when we are in the vicinity of the limit. Going around the four branch points makes a complete journey, that is, we can come back to the branch of $v(t)$ where we started, which implies that adding a point of ramification index 1 to each branch of the excised part of the curve gives us a compact small torus. After cutting off the region containing the preimages of the four branch points and adding a point to each branch, the two branches of the remaining part of the original Seiberg-Witten curve become two Riemann spheres. This can also be seen by taking the Coulomb branch limit of the parameters in Eq. (\ref{eq:SU(2) N_f=2 symmetric SW curve}), which results in two components that have no ramification over $t$, that is, two Riemann spheres. Therefore we can identify a small torus and see nonlocal states becoming massless simultaneously as the cycles around the two of the four branch points vanish as we take the limit. It would be interesting to find out the explicit expression for the small torus as we did in Section \ref{section:SU(3) pure gauge theory}, where we found the algebraic equation that describes the small torus of Argyres-Douglas fixed points, and to compare the small torus with the result of \cite{Argyres:1995xn}.

\begin{figure}[ht]
	\begin{center}
		\includegraphics{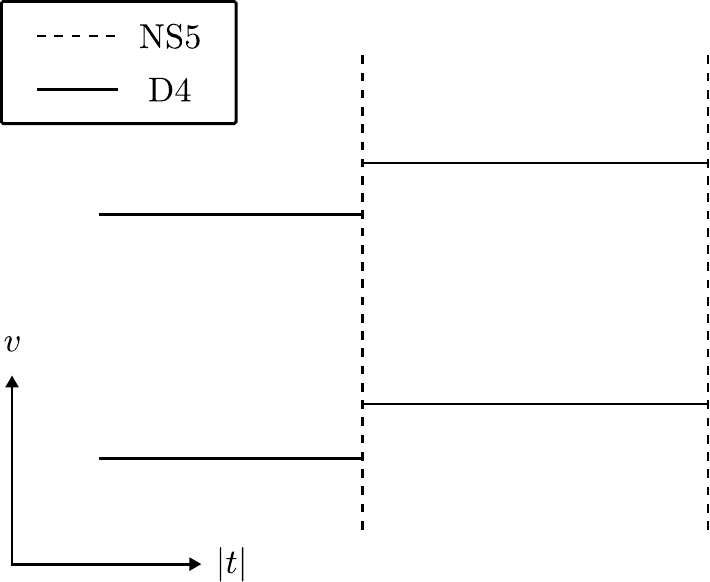}
		\caption{Brane configuration of $SU(2)$ gauge theory with two massive hypermultiplets, with asymmetric distribution of D4-branes}
		\label{figure:SU_2_two_massive_hypers_asymmetric_brane}
	\end{center}
\end{figure}

We have another brane configuration that gives us the same four-dimensional physics, which is shown in Figure \ref{figure:SU_2_two_massive_hypers_asymmetric_brane}. Now the D4-branes that provide massive hypermultiplets are on one side only, thereby losing the symmetry of flipping $t$ to its inverse and swapping $m_1$ and $m_2$.

\begin{figure}[ht]
	\begin{center}
		\includegraphics{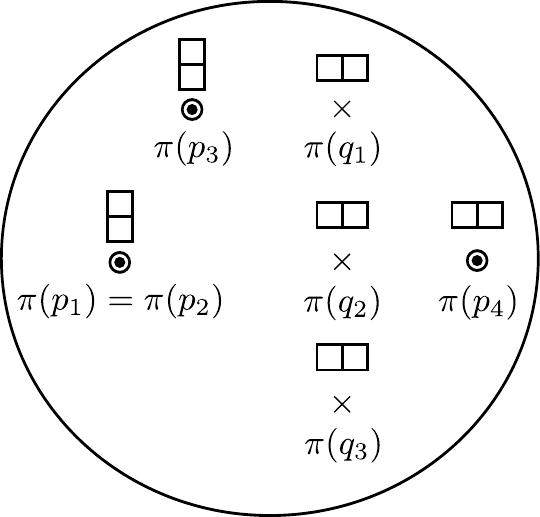}
		\caption{$C_{\text{B}}$ for $SU(2)$ gauge theory with two massive hypermultiplets when the brane configuration is not symmetric}
		\label{figure:SU_2_two_massive_hypers_asymmetric}
	\end{center}
\end{figure}

The corresponding Seiberg-Witten curve $C_{\text{SW}}$ is the zero locus of
\begin{align}
	f(t,v) = {\Lambda}^2 t^2 - (v^2 - u_2)t + (v - m_1)(v - m_2). \label{eq:SU(2) N_f=2 asymmetric SW curve}
\end{align}
After the usual analysis, we can find $C_{\text{B}}$ as shown in Figure \ref{figure:SU_2_two_massive_hypers_asymmetric}. 
Here $\{p_i\}$ are the points on $\mathcal{C}_{\text{SW}}$ such that
\begin{align*}
	\phi(p_1) = (0, m_1),\ \phi(p_2) = (0, m_2),\ \phi(p_3) = (1, \infty),\ \phi(p_4) = (\infty, \infty),
\end{align*}
are the points we add to $C_{\text{SW}}$ to compactify it. Note that $\pi(p_1) = \pi(p_2) = 0$ and $\pi(p_3) = 1$ are not branch points in this case, because each of them has a trivial ramification there as indicated with the corresponding Young tableau. $\pi(p_4) = \infty$ is a branch point. The locations of the other three branch points $\{\pi(q_i)\}$ are given by the three roots $\{t_i\}$ of Eq. (\ref{eq:SU_2_two_massive_hypers_asymmetric}).
\begin{align}
	{\Lambda}^2 t^3 + \left(u_2 - {\Lambda}^2\right)t^2 + (m_1 m_2 - u_2) t + \left( \frac{m_1 - m_2}{2}\right)^2 = 0. \label{eq:SU_2_two_massive_hypers_asymmetric}
\end{align}
Again we see that the locations of $\{\pi(q_i)\}$ depend on the mass parameters as well as the Coulomb branch parameters and the scale. $\{\pi(q_i)\}$ are distringuished from $\pi(p_4)$ in that they are from the ramification points of the noncompact Seiberg-Witten curve $C_{\text{SW}}$. In \cite{Gaiotto:2009hg} there also appears a similar picture of branch points in the analysis of the gauge theory from the asymmetric brane configuration.

\begin{figure}[ht]
	\begin{center}
		\includegraphics{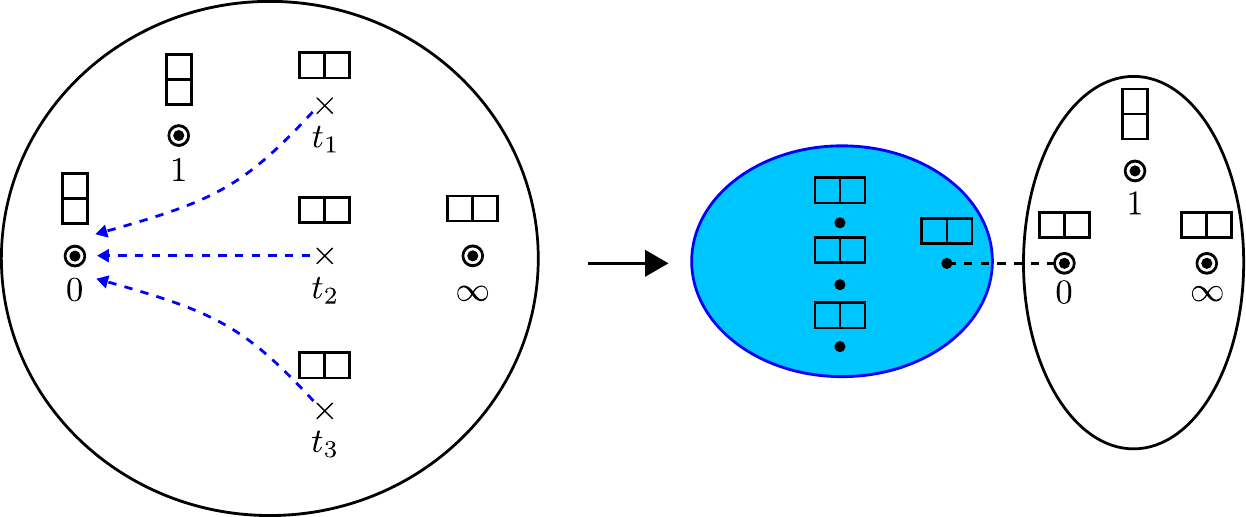}
		\caption{Behaviors of the branch points when $m_1 = m_2 \to {\Lambda}$, $u_2 \to {\Lambda}^2$}
		\label{figure:SU_2_two_massive_hypers_asymmetric_limit}
	\end{center}
\end{figure}

From Eq. (\ref{eq:SU_2_two_massive_hypers_asymmetric}), we can easily identify the limits of the parameters that send $\{\pi(q_i)\}$ to $t=0$. That is,
\begin{enumerate}[(1)]
	\item When $m_1 = m_2 = m$, $t_1 \to 0$.
	\item When $m^2 = u_2$, $t_1$ and $t_2 \to 0$.
	\item When $m = {\Lambda}^2$, $t_1$, $t_2$, and $t_3 \to 0$.
\end{enumerate}
The case of (3) is illustrated in the left side of Figure \ref{figure:SU_2_two_massive_hypers_asymmetric_limit}.
Note that when we take the limit of $m_1 = m_2 \to {\Lambda}$ and $u_2 \to {\Lambda}^2$, the three branch points go to $t = 0$ and we can see that there are nonlocal states that become massless together in the limit. This is the same limit of the parameters as the one in the previous case of different brane configuration, a symmetric brane configuration. Therefore we observe the phenomenon of seemingly different brane configurations giving the same four-dimensional physics. 

However, unlike the previous case of symmetric brane configuration, where there are four branch points with ramification index 2 that are coming together under the limit, here there are only three of them moving toward a point as we take the limit. But note that while in the previous case going around the four branch points once gets us back to where we started, here going around the three branch points once does not complete a roundtrip and we need one more trip to get back to the starting point. This implies that, when excising the part of the Seiberg-Witten curve where the preimages of the three branch points come together, the monodromy around the region corresponds to a point of ramification index 2.
After we cut the curve into two parts, we have one curve with four branch points of ramification index 2, which is a small torus, and the other curve with two branch points of ramification index 2, which is a Riemann sphere. This procedure is illustrated in the right side of Figure \ref{figure:SU_2_two_massive_hypers_asymmetric_limit}. This can also be seen by taking the Coulomb branch limit of the parameters of Eq. (\ref{eq:SU(2) N_f=2 asymmetric SW curve}), which gives us a curve with two ramification points of index 2, the Riemann sphere.

\section{Discussion and outlook}
\label{section:discussion and outlook}
Here we illustrated, through several examples, that when a Seiberg-Witten curve of an $\mathcal{N}=2$ gauge theory has a ramification over a Riemann sphere $C_{\text{B}}$, some of the branch points on $C_{\text{B}}$ can be identified with the punctures of \cite{Gaiotto:2009we} but in general there are additional branch points from the ramification points of the Seiberg-Witten curve, whose locations on $C_{\text{B}}$ depend on various parameters of the theory and therefore can be a useful tool when studying various limits of the parameters, including Argyres-Seiberg duality and the Argyres-Douglas fixed points. Note that interesting phenomena happen when the branch points collide with each other. This is because those cases are exactly when the corresponding Seiberg-Witten curve becomes singular. The merit of utilizing the branch points compared to the direct study of Seiberg-Witten curves is that it becomes more evident and easier to analyze when and how those limits of the parameters occur, as Gaiotto used his punctures and their collisions to investigate various corners of the moduli space of gauge coupling parameters.

\begin{figure}[ht]
	\begin{center}
		\includegraphics{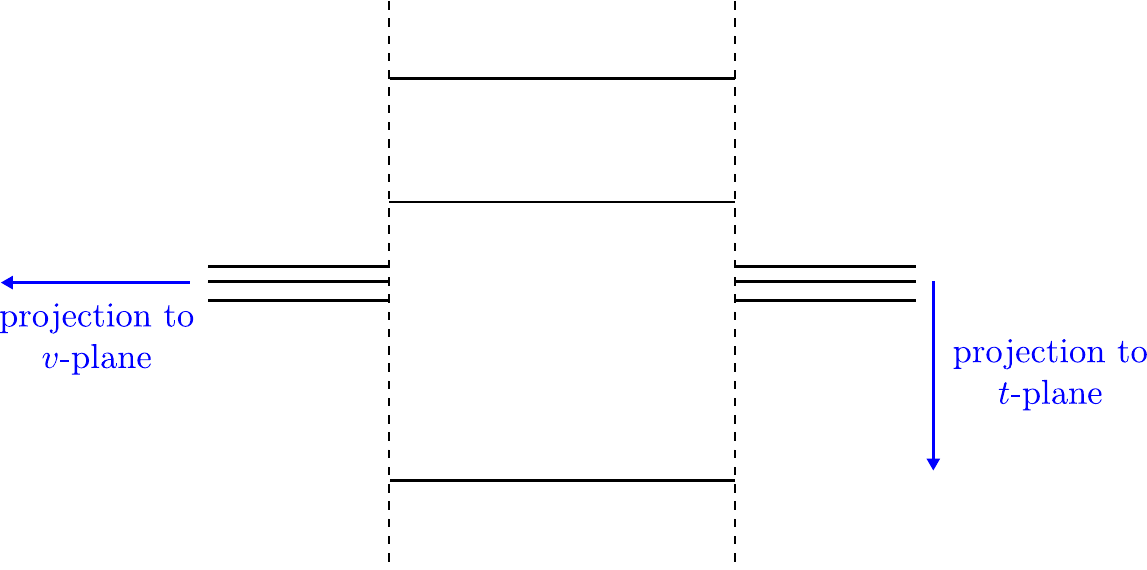}
		\caption{Two different ways of projecting a Seiberg-Witten curve onto a complex plane}
		\label{figure:two_projections}
	\end{center}
\end{figure}

Branch points have played a major role since the inception of the Seiberg-Witten curve. What is different here is that we change the point of view such that we can find branch points in a way that is compatible with the setup of \cite{Gaiotto:2009we}, which enables us to complement and utilize its analysis. This change of the perspective can be illustrated as shown in Figure \ref{figure:two_projections}, which shows a brane configuration of an $SU(3)$ SCFT.

If we want to project the whole Seiberg-Witten curve onto a complex plane, there are two ways: one is projecting the curve onto the $t$-plane, and the other is projecting it onto the $v$-plane. In the original study of \cite{Seiberg:1994rs, Seiberg:1994aj} and in the following extensions of the analysis \cite{Argyres:1994xh, Klemm:1994qs, Argyres:1995wt, Hanany:1995na, Danielsson:1995is, Brandhuber:1995zp, Argyres:1995fw}, the analyses of Seiberg-Witten curves have been done usually by projecting the curve onto the $v$-plane so that it can be seen as a branched two-sheeted cover over the complex plane. Then the branch points are such that the corresponding ramification points on the Seiberg-Witten curve have the same ramification index of 2, because a point on a Seiberg-Witten curve has the ramification index of either 2 or 1 when considering a two-sheeted covering map. 

\begin{figure}
	\centering
	\subfloat[]
		{\label{subfigure:puncture} \includegraphics{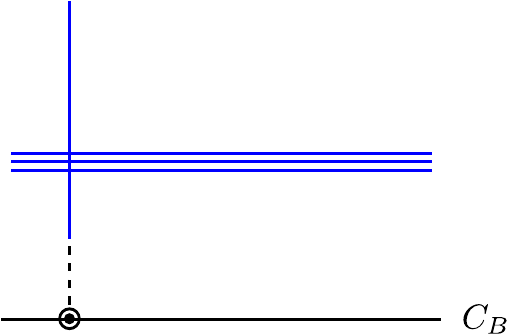}}\quad 
	\subfloat[]
		{\label{subfigure:ramification} \includegraphics{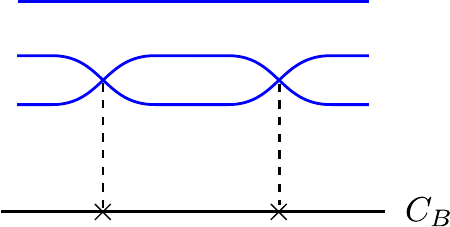}}
	\caption{Configuration of M5-branes around (a) a puncture and (b) ramification points}
	\label{figure:puncture_vs_ramification}
\end{figure}

But here we project the Seiberg-Witten curve onto the $t$-plane such that the curve is a three-sheeted cover over the complex plane. This way of projection, which previously appeared in \cite{Martinec:1995by} and re-popularized by Gaiotto \cite{Gaiotto:2009we}, makes it easier to understand the physical meaning of the branch points. When considering a Seiberg-Witten curve as a two-dimensional subspace of an M5-brane \cite{Klemm:1996bj}, Gaiotto told us that the M5-brane can be described as a deformation of several coincident M5-branes wrapping a Riemann surface plus M5-branes meeting the coincident M5-branes transversely at the location of punctures. From the viewpoint of the coincident M5-branes, a transverse M5-brane is heavy and therefore can be considered as an operator when studying the theory living on the coincident M5-branes. See Figure \ref{subfigure:puncture}, which illustrates the configuration of M5-branes at a puncture and their projection onto $C_{\text{B}}$. Therefore when we project the Seiberg-Witten curve onto the $t$-plane, the branch points that are identified with the punctures can be related to the locations of the transverse M5-branes. 

In comparison to that, the branch points that are not identified with the punctures come from the ramification points of the single noncompact M5-brane, which was the coincident M5-branes before turning on the Coulomb branch parameters of the theory. Nonzero Coulomb branch parameters make them move away from each other, and the result is one smooth but ramified M5-brane whose two-dimensional subspace is interpreted as the Seiberg-Witten curve of the theory. Figure \ref{subfigure:ramification} illustrates two ramification points of a ramified M5-brane and their projection onto $C_{\text{B}}$. If we consider the Seiberg-Witten curve as coming from several sheets of M5-branes, a ramification point of the curve is where those M5-branes come into a contact \cite{Gaiotto:2009hg}. It would be interesting if we can investigate the local physics around these points.

Formulating a cookbook-style procedure of constructing our $C_{\text{B}}$ not from the analysis starting from the equation of a Seiberg-Witten curve but from the punctured Riemann surface of \cite{Gaiotto:2009we} with topological constraints, Coulomb branch parameters, and mass parameters would be interesting. Finding out how many of them are there and what ramification index each of them has will not be a difficult job. For example, when a Seiberg-Witten curve has genus 1 and if we know how many points of nontrivial ramification index we have to add to $C_{\text{SW}}$ to compactify it, say $n$ of them, then there should be $(4 - n)$ additional branch points on $C_{\text{B}}$ because the Riemann-Hurwitz formula requires $C_{\text{B}}$ to have four branch points in this case. We can do a similar job for the other cases. What is difficult is to figure out the dependence of the locations of the branch points on various parameters of the Seiberg-Witten curve, including gauge coupling parameters, Coulomb branch parameters, and mass parameters. If there is a way to see the dependence without the long and tedious analysis we presented here, it will be helpful for pursuing many interesting limits of the parameters.

As we have focused only on the local description near each branch point, there is an ambiguity of how to patch the local descriptions into a global one, because branches can be permuted by the monodromy of the parameters. It would be helpful if we can clear up that ambiguity explicitly.
%





\section*{Acknowledgments}

It is a great pleasure for the author to express sincere thanks to Sergei Gukov who provided precious advice at the various stages of the development of this work, and to John H. Schwarz for enlightening discussions at the finalizing stage of this work, careful reading of this manuscript, and generous support in many ways. The author also thanks Yuji Tachikawa for detailed historical comments. The author  thanks Heejoong Chung, Petr Ho\v{r}ava, Christoph Keller, Sangmin Lee, Sungjay Lee, Yu Nakayama, Jaewon Song, and Piotr Su{\l}kowski for helpful discussions. The author is grateful to the organizers of the 8th Simons Workshop in Mathematics and Physics at the Stony Brook University, where part of this work was completed, for their great hospitality. This work is supported in part by a Samsung Scholarship.

\input{Ramification_Appendix.tex}

\bibliographystyle{JHEP}
\bibliography{Ramification}

\end{document}

%% file: Ramification_Appendix.tex
\appendix

\section{Normalization of a singular algebraic curve}
\label{appendix:normalization}

\begin{figure}[ht]
	\begin{center}
		\includegraphics{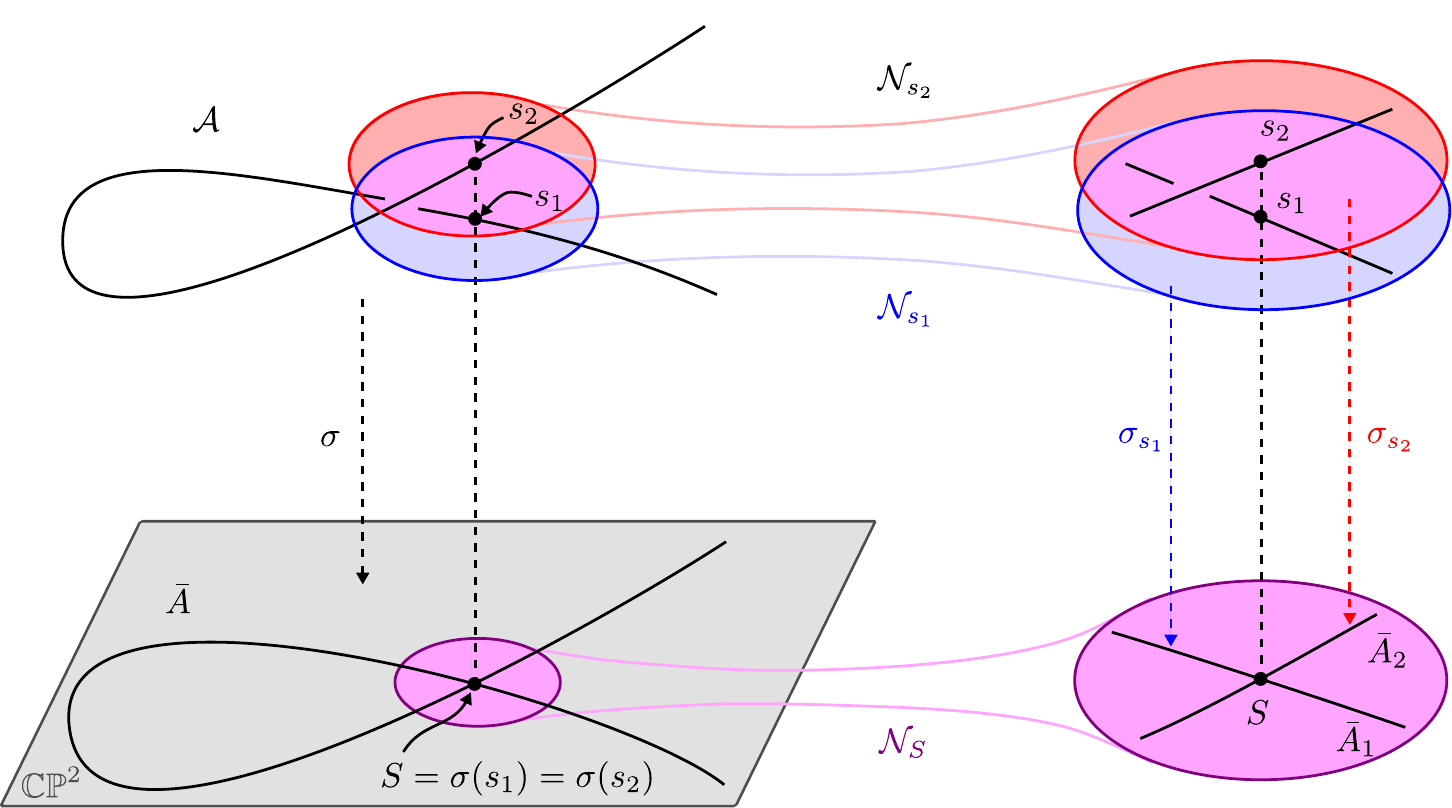}
		\caption{Schematic description of the normalization of a singular curve}
		\label{figure:normalization_schematic}
	\end{center}
\end{figure}

To understand how normalization works, let's try to normalize a curve with a singularity, $\bar{A} \subset \mathbb{CP}^2$. The left side of Figure \ref{figure:normalization_schematic} illustrates how a singularity of $\bar{A}$ is resolved when we normalize it to a smooth curve $\mathcal{A} = \sigma^{-1}(\bar{A})$ by finding a map $\sigma$. 
There are various kinds of singular points, and the case illustrated here is that $\bar{A}$ has two tangents at the singular point $S = \sigma(s_1)=\sigma(s_2)$, which corresponds to two different points $\sigma^{-1}(S) = \{s_1,\ s_2\}$ on $\mathcal{A}$.\footnote{A similar kind of singularity occurs at $(z,w)=(0,0)$ of a curve defined by $zw=0$ in $\mathbb{C}^2$, which can be lifted if we consider embedding the curve into $\mathbb{C}^3$ and moving $z=0$ and $w=0$ complex planes away from each other along the other complex dimension normal to both of them.} Without any normalization, $\bar{A}$ is an irreducible curve that is singular at $S$. After the normalization we get a smooth irreducible curve $\mathcal{A}$.

Finding such $\sigma$ that works over all $\bar{A}$ will not be an easy job, especially because we don't know how to describe $\mathcal{A}$ globally. However if we are interested only in analyzing a local neighborhood of a point on $\bar{A}$, we do not need to find $\sigma$ that maps the whole $\mathcal{A}$ to the entire $\bar{A}$, but finding a local normalization \cite{Griffiths} of $\bar{A}$ near the point will be good enough for that purpose. What is good about this local version of normalization is that we know how to describe $\mathcal{A}$ locally. That is, because $\mathcal{A}$ is a Riemann surface, we can choose a local coordinate $s \in \mathbb{C}$ on $\mathcal{A}$ such that $s_i = 0$. Then a local normalization is described by a map $\sigma_{s_i}$ from the neighborhood of $s_i \in \mathcal{A}$ to the neighborhood of $S \in \bar{A}$.
\begin{align*}
	\sigma_{s_i}: \mathcal{N}_{s_{i}} \to \mathcal{N}_{S},\ s \mapsto (x(s),y(s)),
\end{align*}
where $(x,y)$ is a coordinate system of $\mathbb{C}^2 \subset \mathbb{CP}^2$ such that $S = (0,0)$. Or if we see $\sigma_{s_i}$ as a map into a subset of $\mathbb{CP}^2$ when $S = [X_S,Y_S,Z_S] = [X_S/Z_S, Y_S/Z_S, 1]$, 
\begin{align*}
	\sigma_{s_i}: \mathcal{N}_{s_{i}} \to \mathcal{N}_{S},\ s \mapsto [X_S/Z_S + x(s),\ Y_S/Z_S + y(s),\ 1].
\end{align*}
We can sew up the local normalizations to get a global normalization if we have enough of them to cover the whole curve. 

Now let's get back to the case of Figure \ref{figure:normalization_schematic} and find its local normalizations. Schematic descriptions of the local normalizations are shown in the right side of Figure \ref{figure:normalization_schematic}. When we zoom into the neighborhood $\mathcal{N}_{S}$ of the singular point $S$ on $\bar{A}$, we see a reducible curve, called the local analytic curve \cite{Griffiths} of $\bar{A}$ at $S$, with two irreducible components $\{\bar{A}_1,\bar{A}_2\}$, where each component $\bar{A}_i$ is coming from a part of $\mathcal{A}$. By choosing $\mathcal{N}_{S}$ as small as possible, we can get a good approximation of $\bar{A}$ at $S$ by the local analytic curve $f_{S}(x,y)=0$. Because we have two irreducible component for the local analytic curve illustated here, we can factorize $f_{S}(x,y)$ into its irreducible components $f_{s_i}(x,y)$, i.e. $f_{S}(x,y) = f_{s_1}(x,y) f_{s_2}(x,y)$, each giving us the local description of the component. Then we find a local normalization $\sigma_{s_i}(s) = (x(s),y(s))$ for each component defined as the zero locus of $f_{s_i}(x(s),y(s)) $.

\section{Calculation of local normalizations}
Calculation of a local normalization of a curve near a point is done here by finding a Puiseux expansion \cite{Kirwan} of the curve at the point. Puiseux expansion is essentially a convenient way to get a good approximation of a curve in $\mathbb{CP}^2$ around a point $P$ on the curve. That is, for a local analytic curve defined as $f_P(x,y)=0$, the solutions of the equation, which describes the different branches of the curve at $P$, is called Puiseux expansions of the curve at $P$. 

When the local analytic curve is irreducible, as we go around $P$ the branches of the local analytic curve at $P$ are permuted among themselves transitively. But when it is reducible, for example into two components like the case we saw in Appendix \ref{appendix:normalization}, the permutations happen only among the branches of each component.

\subsection{$SU(2)$ SCFT}
\label{appendix:SU(2)}
We showed in Section \ref{section:SU(2) SCFT}
how to compactify the Seiberg-Witten curve of $SU(2)$ SCFT. So let's start with the compactified curve, $\bar{C}_{\text{SW}}$, that is defined as the zero locus of
\begin{align*}
	F(X,Y,Z) = (X-Z)(X-t_1 Z)Y^2 - u X Z^3
\end{align*}
in $\mathbb{CP}^2$. We want to get the local normalizations near 
\begin{enumerate}[(1)]
	\item $\{p_i \in \mathcal{C}_{\text{SW}}\}$, where $\{\phi(p_i)\}$ are the points we add to $C_{\text{SW}}$ to compactify it,
	\item $\{q_i \in \mathcal{C}_{\text{SW}}|dt(q_i) = 0\} \Leftrightarrow \{q_i \in \mathcal{C}_{\text{SW}}\ |\ (\partial f / \partial v)(t(q_i),v(q_i)) = 0\}$,
	\item $\{r_i \in \mathcal{C}_{\text{SW}}|v(r_i) = 0\}$.
\end{enumerate}
The corresponding points on $\bar{C}_{\text{SW}}$ are	
\begin{align*}
	\sigma(p_1) &= [0,0,1],\\
	\sigma(p_2) = \sigma(p_3) &= [0,1,0],\\
	\sigma(p_4) &= [1,0,0]
\end{align*}
from (1). (2) and (3) do not give us any other candidate.

\begin{enumerate}

\item Near $\sigma(p_1) = [0, 0, 1]$, let's denote a small deviation from $[0,0,1]$ by $[x,y,1]$. Along $\bar{C}_{\text{SW}}$ $x$ and $y$ satisfy
\begin{align}
	F(x,y,1) = (x-1)(x-t_1)y^2 -ux = 0. \label{eq:deviation from [0,0,1]}
\end{align}
From this polynomial we can get the corresponding Newton polygon. Here is how we get one. First we mark a point at $(a,b) \in \mathbb{Z}^2$ if we have in the polynomial a term $x^a y^b$ with nonzero coefficient. We do this for every term in the polynomial and get several points in the $\mathbb{Z}^2$-plane. For instance, the polynomial (\ref{eq:deviation from [0,0,1]}) gives the  points in Figure \ref{figure:SU_2_SCFT_at_0_0_1}, where the horizontal axis corresponds to the exponent of $x$ and the vertical one to that of $y$ for a term that is represented by a point.
\begin{figure}[ht]
	\begin{center}
		\includegraphics{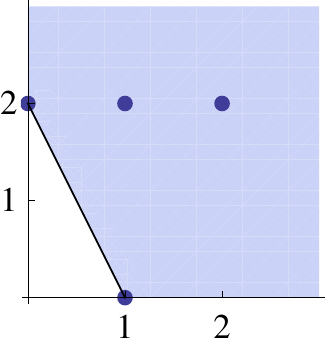}
		\caption{Newton polygon of $F(x,y,1)$}
		\label{figure:SU_2_SCFT_at_0_0_1}
	\end{center}
\end{figure}
Now we connect some of the points with lines so that the lines with the two axes make a polygon that contains all the points and is convex to the origin. This is the Newton polygon of the polynomial.

Using this Newton polygon, we can find Puiseux expansions at $\sigma(p_1)$. Here we will describe just how we can get the Puiseux expansions using the data we have at hand. The underlying principle why this procedure works is illustrated in \cite{Kirwan}, for example. First we pick a line segment that corresponds to the steepest slope and collect the terms corresponding to the points on that edge to make a new polynomial. Then the zero locus of the polynomial is the local representation of $\bar{C}_{\text{SW}}$ near $[0,0,1]$. In this case, the polynomial is
\begin{align*}
	t_1 y^2 - u x.
\end{align*}
The zero locus of this polynomial is an approximation of $\bar{C}_{\text{SW}}$ at $x=y=0$, i.e. the local analytic curve at $[0,0,1]$. We can get a better approximation by including ``higher-order'' terms, but this is enough for now. The solutions of this polynomial,
\begin{align*}
	y(x) = \pm \sqrt{\frac{ux}{t_1}},
\end{align*} 
are the Puiseux expansions of $y$ in $x$ at $x=y=0$. We can see that there are two branches of $y(x)$, that the two branches are coming together at $x=y=0$, and that the monodromy around $x=0$ permutes the two branches with each other. 

To get a local normalization near the point, note that
\begin{align*}
		\sigma_{p_1}: s \mapsto [x, y, 1] = [s^2, a_0 s, 1],\ a_0 = \sqrt{u/t_1}
\end{align*}
maps a neighborhood of $s = 0$ to the two branches. Therefore $\sigma_{p_1}$ is a good local normalization when we consider $s$ as a coordinate patch for $\mathcal{C}_{\text{SW}}$ where $p_1$ is located at $s=0$. 

Now we have a local normalization $\sigma_{p_1}$ near $p_1$. Let's use this to calculate the ramification index $\nu_{p_1}(\pi)$. Remember that the local description of $\pi:\mathcal{C}_{\text{SW}} \to C_B$ near $p_1$ is realized in Section \ref{section:SU(2) SCFT}
as
\begin{align*}
	\pi_{p_1}(s) = \frac{X(s)}{Z(s)}.
\end{align*}
Near $s=0$,
\begin{align*}
	\pi_{p_1}(s) - \pi_{p_1}(0) = \frac{x(s)}{1} - 0 = s^2.
\end{align*}
The exponent of this map is the ramification index at $s=0$. That is, $\nu_{p_1}(\pi) = 2$.

We can also calculate the degree of $(\omega)$ at $p_1$ using the local normalization. Remember that $(\omega)$ is the Seiberg-Witten differential pulled back by $\sigma$ onto $\mathcal{C}_{\text{SW}}$.
\begin{align*}
	\omega = \sigma^*(\lambda) = \sigma^*\left(\frac{Y/Z}{X/Z}d\left(\frac{X}{Z}\right)\right).
\end{align*}
Near $s=0$, this becomes
\begin{align*}
	\omega_{p_1} = \frac{y(s)}{x(s)}d(x(s)) = \frac{a_0 s}{s^2} \cdot d(s^2) = 2 a_0 ds.
\end{align*}
Therefore $\omega$ has neither pole nor zero of any order at $p_1$, which implies $\nu_{p_1}(\omega) = 0$.
\item Near $\sigma(p_2) = \sigma(p_3) = [0, 1, 0]$, let's denote a deviation from $[0, 1, 0]$ by $[x, 1, z]$. Then along $\bar{C}_{\text{SW}}$ $x$ and $z$ satisfy
\begin{align*}
	F(x, 1, z) = (x - z)(x - t_1 z) - u x z^3 =0.
\end{align*}
The Newton polygon of this polynomial is shown in Figure \ref{figure:SU_2_SCFT_at_0_1_0_first}.
\begin{figure}[ht]
	\begin{center}
		\includegraphics{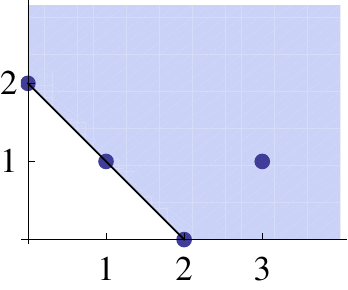}
		\caption{Newton polygon of $F(x,1,z)$}
		\label{figure:SU_2_SCFT_at_0_1_0_first}
	\end{center}
\end{figure}
We collect the terms corresponding to the points on the edge to get a polynomial 
\begin{align*}
	x^2+x z \left(-1-t_1\right)+z^2 t_1 = (x - z)(x - t_1 z),
\end{align*}
whose zero locus is the local analytic curve of $\bar{C}_{\text{SW}}$ at $[0, 1, 0]$. Note that this polynomial is reducible and has two irreducible components. This is the situation described in Figure \ref{figure:normalization_schematic}. Therefore we can see that $[0, 1, 0]$ has two preimages $p_2$ and $p_3$ on $\mathcal{C}_{\text{SW}}$ by $\sigma$. But this local description of the curve is not accurate enough for us to calculate $R_\pi$ or $(\omega)$. To see why this is not enough, let's focus on one of the two components, $x - t_1 z$. This gives us the following local normalization near $p_3$.
\begin{align*}
	\sigma_{p_3}: s \mapsto [x, 1, z] = [t_1 s, 1, s].
\end{align*}
From this normalization we get
\begin{align*}
	\pi_{p_3}(s) &= \frac{x(s)}{z(s)} = t_1,
\end{align*}
which maps the neighborhood of $p_3$ on $\mathcal{C}_{\text{SW}}$ to a single point $t_1$ on $C_B$. Also,
\begin{align*}
	\omega_{p_3} = \frac{1}{x(s)} d\left( \frac{x(s)}{z(s)} \right) = \frac{1}{t_1 s} d(t_1) = 0,
\end{align*}
which does not make sense. The reason for these seemingly inconsistent results is because the local analytic curve we have now is not accurate enough to capture the true nature of $\bar{C}_{\text{SW}}$. Therefore we need to include ``higher-order'' terms of the Puiseux expansion. To do this we first pick one of the two components that we want to improve our approximation. Let's stick with $x - t_1 z$. The idea is to get a better approximation by including more terms of higher order. That is, we add to the previous Puiseux expansion
\begin{align*}
	x(z) = t_1 z
\end{align*}
one more term
\begin{align*}
	x(z) = z (t_1 + x_1(z))
\end{align*}
and then find such $x_1(z)$ that gives us a better approximation of the branch of $\bar{C}_{\text{SW}}$. For that purpose we put this $x(z)$ into $F(x, 1, z)$. Then we get
\begin{align*}
	F(z(t_1 + x_1), 1, z) = z^2 F_1 (x_1, z),
\end{align*}
where we factored out $z^2$ that is the common factor of every term in $F$. Now we draw the Newton polygon of $F_1(x_1,z)$ and do the same job as we have done so far. The Newton polygon is shown in Figure \ref{figure:SU_2_SCFT_at_0_1_0_second}.
\begin{figure}[ht]
	\begin{center}
		\includegraphics{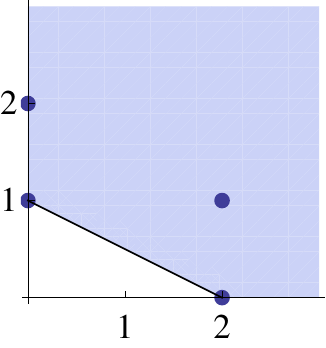}
		\caption{Newton polygon of $F_1(x_1,z)$}
		\label{figure:SU_2_SCFT_at_0_1_0_second}
	\end{center}
\end{figure}
Collecting the terms on the line segment gives
\begin{align*}
	(t_1 - 1) x_1 - u t_1 z^2.
\end{align*}
Setting this to zero gives $x_1(z)$, and by putting it back to $x(z)$, we get
\begin{align*}
	x = z (t_1 + x_1(z)) = t_1 z + \frac{t_1 u}{t_1 - 1} z^3.
\end{align*}
We now have an improved Puiseux expansion. If we want to do even better, we can iterate this process. But, as we will see below, this is enough for us for now, so we will stop here.

For the other irreducible component, $x - z$, we do a similar calculation and get the same Newton polygon and the following Puiseux expansion.
\begin{align*}
	x = z (1 + x_1(z)) = z + \frac{u}{1 - t_1} z^3.
\end{align*}

These expansions give us the following local normalizations
\begin{align*}
	\sigma_{p_i}: s \mapsto [x, 1, z] = [b_0 s + b_1 s^3, 1, s],
\end{align*}
where $b_0$ and $b_1$ are
\begin{align*}
	b_0 = 1,\ b_1 = \frac{u}{1 - t_1}
\end{align*}
at $p_2$ and
\begin{align*}
	b_0 = t_1,\ b_1 = \frac{t_1 u}{t_1 - 1}
\end{align*}
at $p_3$. From each of these local normalizations we get, near each $p_i$,
\begin{align*}
	\pi_{p_i}(s) - \pi_{p_i}(0) = \frac{x(s)}{z(s)} - b_0 \propto s^2 \Rightarrow \nu_{p_2}(\pi) = \nu_{p_3}(\pi) = 2,
\end{align*}
and 
\begin{align*}
	\omega_{p_i}  = \frac{1}{x(s)} d\left(\frac{x(s)}{z(s)}\right) \propto ds \Rightarrow \nu_{p_2}(\omega) = \nu_{p_3}(\omega) = 0.
\end{align*}
\item Next, consider $\sigma(p_4) = [1, 0, 0]$. We start by denoting the deviations from $[1, 0, 0]$ as $[1, y, z]$. Then $y$ and $z$ satisfy
\begin{align*}
	F(1, y, z) = (1-z)(1-t_1 z)y^2 - u z^3 = 0,
\end{align*}
whose Newton polygon is shown in Figure \ref{figure:SU_2_SCFT_at_1_0_0}.
\begin{figure}[ht]
	\begin{center}
		\includegraphics{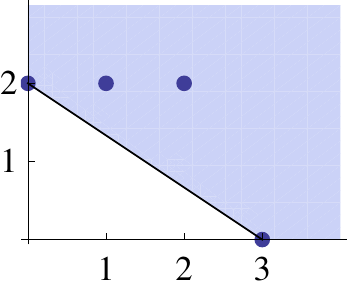}
		\caption{Newton polygon of $F(1,y,z)$}
		\label{figure:SU_2_SCFT_at_1_0_0}
	\end{center}
\end{figure}
This gives us a polynomial
\begin{align*}
	y^2 - u z^3,
\end{align*}
whose zero locus is the local analytic curve of $\bar{C}_{\text{SW}}$ at $[1,0,0]$. The corresponding local normalization is
\begin{align*}
	\sigma_{p_4}: s \mapsto [1, y, z] = [1, c_0 s^3, s^2],\ c_0 = \sqrt{u}.
\end{align*}
Using this local normalization, we get
\begin{align*}
	\frac{1}{\pi_{p_4}(s)} - \frac{1}{\pi_{p_4}(0)} = \frac{z(s)}{1}  - \frac{1}{\infty} \propto s^2 \Rightarrow \nu_{p_4}(\pi) = 2,
\end{align*}
where we took a reciprocal of $\pi_{p_4}(s)$ because $\pi_{p_4}(s=0) = \pi(p_4) = \infty$. And we also find
\begin{align*}
	\omega_{p_4} = y(s) d\left(\frac{1}{z(s)}\right) \propto ds \Rightarrow \nu_{p_4}(\omega) = 0.
\end{align*}

\end{enumerate}

As we have found out in Sections \ref{section:SU(2) SCFT}, for the Seiberg-Witten curve of $SU(2)$ SCFT, $\{p_1,\ \ldots,\ p_4\}$ are all the points that we need to investigate. Therefore we have all the local normalizations we need to construct $R_\pi$ and $\omega$. From the results of this subsection, we have
\begin{align*}
	R_\pi = 1 \cdot [p_1] + 1 \cdot [p_2] + 1 \cdot [p_3] + 1 \cdot [p_4]
\end{align*}
and
\begin{align*}
	(\omega)=0.
\end{align*}


\subsection{$SU(2) \times SU(2)$ SCFT}
\label{appendix:SU(2) times SU(2)}
The corresponding Seiberg-Witten curve $C_{\text{SW}}$ is the zero locus of
\begin{align*}
	f(t,v) = (t-1) (t-t_1) (t-t_2) v^2 - u_1 t^2 - u_2 t.
\end{align*}
We embed this into $\mathbb{CP}^2$ to compactify it to $\bar{C}_{\text{SW}}$, the zero locus of
\begin{align*}
	F(X, Y, Z) = (X - Z) (X - t_1 Z) (X - t_2 Z) Y^2 - u_1 X^2 Z^3 -u_2 X Z^4.
\end{align*}
in $\mathbb{CP}^2$. Now we want to get the local normalizations near
\begin{enumerate}[(1)]
	\item $\{p_i \in \mathcal{C}_{\text{SW}}\}$, where $\{\phi(p_i)\}$ are the points we add to $C_{\text{SW}}$ to compactify it,
	\item $\{q_i \in \mathcal{C}_{\text{SW}}|dt(q_i) = 0\} \Leftrightarrow \{q_i \in \mathcal{C}_{\text{SW}}\ |\ (\partial f / \partial v)(t(q_i),v(q_i)) = 0\}$,
	\item $\{r_i \in \mathcal{C}_{\text{SW}}|v(r_i) = 0\}$.
\end{enumerate}
The corresponding points on $\bar{C}_{\text{SW}}$ are
\begin{align*}
	\sigma(p_1) &= [0, 0, 1],\\
	\sigma(p_2) = \sigma(p_3) = \sigma(p_4) &= [0, 1, 0],\\
	\sigma(p_5) &= [1, 0, 0]
\end{align*}
from (1), and
\begin{align*}
	\sigma(q) = [\rho, 0, 1],\ \rho = -u_2/u_1
\end{align*}
from (2). (3) does not give us any other candidate.

\begin{enumerate}
\item Near $\sigma(p_1)=[0,0,1]$, the Newton polygon of $F(x, y, 1)$ is shown in Figure \ref{figure:SU_2_X_SU_2_SCFT_at_0_0_1}.
\begin{figure}[ht]
	\begin{center}
		\includegraphics{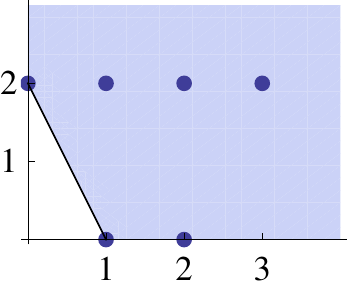}
		\caption{Newton polygon of $F(x, y, 1)$}
		\label{figure:SU_2_X_SU_2_SCFT_at_0_0_1}
	\end{center}
\end{figure}
This gives us a polynomial
\begin{align*}
	t_1 t_2 y^2 + u_2 x,
\end{align*}
whose zero locus is the local analytic curve of $\bar{C}_{\text{SW}}$ at $[0,0,1]$. The local normalization near $p_1$ is
\begin{align*}
	\sigma_{p_1}: s \mapsto [x, y, 1] = [s^2, a_0 s, 1],\ a_0 = \sqrt{-u_2 / (t_1 t_2)},
\end{align*}
from which we can get
\begin{alignat*}{3}
	&\pi_{p_1}(s) - \pi_{p_1}(0) = \frac{x(s)}{1} - 0 \propto s^2 &\ \Rightarrow\ & \nu_{p_1}(\pi) = 2, \\
	&\omega_{p_1} = \frac{y(s)}{x(s)}d(x(s)) \propto ds &\ \Rightarrow\ & \nu_{p_1}(\omega) = 0.
\end{alignat*}
\item Near $\sigma(p_2)=\sigma(p_3)=\sigma(p_4)= [0, 1, 0]$, the Newton polygon of $F(x, 1, z)$ is shown in Figure \ref{figure:SU_2_X_SU_2_SCFT_at_0_1_0_first}.
\begin{figure}[ht]
	\begin{center}
		\includegraphics{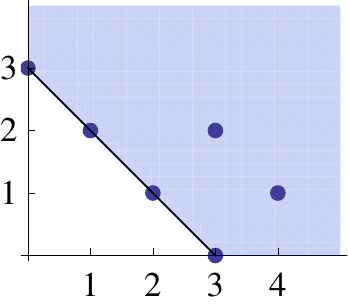}
		\caption{Newton polygon of $F(x, 1, z)$}
		\label{figure:SU_2_X_SU_2_SCFT_at_0_1_0_first}
	\end{center}
\end{figure}
This gives us 
\begin{align*}
	x^3+x^2 z \left(-1-t_1-t_2\right)-z^3 t_1 t_2+x z^2 \left(t_1+t_2+t_1 t_2\right) = (x - z)(x - t_1 z)(x - t_2 z),
\end{align*}
whose zero locus is the local analytic curve of $\bar{C}_{\text{SW}}$ at $[0,1,0]$. We see that it has three irreducible components, and that each component needs a higher-order term to calculate $\nu_{p_i}(\pi)$ and $\nu_{p_i}(\omega)$. We pick a component
\begin{align*}
	x = b_0 z.
\end{align*}
By denoting the higher-order term as $x_1(z)$, now $x(z)$ is
\begin{align*}
	x = z(b_0 + x_1(z)),\ b_0 = 
	\begin{cases}
		1\ \mathrm{at}\ p_2, \\
		t_1\ \mathrm{at}\ p_3, \\
		t_2\ \mathrm{at}\ p_4.
	\end{cases}
\end{align*}
and by putting this back into $F(x, 1, z)$, we get
\begin{align*}
	F(x, 1, z) = z^3 F_1 (x_1, z).
\end{align*}
The Newton polygon of $F_1 (x_1,z)$ is shown in Figure \ref{figure:SU_2_X_SU_2_SCFT_at_0_1_0_second}.
\begin{figure}[ht]
	\begin{center}
		\includegraphics{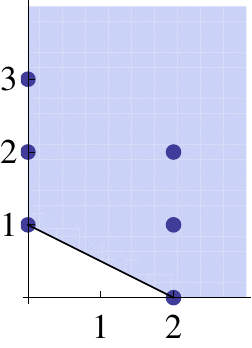}
		\caption{Newton polygon of $F_1 (x_1,z)$}
		\label{figure:SU_2_X_SU_2_SCFT_at_0_1_0_second}
	\end{center}
\end{figure}
This gives us a polynomial
\begin{align*}
	x_1 - b_1 z^2,\ b_1 = 
	\begin{cases}
		\frac{u_1 + u_2}{(1 - t_1)(1 - t_2)}\ \mathrm{at}\ p_2, \\
		\frac{t_1(t_1 u_1 + u_2)}{(t_1-1)(t_1 - t_2)}\ \mathrm{at}\ p_3, \\
		\frac{t_2(t_2 u_1 + u_2)}{(t_2 - 1)(t_2 - t_1)}\ \mathrm{at}\ p_4.
	\end{cases}
\end{align*}
Therefore the Puiseux expansion at each $p_i$ is
\begin{align*}
	x = z(b_0 + x_1(z)) = b_0 z + b_1 z^3.
\end{align*}
The local normalization near each $p_i$ is
\begin{align*}
	\sigma_{p_i}: s \mapsto [x, 1, z] = [b_0 s + b_1 s^3, 1, s],
\end{align*}
from which we can get
\begin{alignat*}{3}
	&\pi_{p_i}(s) - \pi_{p_i}(0) = \frac{x(s)}{z(s)} - b_0 \propto s^2 &\ \Rightarrow\ & \nu_{p_i}(\pi) = 2, \\
	&\omega_{p_i} = \frac{1}{x(s)}d\left(\frac{x(s)}{z(s)}\right) \propto ds &\ \Rightarrow\ & \nu_{p_i}(\omega) = 0.
\end{alignat*}
\item Near $\sigma(p_5)= [1, 0, 0]$, the Newton polygon of $F(1, y, z)$ is shown in Figure \ref{figure:SU_2_X_SU_2_SCFT_at_1_0_0}.
\begin{figure}[ht]
	\begin{center}
		\includegraphics{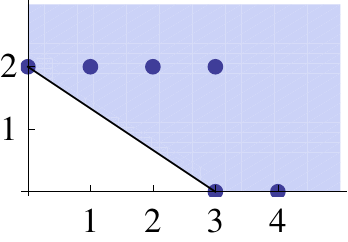}
		\caption{Newton polygon of $F(1, y, z)$}
		\label{figure:SU_2_X_SU_2_SCFT_at_1_0_0}
	\end{center}
\end{figure}
This gives us 
\begin{align*}
	y^2 - u_1 z^3
\end{align*}
as the local analytic curve of $\bar{C}_{\text{SW}}$ at $[1,0,0]$. The local normalization near $p_5$ is
\begin{align*}
	\sigma_{p_5}: s \mapsto [1, y, z] = [1, c_0 s^3, s^2],\ c_0 = \sqrt{u_1},
\end{align*}
from which we can get
\begin{alignat*}{3}
	&\frac{1}{\pi_{p_5}(s)} - \frac{1}{\pi_{p_5}(0)} = \frac{z(s)}{1} - \frac{1}{\infty} \propto s^2 &\ \Rightarrow\ & \nu_{p_5}(\pi) = 2, \\
	&\omega_{p_5} = y(s) d\left(\frac{1}{z(s)}\right) \propto ds &\ \Rightarrow\ & \nu_{p_5}(\omega) = 0.
\end{alignat*}
\item Near $\sigma(q)= [\rho, 0, 1]$, the Newton polygon of $F( \rho + x, y, 1)$ is shown in Figure \ref{figure:SU_2_X_SU_2_SCFT_at_rho_0_1}.
\begin{figure}[ht]
	\begin{center}
		\includegraphics{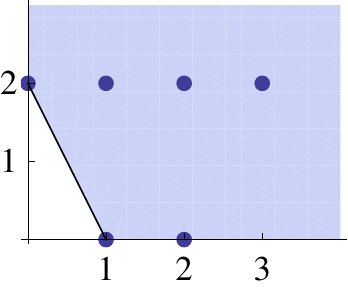}
		\caption{Newton polygon of $F( \rho + x, y, 1)$}
		\label{figure:SU_2_X_SU_2_SCFT_at_rho_0_1}
	\end{center}
\end{figure}
This gives us a polynomial
\begin{align*}
	u_2 x - (\rho - 1)(\rho - t_1)(\rho - t_2) y^2,
\end{align*}
whose zero locus is the local analytic curve of $\bar{C}_{\text{SW}}$ at $[\rho, 0, 1]$. The local normalization near $q$ is
\begin{align*}
	\sigma_q: s \mapsto [\rho + x, y, 1] = [\rho + s^2, d_0 s, 1],\ d_0 = \sqrt{\frac{u_2}{(\rho - 1)(\rho - t_1)(\rho - t_2)}},
\end{align*}
from which we can get
\begin{alignat*}{3}
	&\pi_q(s) - \pi_q(0) = \frac{\rho + x(s)}{1} - \rho \propto s^2 &\ \Rightarrow\ & \nu_{q}(\pi) = 2, \\
	&\omega_q = \frac{d_0 s}{\rho}d(x(s)) \propto s^2 ds &\ \Rightarrow\ & \nu_{q}(\omega) = 2.
\end{alignat*}
\end{enumerate}
From these results we can find out
\begin{gather*}
	R_\pi = 1 \cdot [p_1] + 1 \cdot [p_2] + 1 \cdot [p_3] + 1 \cdot [p_4] + 1 \cdot [p_5] + 1 \cdot [q],\\
	(\omega) = 2 \cdot [q].
\end{gather*}


\subsection{$SU(3)$ SCFT}
\label{appendix:SU(3)}
The Seiberg-Witten curve $C_{\text{SW}}$ is the zero locus of
\begin{align*}
	f(t,v) = (t-1)(t-t_1)v^3 - u_2 t v - u_3 t.
\end{align*}
We embed $C_{\text{SW}}$ into $\mathbb{CP}^2$ to compactify it to $\bar{C}_{\text{SW}}$, which is the zero locus of
\begin{align*}
	F(X, Y, Z) = (X - Z)(X - t_1 Z)Y^3 - u_2 XYZ^3 - u_3 XZ^4
\end{align*}
in $\mathbb{CP}^2$. We want to get the local normalizations near
\begin{enumerate}[(1)]
	\item $\{p_i \in \mathcal{C}_{\text{SW}}\}$, where $\{\phi(p_i)\}$ are the points we add to $C_{\text{SW}}$ to compactify it,
	\item $\{q_i \in \mathcal{C}_{\text{SW}}|dt(q_i) = 0\} \Leftrightarrow \{q_i \in \mathcal{C}_{\text{SW}}\ |\ (\partial f / \partial v)(t(q_i),v(q_i)) = 0\}$,
	\item $\{r_i \in \mathcal{C}_{\text{SW}}|v(r_i) = 0\}$.
\end{enumerate}
The corresponding points on $\bar{C}_{\text{SW}}$ are
\begin{align*}
	\sigma(p_1) &= [0,0,1],\\
	\sigma(p_2) = \sigma(p_3) &= [0,1,0],\\
	\sigma(p_4) &= [1,0,0]
\end{align*}
from (1), and
\begin{gather*}
	\sigma(q_\pm) = [t_\pm, v_0, 1],\ 
	t_\pm = \frac{1 + t_1 + \rho}{2} \pm \sqrt{\left(\frac{1 + t_1 + \rho}{2}\right)^2 - t_1},\ \rho = \frac{(u_2 / 3)^3}{(u_3 / 2)^2},\ v_0 = -\frac{(u_3 / 2)}{(u_2 / 3)}
\end{gather*}
from (2). (3) does not give us any other candidate.

\begin{enumerate}
\item Near $\sigma(p_1)= [0,0,1]$, the Newton polygon of $F(x, y, 1)$ is shown in Figure \ref{figure:SU_3_SCFT_at_0_0_1}.
\begin{figure}[ht]
	\begin{center}
		\includegraphics{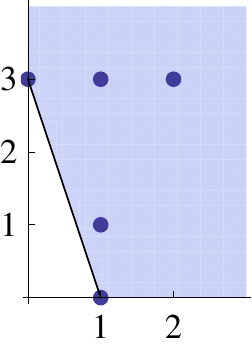}
		\caption{Newton polygon of $F(x,y,1)$}
		\label{figure:SU_3_SCFT_at_0_0_1}
	\end{center}
\end{figure}
This gives us a polynomial
\begin{align*}
	t_1 y^3 - u_3 x,
\end{align*}
whose zero locus is the local analytic curve of $\bar{C}_{\text{SW}}$ at $[0,0,1]$. The local normalization near $p_1$ is
\begin{align*}
	\sigma_{p_1}: s \mapsto [x, y, 1] = [s^3, a_0 s, 1],\ a_0 = \sqrt[3]{u_3 / t_1},
\end{align*}
from which we can get
\begin{alignat*}{3}
	&\pi_{p_1}(s) - \pi_{p_1}(0) = \frac{x(s)}{1} - 0 \propto s^3 &\ \Rightarrow\ & \nu_{p_1}(\pi) = 3, \\
	&\omega_{p_1} = \frac{y(s)}{x(s)}d(x(s)) \propto ds &\ \Rightarrow\ & \nu_{p_1}(\omega) = 0.
\end{alignat*}
\item Near $\sigma(p_2)=\sigma(p_3)= [0,1,0]$, the Newton polygon of $F(x, 1, z)$ is shown in Figure \ref{figure:SU_3_SCFT_at_0_1_0_first}.
\begin{figure}[ht]
	\begin{center}
		\includegraphics{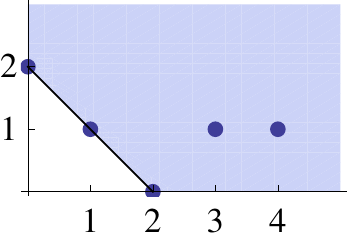}
		\caption{Newton polygon of $F(x,1,z)$}
		\label{figure:SU_3_SCFT_at_0_1_0_first}
	\end{center}
\end{figure}
This gives us 
\begin{align*}
	x^2 - (1 + t_1)xz + t_1 z^2 = (x - z)(x - t_1 z),
\end{align*}
whose zero locus is the local analytic curve of $\bar{C}_{\text{SW}}$ at $[0,1,0]$. We see that it has two irreducible components, and that each component needs a higher-order term to describe $\bar{C}_{\text{SW}}$ up to the accuracy to calculate $\nu_{p_1}(\pi)$ and $\nu_{p_1}(\omega)$. We pick a component
\begin{align*}
	x = b_0 z,\ b_0 = 
	\begin{cases}
		1\ \mathrm{at}\ p_2, \\
		t_1\ \mathrm{at}\ p_3.
	\end{cases}
\end{align*}
By denoting the higher-order term as $x_1(z)$, now $x(z)$ is
\begin{align*}
	x = z(b_0 + x_1(z)),
\end{align*}
and by putting this back into $F(x, 1, z)$, we get
\begin{align*}
	F(x, 1, z) = z^2 F_1 (x_1, z).
\end{align*}
The Newton polygon of $F_1 (x_1,z)$ is shown in Figure \ref{figure:SU_3_SCFT_at_0_1_0_second}.
\begin{figure}[ht]
	\begin{center}
		\includegraphics{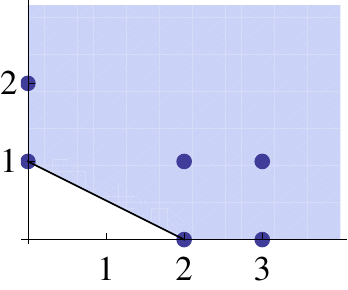}
		\caption{Newton polygon of $F_1 (x_1,z)$}
		\label{figure:SU_3_SCFT_at_0_1_0_second}
	\end{center}
\end{figure}
This gives us a polynomial
\begin{align*}
	x_1 - b_1 z^2,\ b_1 = 
	\begin{cases}
		\frac{u_2}{1-t_1}\ \mathrm{at}\ p_2, \\
		\frac{t_1 u_2}{t_1 - 1}\ \mathrm{at}\ p_3. 
	\end{cases}
\end{align*}
Therefore the Puiseux expansion at each $p_i$ is
\begin{align*}
	x = z(b_0 + x_1(z)) = b_0 z + b_1 z^3.
\end{align*}
The local normalization near each $p_i$ is
\begin{align*}
	\sigma_{p_i}: s \mapsto [x, 1, z] = [b_0 s + b_1 s^3, 1, s],
\end{align*}
from which we can get
\begin{alignat*}{3}
	&\pi_{p_i}(s) - \pi_{p_i}(0) = \frac{x(s)}{z(s)} - b_0 \propto s^2 &\ \Rightarrow\ & \nu_{p_i}(\pi) = 2, \\
	&\omega_{p_i} = \frac{1}{x(s)}d\left(\frac{x(s)}{z(s)}\right) \propto ds &\ \Rightarrow\ & \nu_{p_i}(\omega) = 0.
\end{alignat*}
\item Near $\sigma(p_4)= [1,0,0]$, the Newton polygon of $F(1, y, z)$ is shown in Figure \ref{figure:SU_3_SCFT_at_1_0_0}.
\begin{figure}[ht]
	\begin{center}
		\includegraphics{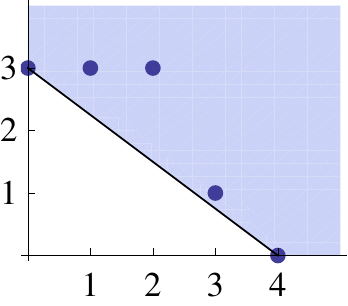}
		\caption{ }
		\label{figure:SU_3_SCFT_at_1_0_0}
	\end{center}
\end{figure}
This gives us 
\begin{align*}
	y^3 - u_3 z^4
\end{align*}
as the local analytic curve of $\bar{C}_{\text{SW}}$ at $[1,0,0]$. The local normalization near $p_4$ is
\begin{align*}
	\sigma_{p_4}: s \mapsto [1, y, z] = [1, c_0 s^4, s^3],\ c_0 = \sqrt[3]{u_3},
\end{align*}
from which we can get
\begin{alignat*}{3}
	&\frac{1}{\pi_{p_4}(s)} - \frac{1}{\pi_{p_4}(0)} = \frac{z(s)}{1} - \frac{1}{\infty} \propto s^3 &\ \Rightarrow\ & \nu_{p_4}(\pi) = 3, \\
	&\omega_{p_4} = y(s) d\left(\frac{1}{z(s)}\right) \propto ds &\ \Rightarrow\ & \nu_{p_4}(\omega) = 0.
\end{alignat*}
\item Near $\sigma(q_\pm)= [t_\pm, v_0, 1]$, the Newton polygon of $F(t_\pm + x, v_0 + y, 1)$ is shown in Figure \ref{figure:SU_3_SCFT_at_tp_v0_1}.
\begin{figure}[ht]
	\begin{center}
		\includegraphics{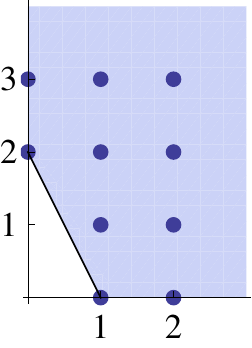}
		\caption{Newton polygon of $F(t_\pm + x, v_0 + y, 1)$}
		\label{figure:SU_3_SCFT_at_tp_v0_1}
	\end{center}
\end{figure}
This gives us a polynomial
\begin{align*}
	\frac{1}{\rho}\left(\frac{1+t_1+\rho}{2} - t_\pm\right)x - \left(\frac{3 t_\pm}{2v_0^2}\right) y^2,
\end{align*}
whose zero locus is the local analytic curve of $\bar{C}_{\text{SW}}$ at $[t_\pm, v_0, 1]$. The local normalization near $q_\pm$ is
\begin{gather*}
	\sigma_{q_\pm}: s \mapsto [t_\pm + x, v_0 + y, 1] = [t_\pm + s^2, v_0 + d_0 s, 1],\ 
	d_0 = v_0 \sqrt{\frac{2}{3 \rho}\left(\frac{1+t_1+\rho}{2 t_\pm} - 1 \right)},
\end{gather*}
from which we can get
\begin{alignat*}{3}
	&\pi_{q_\pm}(s) - \pi_{q_\pm}(0) = \frac{t_\pm + x(s)}{1} - t_\pm \propto s^2 &\ \Rightarrow\ & \nu_{q_\pm}(\pi) = 2, \\
	&\omega_{q_\pm} = \frac{v_0}{t_\pm}d(x(s)) \propto s ds &\ \Rightarrow\ & \nu_{q_\pm}(\omega) = 1.
\end{alignat*}
\end{enumerate}
From these results we get
\begin{gather*}
	R_\pi = 2\cdot[p_1] + 1\cdot[p_2] + 1\cdot[p_3] + 2\cdot[p_4] + 1\cdot[q_+] + 1\cdot[q_-], \\
	(\omega) = 1 \cdot [q_+] + 1 \cdot [q_-].
\end{gather*}

\subsection{$SU(3)$ pure gauge theory}
\label{appendix:SU(3) pure gauge theory}
The Seiberg-Witten curve $C_{\text{SW}}$ is the zero locus of
\begin{align*}
	f(t,v) = t^2 + (v^3 - u_2 v - u_3) t + \Lambda^6.
\end{align*} 
To avoid cluttered notations, let's rescale the variables in the following way:
\begin{align}
	\frac{t}{\Lambda^3} \to t,\ \frac{v}{\Lambda} \to v,\ \frac{u_k}{\Lambda^k} \to u_k. 
\end{align}
It is easy to restore the scale if needed, just reversing the direction of the rescaling. Then the equation that we start the usual analysis with is
\begin{align*}
	f(t,v) = t^2 + (v^3 - u_2 v - u_3) t + 1 = tv^3 - u_2 t v + (t^2 -u_3 t +1)
\end{align*}
whose zero locus defines $C_{\text{SW}}$. We embed $C_{\text{SW}}$ into $\mathbb{CP}^2$ to compactify it to $\bar{C}_{\text{SW}}$, the zero locus of
\begin{align*}
	F(X, Y, Z) = XY^3 -u_2 XYZ^2 + (X^2Z^2 - u_3 XZ^3 + Z^4).
\end{align*}
in $\mathbb{CP}^2$. We want to get the local normalizations near
\begin{enumerate}[(1)]
	\item $\{p_i \in \mathcal{C}_{\text{SW}}\}$, where $\{\phi(p_i)\}$ are the points we add to $C_{\text{SW}}$ to compactify it,
	\item $\{q_i \in \mathcal{C}_{\text{SW}}|dt(q_i) = 0\} \Leftrightarrow \{q_i \in \mathcal{C}_{\text{SW}}\ |\ (\partial f / \partial v)(t(q_i),v(q_i)) = 0\}$,
	\item $\{r_i \in \mathcal{C}_{\text{SW}}|v(r_i) = 0\}$.
\end{enumerate}
The corresponding points on $\bar{C}_{\text{SW}}$ are
\begin{align*}
	\sigma(p_1) &= [0, 1, 0],\\
	\sigma(p_2) &= [1, 0, 0]
\end{align*}
from(1),
\begin{align*}
	\sigma(q_{ab}) = [t_{2ab}, v_{2a}, 1],\ a,b = \pm1,\ t_{2ab} = \left( {v_{2a}}^3 + \frac{u_3}{2} \right) + b \sqrt{\left( {v_{2a}}^3 + \frac{u_3}{2} \right)^2 - 1},\ v_{2a} = a\sqrt{\frac{u_2}{3}}.
\end{align*}
from(2), and
\begin{align*}
	\sigma(r_{\pm}) = [t_{3\pm}, 0, 1],\ t_{3\pm}= \frac{u_3}{2} \pm \sqrt{\left(\frac{u_3}{2}\right)^2 - 1}
\end{align*}
from(3).

\begin{enumerate}
\item Near $\sigma(p_1)= [0, 1, 0]$, the Newton polygon of $F(x, 1, z)$ is shown in Figure \ref{figure:SU_3_no_matter_at_0_1_0}.
\begin{figure}[ht]
	\begin{center}
		\includegraphics{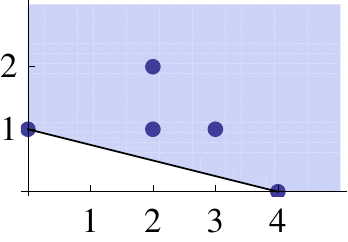}
		\caption{Newton polygon of $F(x, 1, z)$}
		\label{figure:SU_3_no_matter_at_0_1_0}
	\end{center}
\end{figure}
This gives us a polynomial
\begin{align*}
	x + z^4,
\end{align*}
whose zero locus is the local analytic curve of $\bar{C}_{\text{SW}}$ at $[0, 1, 0]$. The local normalization near $p_1$ is
\begin{align*}
	\sigma_{p_1}: s \mapsto [x, 1, z] = [-s^4, 1, s],
\end{align*}
from which we can get
\begin{alignat*}{3}
	&\pi_{p_1}(s) - \pi_{p_1}(0) = \frac{x(s)}{z(s)} - 0 \propto s^3 &\ \Rightarrow\ & \nu_{p_1}(\pi) = 3, \\
	&\omega_{p_1} = \frac{1}{x(s)}d\left(\frac{x(s)}{z(s)}\right) \propto \frac{ds}{s^2} &\ \Rightarrow\ & \nu_{p_1}(\omega) = -2.
\end{alignat*}
\item Near $p_2= [1, 0, 0]$, the Newton polygon of $F(1, y, z)$ is shown in Figure \ref{figure:SU_3_no_matter_at_1_0_0}.
\begin{figure}[ht]
	\begin{center}
		\includegraphics{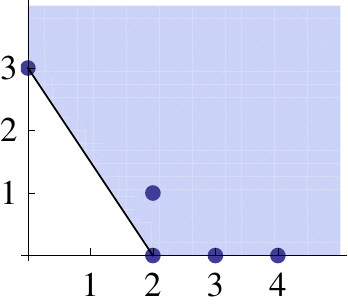}
		\caption{Newton polygon of $F(1, y, z)$}
		\label{figure:SU_3_no_matter_at_1_0_0}
	\end{center}
\end{figure}
This gives us 
\begin{align*}
	y^3 + z^2
\end{align*}
as the local analytic curve of $\bar{C}_{\text{SW}}$ at $[1, 0, 0]$. The local normalization near $p_2$ is
\begin{align*}
	\sigma_{p_2}: s \mapsto [1, y, z] = [1, -s^2, s^3],
\end{align*}
from which we can get
\begin{alignat*}{3}
	&\frac{1}{\pi_{p_2}(s)} - \frac{1}{\pi_{p_2}(0)} = \frac{z(s)}{1} - \frac{1}{\infty} \propto s^3 &\ \Rightarrow\ & \nu_{p_2}(\pi) = 3, \\
	&\omega_{p_2} = y(s) d\left(\frac{1}{z(s)}\right) \propto \frac{ds}{s^2} &\ \Rightarrow\ & \nu_{p_2}(\omega) = -2.
\end{alignat*}
\item Near $q_{ab}= [t_{2ab}, v_{2a}, 1]$, the Newton polygon of $F(t_{2ab} + x, v_{2a} + y, 1)$ is shown in Figure \ref{figure:SU_3_no_matter_at_t2ab_v2a_1}.
\begin{figure}[ht]
	\begin{center}
		\includegraphics{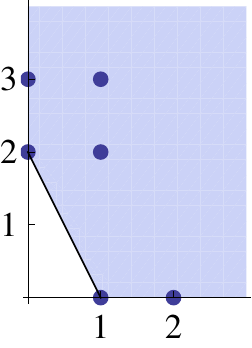}
		\caption{Newton polygon of $F(t_{2ab} + x, v_{2a} + y, 1)$}
		\label{figure:SU_3_no_matter_at_t2ab_v2a_1}
	\end{center}
\end{figure}
This gives us a polynomial
\begin{align*}
	\left(2b \sqrt{\left( v_{2a}^3 + \frac{u_3}{2} \right)^2-1} \right) x + 3 v_{2a} t_{2ab} y^2,
\end{align*}
whose zero locus is the local analytic curve of $\bar{C}_{\text{SW}}$ at $[t_{2ab}, v_{2a}, 1]$. The local normalization near $q_{ab}$ is
\begin{gather*}
	\sigma_{q_{ab}}: s \mapsto [t_{2ab} + x, v_{2a} + y, 1] = [t_{2ab} + s^2, v_{2a} + c_0 s, 1],\\ 
	c_0 = \sqrt{-\frac{2b}{3 v_{2a} t_{2ab}} \sqrt{\left( v_{2a}^3 + \frac{u_3}{2} \right)^2-1}}.
\end{gather*}
from which we can get
\begin{alignat*}{3}
	&\pi_{q_{ab}}(s) - \pi_{q_{ab}}(0) = \frac{t_{2ab} + x(s)}{1} - t_{2ab} \propto s^2 &\ \Rightarrow\ & \nu_{q_{ab}}(\pi) = 2, \\
	&\omega_{q_{ab}} = \frac{v_{2a}}{t_{2ab}}d(x(s)) \propto s ds &\ \Rightarrow\ & \nu_{q_{ab}}(\omega) = 1.
\end{alignat*}
\item Near $r_{\pm}= [t_{3\pm}, 0, 1]$, the Newton polygon of $F(t_{3\pm} + x, y, 1)$ is shown in Figure \ref{figure:SU_3_no_matter_at_t3a_0_1}.
\begin{figure}[ht]
	\begin{center}
		\includegraphics{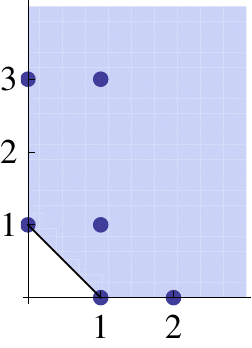}
		\caption{Newton polygon of $F(t_{3\pm} + x, y, 1)$}
		\label{figure:SU_3_no_matter_at_t3a_0_1}
	\end{center}
\end{figure}
This gives us a polynomial
\begin{align*}
	2\left(t_{3\pm} - \frac{u_3}{2}\right)x-u_2 t_{3\pm}y,
\end{align*}
whose zero locus is the local analytic curve of $\bar{C}_{\text{SW}}$ at $[t_{3\pm}, 0, 1]$. The local normalization near $r_{\pm}$ is
\begin{align*}
	\sigma_{r_{\pm}}: s \mapsto [t_{3\pm} + x, y, 1] = [t_{3\pm} + s, d_0 s, 1],\ d_0 = \frac{1}{u_2}\left(2 - \frac{u_3}{t_{3\pm}}\right).
\end{align*}
from which we can get
\begin{alignat*}{3}
	&\pi_{r_{\pm}}(s) - \pi_{r_{\pm}}(0) = \frac{t_{3\pm} + x(s)}{1} - t_{3\pm} \propto s &\ \Rightarrow\ & \nu_{r_\pm}(\pi) = 1, \\
	&\omega_{r_{\pm}} = \frac{y(s)}{t_{3\pm}}d(x(s)) \propto s ds &\ \Rightarrow\ & \nu_{r_{\pm}}(\omega) = 1.
\end{alignat*}

\end{enumerate}

From these results we can find out
\begin{gather*}
	R_\pi = 2 \cdot [p_1] + 2 \cdot [p_2] + 1 \cdot [q_{++}] + 1 \cdot [q_{+-}] + 1 \cdot [q_{-+}] + 1 \cdot [q_{--}],\\
	(\omega) = -2 \cdot [p_1] -2 \cdot [p_2] + 1 \cdot [q_{++}] + 1 \cdot [q_{+-}] + 1 \cdot [q_{-+}] + 1 \cdot [q_{--}] + 1 \cdot [r_+] + 1 \cdot [r_-].
\end{gather*}

%% file: Ramification.bbl
\providecommand{\href}[2]{#2}\begingroup\raggedright\begin{thebibliography}{10}

\bibitem{Witten:1997sc}
E.~Witten, {\it {Solutions of four-dimensional field theories via M- theory}},
  {\em Nucl. Phys.} {\bf B500} (1997) 3--42,
  [\href{http://xxx.lanl.gov/abs/hep-th/9703166}{{\tt hep-th/9703166}}].

\bibitem{Seiberg:1994rs}
N.~Seiberg and E.~Witten, {\it {Monopole Condensation, And Confinement In N=2
  Supersymmetric Yang-Mills Theory}},  {\em Nucl. Phys.} {\bf B426} (1994)
  19--52, [\href{http://xxx.lanl.gov/abs/hep-th/9407087}{{\tt
  hep-th/9407087}}].

\bibitem{Seiberg:1994aj}
N.~Seiberg and E.~Witten, {\it {Monopoles, duality and chiral symmetry breaking
  in N=2 supersymmetric QCD}},  {\em Nucl. Phys.} {\bf B431} (1994) 484--550,
  [\href{http://xxx.lanl.gov/abs/hep-th/9408099}{{\tt hep-th/9408099}}].

\bibitem{Gaiotto:2009we}
D.~Gaiotto, {\it {N=2 dualities}},
  \href{http://xxx.lanl.gov/abs/0904.2715}{{\tt arXiv:0904.2715}}.

\bibitem{Argyres:2007cn}
P.~C. Argyres and N.~Seiberg, {\it {S-duality in N=2 supersymmetric gauge
  theories}},  {\em JHEP} {\bf 12} (2007) 088,
  [\href{http://xxx.lanl.gov/abs/0711.0054}{{\tt arXiv:0711.0054}}].

\bibitem{Argyres:1995jj}
P.~C. Argyres and M.~R. Douglas, {\it {New phenomena in SU(3) supersymmetric
  gauge theory}},  {\em Nucl. Phys.} {\bf B448} (1995) 93--126,
  [\href{http://xxx.lanl.gov/abs/hep-th/9505062}{{\tt hep-th/9505062}}].

\bibitem{Kirwan}
F.~Kirwan, {\em Complex Algebraic Curves}.
\newblock Cambridge University Press, 1992.

\bibitem{Griffiths}
P.~A. Griffiths, {\em Introduction to Algebraic Curves}.
\newblock American Mathematical Society, 1989.

\bibitem{Gaiotto:2009hg}
D.~Gaiotto, G.~W. Moore, and A.~Neitzke, {\it {Wall-crossing, Hitchin Systems,
  and the WKB Approximation}},  \href{http://xxx.lanl.gov/abs/0907.3987}{{\tt
  arXiv:0907.3987}}.

\bibitem{Fayyazuddin:1997by}
A.~Fayyazuddin and M.~Spalinski, {\it {The Seiberg-Witten differential from
  M-theory}},  {\em Nucl. Phys.} {\bf B508} (1997) 219--228,
  [\href{http://xxx.lanl.gov/abs/hep-th/9706087}{{\tt hep-th/9706087}}].

\bibitem{Henningson:1997hy}
M.~Henningson and P.~Yi, {\it {Four-dimensional BPS-spectra via M-theory}},
  {\em Phys. Rev.} {\bf D57} (1998) 1291--1298,
  [\href{http://xxx.lanl.gov/abs/hep-th/9707251}{{\tt hep-th/9707251}}].

\bibitem{Mikhailov:1997jv}
A.~Mikhailov, {\it {BPS states and minimal surfaces}},  {\em Nucl. Phys.} {\bf
  B533} (1998) 243--274, [\href{http://xxx.lanl.gov/abs/hep-th/9708068}{{\tt
  hep-th/9708068}}].

\bibitem{Hollowood:1997pp}
T.~J. Hollowood, {\it {Strong coupling N = 2 gauge theory with arbitrary gauge
  group}},  {\em Adv. Theor. Math. Phys.} {\bf 2} (1998) 335--355,
  [\href{http://xxx.lanl.gov/abs/hep-th/9710073}{{\tt hep-th/9710073}}].

\bibitem{Argyres:1995xn}
P.~C. Argyres, M.~R. Plesser, N.~Seiberg, and E.~Witten, {\it {New N=2
  Superconformal Field Theories in Four Dimensions}},  {\em Nucl. Phys.} {\bf
  B461} (1996) 71--84, [\href{http://xxx.lanl.gov/abs/hep-th/9511154}{{\tt
  hep-th/9511154}}].

\bibitem{Argyres:1994xh}
P.~C. Argyres and A.~E. Faraggi, {\it {The vacuum structure and spectrum of N=2
  supersymmetric SU(n) gauge theory}},  {\em Phys. Rev. Lett.} {\bf 74} (1995)
  3931--3934, [\href{http://xxx.lanl.gov/abs/hep-th/9411057}{{\tt
  hep-th/9411057}}].

\bibitem{Klemm:1994qs}
A.~Klemm, W.~Lerche, S.~Yankielowicz, and S.~Theisen, {\it {Simple
  singularities and N=2 supersymmetric Yang-Mills theory}},  {\em Phys. Lett.}
  {\bf B344} (1995) 169--175,
  [\href{http://xxx.lanl.gov/abs/hep-th/9411048}{{\tt hep-th/9411048}}].

\bibitem{Argyres:1995wt}
P.~C. Argyres, M.~R. Plesser, and A.~D. Shapere, {\it {The Coulomb phase of N=2
  supersymmetric QCD}},  {\em Phys. Rev. Lett.} {\bf 75} (1995) 1699--1702,
  [\href{http://xxx.lanl.gov/abs/hep-th/9505100}{{\tt hep-th/9505100}}].

\bibitem{Hanany:1995na}
A.~Hanany and Y.~Oz, {\it {On the Quantum Moduli Space of Vacua of $N=2$
  Supersymmetric $SU(N_c)$ Gauge Theories}},  {\em Nucl. Phys.} {\bf B452}
  (1995) 283--312, [\href{http://xxx.lanl.gov/abs/hep-th/9505075}{{\tt
  hep-th/9505075}}].

\bibitem{Danielsson:1995is}
U.~H. Danielsson and B.~Sundborg, {\it {The Moduli space and monodromies of N=2
  supersymmetric SO(2r+1) Yang-Mills theory}},  {\em Phys. Lett.} {\bf B358}
  (1995) 273--280, [\href{http://xxx.lanl.gov/abs/hep-th/9504102}{{\tt
  hep-th/9504102}}].

\bibitem{Brandhuber:1995zp}
A.~Brandhuber and K.~Landsteiner, {\it {On the monodromies of N=2
  supersymmetric Yang-Mills theory with gauge group SO(2n)}},  {\em Phys.
  Lett.} {\bf B358} (1995) 73--80,
  [\href{http://xxx.lanl.gov/abs/hep-th/9507008}{{\tt hep-th/9507008}}].

\bibitem{Argyres:1995fw}
P.~C. Argyres and A.~D. Shapere, {\it {The Vacuum Structure of N=2 SuperQCD
  with Classical Gauge Groups}},  {\em Nucl. Phys.} {\bf B461} (1996) 437--459,
  [\href{http://xxx.lanl.gov/abs/hep-th/9509175}{{\tt hep-th/9509175}}].

\bibitem{Martinec:1995by}
E.~J. Martinec and N.~P. Warner, {\it {Integrable systems and supersymmetric
  gauge theory}},  {\em Nucl. Phys.} {\bf B459} (1996) 97--112,
  [\href{http://xxx.lanl.gov/abs/hep-th/9509161}{{\tt hep-th/9509161}}].

\bibitem{Klemm:1996bj}
A.~Klemm, W.~Lerche, P.~Mayr, C.~Vafa, and N.~P. Warner, {\it {Self-Dual
  Strings and N=2 Supersymmetric Field Theory}},  {\em Nucl. Phys.} {\bf B477}
  (1996) 746--766, [\href{http://xxx.lanl.gov/abs/hep-th/9604034}{{\tt
  hep-th/9604034}}].

\end{thebibliography}\endgroup
